\documentclass{article}                     
\usepackage{graphicx}
\usepackage[dvips]{color}
\usepackage{graphicx}
\usepackage{amsmath}
\usepackage{amssymb}
\usepackage{amsfonts}
\def\beq{\begin{equation}}
\def\eeq{\end{equation}}
\def\beqa{\begin{eqnarray}}
\def\eeqa{\end{eqnarray}}

\def\X{{\sf X}}

\def\A{{\sf A}}

\begin{document}

\title{Spin in relativistic quantum theory\thanks{During the
preparation of this paper Walter Gl\"ockle passed away.  
We dedicate this paper to Walter, who was a great friend and
collaborator.}
}



\author{W. N. Polyzou, \\
Department of Physics and Astronomy,
The University of Iowa, \\
Iowa City, IA 52242 \\
W. Gl\"ockle, \\
Institut f\"ur theoretische Physik II,
Ruhr-Universit\"at Bochum, \\
D-44780 Bochum, Germany\\
H. Wita{\l}a\\
M. Smoluchowski Institute of Physics, Jagiellonian
University, \\
PL-30059 Krak\'ow, Poland\\
}

\date{today}

\maketitle

\begin{abstract}
We discuss the role of spin in Poincar\'e invariant formulations of
quantum mechanics.
\end{abstract}


\section{Introduction}
\label{intro}

In this paper we discuss the role of spin in relativistic few-body
models.  The new feature with spin in relativistic quantum mechanics
is that sequences of rotationless Lorentz transformations that map
rest frames to rest frames can generate rotations.  To get a
well-defined spin observable one needs to define it in one preferred
frame and then use specific Lorentz transformations to relate the spin
in the preferred frame to any other frame.  Different choices of the
preferred frame and the specific Lorentz transformations lead to an
infinite number of possible observables that have all of the
properties of a spin.  This paper provides a general discussion of the
relation between the spin of a system and the spin of its elementary
constituents in relativistic few-body systems.

In section \ref{section2} we discuss the Poincar\'e group, which is the group
relating inertial frames in special relativity.  Unitary
representations of the Poincar\'e group preserve quantum probabilities
in all inertial frames, and define the relativistic dynamics.  In
section \ref{section3} we construct a large set of abstract operators out of the
Poincar\'e generators, and determine their commutation relations and
transformation properties.  In section \ref{section4} 
we identify complete sets of
commuting observables, including a large class of spin operators.  We
use the Poincar\'e commutation relations to determine the eigenvalue
spectrum of these operators.  Representations of the physical Hilbert
space are constructed as square integrable functions of these
commuting observables over their spectra.   The transformation properties
of these operators are used to construct unitary representations of the 
Poincar\'e group and its infinitesimal generators on this space.
This construction gives irreducible representations of the 
Poincar\'e group.

The commuting observables introduced in section \ref{section4} include
a large class of spin observables.  All of these operators are
functions of the Poincar\'e generators, are Hermetian, satisfy $SU(2)$
commutation relations, commute with the four momentum, and have a
square that is the spin Casimir operator of the Poincar\'e group.  In
section \ref{section5} we discuss the most important examples of spin
operators and how they are related.  These are the helicity,
canonical, and light-front spins.  In section \ref{section6} we
discuss the problem of adding angular momenta; specifically how
single-particle spins and orbital angular momenta are added in
relativistic systems to obtain the total spin of the system. This is
the problem of constructing Clebsch-Gordan coefficients for the
Poincar\'e group.  We consider the implications doing this using
different spin and orbital observables.  We show that the coupling is,
up to some overall rotations on the initial and final spins,
independent of the spin observables used in the coupling.  In section
\ref{section7} we discuss the relation between two and four component
spinors and their relation to field theory.  We show in general how
boosts transform Poincar\'e covariant spinors to Lorentz covariant
spinors, and how this role is played by the $u$ and $v$ Dirac spinors
in the spin $1/2$ case.  In section \ref{section8} we argue that there
is no loss of generality in working with models with a non-interacting
spin (Bakamjian-Thomas models) by showing that any model is related to
a Bakamjian-Thomas model \cite{Bakamjian:1953kh} by an $S$-matrix
preserving unitary transformation.  In section \ref{section9} we
consider aspects of the relativistic three-nucleon problem.  We show
how relativistic invariance can be realized by requiring invariance
with respect to rotations in the rest frame of the three-nucleon
system. We also argue that $S$-matrix cluster properties is an
important additional constraint on the treatment of the spin.  Finally
in section \ref{section10} we discus the relation between the
different types of spins and experimental observables.

\section{The Poincar\'e group}
\label{section2}

The Poincar\'e group is the group of space-time transformations that
relate different inertial frames in the theory of special relativity.
In a relativistically invariant quantum theory the Poincar\'e group is
a symmetry group of the theory \cite{Wigner:1939cj}.

The Poincar\'e group is the group of point transformations that
preserve the proper time, $\tau_{ab}$, or proper distance, $d_{ab}$,
between any two events with space-time coordinates $x_a^{\mu}$ and
$x_b^{\mu}$,
\beq
-\tau_{ab}^2 = d_{ab}^2 =
(x_a-x_b)^{\mu} (x_a-x_b)^{\nu}\eta_{\mu \nu} = (x_a-x_b)^2, 
\label{b.1}
\eeq
where $\eta_{\alpha \beta}$ is the Minkowski metric with signature
$(-,+,+,+)$ and repeated
4-vector indices are assumed to be summed from 0 to 3.

The most general point transformation, $x^{\mu}\to x^{\prime\mu} = f^{\mu}(x)$,
satisfying (\ref{b.1}) has the form
\beq
x^{\mu} \to x^{\mu'} = \Lambda^{\mu}{}_{\nu} x^{\nu} + a^{\mu},
\label{b.2}
\eeq
where $a^{\mu}$ is a constant $4$-vector, 
$x^{\mu}$ is $x_a^{\mu}$ or $x^{\mu}_b$ and
the Lorentz transformation, $\Lambda^{\mu}{}_{\nu}$, is a
constant matrix satisfying
\begin{equation}
\eta_{\mu \nu} \Lambda^{\mu}{}_{\alpha} \Lambda^{\nu}{}_{\beta}
=\eta_{\alpha \beta}.
\label{b.3}
\end{equation}

The full Poincar\'e group includes Lorentz transformations,
$\Lambda^{\mu}{}_{\nu}$, that are not continuously connected to the
identity.  These transformations involve discrete space reflections,
time reversals, or both.  Since time reversal and space reflection are
not symmetries of the weak interactions, the symmetry group associated
with special relativity is the subgroup of Poincar\'e transformations
that is continuously connected to the identity.  This subgroup
contains the active transformations that can be experimentally realized.
In this paper the term Poincar\'e group refers to
this subgroup.

It is sometimes useful to represent Poincar\'e transformations using
the group of complex $2\times 2$ matrices with unit determinant
\cite{wightman_2},
$SL(2,\mathbb{C})$.  In this representation real four vectors are
represented by $2\times 2$ Hermetian matrices.  A basis for the $2
\times 2$ Hermetian matrices (over the real numbers) 
are the identity and the three Pauli spin
matrices
\beq
\sigma_{\mu} := (I, \sigma_1,\sigma_2, \sigma_3).
\label{b.4}
\eeq
There is a 1-1 correspondence between real four vectors and $2 \times 2$
Hermetian matrices given by
\beq
\X := x^{\mu} \sigma_{\mu} =
\left (
\begin{array}{cc}
x^0 + x^3 & x^1 -i x^2 \\
x^1 + i x^2 & x^0 - x^3 \\
\end{array} 
\right )
\qquad x^{\mu} = {\frac {1} {2}} \mbox{Tr}(\sigma_{\mu}\X).
\label{b.5}
\eeq
The determinant of $\X$ is the square of the proper time of the vector
\beq
\tau^2 = \mbox{det}(\X) = -\eta_{\mu\nu} x^{\mu}x^{\nu} .
\label{b.6}
\eeq
The most general linear transformation that preserves  both the
Hermiticity and the determinant of $\X$ has the form
\beq
\X \to \X' = \Lambda \X \Lambda^{\dagger}
\label{b.7}
\eeq
where $\Lambda$ are complex $2 \times 2$ matrices with
$\mbox{det}(\Lambda ) =1$.  We have used the notation $\Lambda$ for
these $2\times 2$ matrices because they are related to the $4\times 4$
Lorentz transformation $\Lambda^{\mu}{}_{\nu}$ by
\beq
\Lambda^{\mu}{}_{\nu} = {\frac {1} {2}} \mbox{Tr} (\sigma_{\mu} \Lambda \sigma_{\nu}
\Lambda^{\dagger}).
\label{b.8}
\eeq
This is a 2 to 1 correspondence because both $\Lambda$ and $-\Lambda$
result in the same $\Lambda^{\mu}{}_{\nu}$ in (\ref{b.8}).  This
relation between $SL(2,\mathbb{C})$ and the Lorentz group is the same
2 to 1 correspondence that one has in relating $SU(2)$ rotations to
$SO(3)$ rotations.  It emerges when the $SL(2,\mathbb{C})$ matrices
are restricted to the $SU(2)$ subgroup.  In the $2\times 2$ matrix
representation a Poincar\'e transformation has the form
\beq
\X \to \X' = \Lambda \X \Lambda^{\dagger} + \A \qquad \A=\A^{\dagger} 
\label{b.9}
\eeq
where 
\beq
\A= a^{\mu} \sigma_{\mu} \qquad a^{\mu} = \frac{1}{2}\mbox{Tr}(\A\sigma_{\mu}).
\label{b.10}
\eeq

Elements of the Poincar\'e group are 
pairs $(\Lambda^{\mu}{}_{\nu},a^{\mu})$ or equivalently 
$(\Lambda ,A)$.  The group product is  
\beq
(\Lambda_2^{\mu}{}_{\nu},a_2^{\mu})(\Lambda_1^{\mu}{}_{\nu},a_1^{\mu})
=
(\Lambda_2^{\mu}{}_{\alpha}\Lambda_1^{\alpha}{}_{\nu}  ,\Lambda_2^{\mu}{}_{\nu}
a_1^{\nu} + a_2^{\mu})
\label{b.11}
\eeq
or equivalently
\beq
(\Lambda_2,\A_2) (\Lambda_1,\A_1) = (\Lambda_2 \Lambda_1,\Lambda_2 
\A_1\Lambda_2^{\dagger}  + \A_2).
\label{b.12}
\eeq
The identity is $(\eta^{\mu}{}_{\nu},0)$ or $(I,0)$, and the inverse is
\beq
((\Lambda^{-1})^{\mu}{}_{\nu}  ,- (\Lambda^{-1})^{\mu}{}_{\nu} a^{\nu})
\label{b.13}
\eeq 
or
\beq
(\Lambda^{-1} ,- \Lambda^{-1} \A \Lambda^{\dagger -1}) .
\label{b.14}
\eeq

The most general $2 \times 2$ matrix with unit
determinant has the form
\beq
\Lambda (\mathbf{z})  = e^{\mathbf{z} \cdot \pmb{\sigma}} = \cosh (z) + 
\frac{1}{z} \sinh (z)\mathbf{z} \cdot \pmb{\sigma}  ~,
\label{b.15}
\eeq
where $\mathbf{z}$ is a complex 3-vector and $z = \sqrt{ \sum_k
z_k^2}$ is a complex scalar.  The branch of the square root does not
matter because both $\frac{1}{z} \sinh (z)$ and $\cosh (z)$ are 
even in $z$.
This representation can be understood by noting that the $\sigma_{\mu}$
are a basis (over the complex numbers) 
for all complex matrices and $\det (e^{z^{\mu}
  \sigma_{\mu}}) = e^{z^{\mu} \mbox{Tr} (\sigma_{\mu})} = e^{2z^0}$ which is
$1$ for $z^0=0$.

It is easy to see that the matrices $\Lambda (\mathbf{z})$
correspond to Lorentz transformations that are continuously connected to 
the identity because
\beq
\Lambda (\lambda\mathbf{z} ) = e^{\lambda \mathbf{z} \cdot \pmb{\sigma}}
\label{b.16}
\eeq
is a Lorentz transformation for all $\lambda$ and it continuously
approaches the identity as $\lambda$ varies between $1$ and $0$.  The
$SO(3,1)$ Lorentz transformation constructed by using (\ref{b.16}) in
(\ref{b.8}) is also continuously connected to the identity.

When $\mathbf{z}=\pmb{\rho}/2$ is a real vector, $\Lambda
(\pmb{\rho}/2)$ is a positive matrix (Hermetian with positive
eigenvalues).  It corresponds to a rotationless Lorentz transformation
in the direction $\hat{\pmb{\rho}}$ with rapidity $\rho$:
\beq
e^{\mathbf{z} \cdot \pmb{\sigma}} \to
e^{{\frac {1} {2}} \pmb{\rho} \cdot \pmb{\sigma}}
= \cosh (\rho/2)
+ \hat{\pmb{\rho}}\cdot \pmb{\sigma}
\sinh (\rho/2).
\label{b.17}
\eeq
Using Eq. (\ref{b.17}) in (\ref{b.8}) leads to the four-vector form of this 
transformation
\begin{eqnarray}
\mathbf{x}' &=& \mathbf{x} + \hat{\pmb{\rho}} \sinh (\rho) x^0 + 
\hat{\pmb{\rho}} ( \cosh (\rho) -1) ( \hat{\pmb{\rho}} \cdot \mathbf{x})\cr
x^{0'} &=& \cosh (\rho) x^0 + \sinh (\rho) 
( \hat{\pmb{\rho}} \cdot \mathbf{x})\cr
\rho &= &  \sqrt{ \pmb{\rho}^2}. 
\label{b.18}
\end{eqnarray}
If we make the identifications $\mathbf{p}/m = \hat{\pmb{\rho}}\sinh
(\rho) = \gamma \pmb{\beta}$ and $p^0/m = \cosh (\rho)= \gamma$ this
Lorentz transformation can also be parameterized by the transformed
momentum of a mass $m$ particle initially at rest as
\beq
(\Lambda_c(p))^{\mu}{}_{\nu} = 
\left (
\begin{array}{cc}
{\frac {p^0} {m}} & {\frac {\mathbf{p}} {m}} \\
{\frac {\mathbf{p}} {m}} & \delta_{ij} + {\frac {p_i p_j}  {m(m+p^0)}} \\
\end{array}
\right ).
\label{b.19}
\eeq
Equations (\ref{b.17},\ref{b.18}) and (\ref{b.19}) are different ways
of parameterizing the same Lorentz transformation.  We refer to this
transformation as a rotationless or canonical Lorentz boost, hence the
subscript $c$.

When $\mathbf{z}= i \pmb{\theta}/2$ is an imaginary
vector, $\Lambda (\mathbf{z})$ is unitary and corresponds to a rotation
about the $\hat{\pmb{\theta}}$ axis through an angle $\theta$
\beq
e^{\mathbf{z} \cdot \pmb{\sigma}} \to
e^{{\frac {i}  {2}} \pmb{\theta} \cdot \pmb{\sigma}}
= \cos (\theta/2)
+ i \hat{\pmb{\theta}} \cdot \pmb{\sigma}
\sin (\theta/2).
\label{b.20}
\eeq
Again using Eq.(\ref{b.20}) in (\ref{b.8}) leads to
\beq
\mathbf{x}'  =  \cos (\theta) \mathbf{x} + \sin ( \theta) ( \mathbf{x} \times 
\hat{\pmb{\theta}}) + (1- \cos (\theta)) \hat{\pmb{\theta}} 
( \hat{\pmb{\theta}} \cdot \mathbf{x}).
\label{b.21}
\eeq
A rotation about any axis by $2 \pi$ transforms $\Lambda (\mathbf{z})$
to $-\Lambda (\mathbf{z})$ which corresponds to the same
$\Lambda^{\mu}{}_{\nu}$ in (\ref{b.8}).

Since any $SL(2,\mathbb{C})$ matrix $A$ has a polar decomposition: 
\beq
A= PU \qquad P=(AA^{\dagger})^{1/2} \qquad U = (AA^{\dagger})^{-1/2}A   
\label{b.22}
\eeq
or
\beq
A= U'P' \qquad P'=(A^{\dagger}A)^{1/2} \qquad U' = A (A^{\dagger}A)^{-1/2}   
\label{b.23}
\eeq
into the product of a positive Hermetian matrix $P$ and a unitary matrix $U$, 
every Lorentz transformation can be decomposed into the product of a
canonical boost and a rotation, in either order.  The boost and 
rotation are the matrices $P(P')$ and $U(U')$, respectively, 
in (\ref{b.22}-\ref{b.23}).  

In what follows we use the notation ${\cal P}$ to refer to both the
group of Poincar\'e transformations connected to the identity and the
group inhomogeneous $SL(2,\mathbb{C})$.  ${\cal P}$ is a ten parameter
group; six parameters are needed to fix the complex 3-vector
$\mathbf{z}$ in $\Lambda (\mathbf{z})$, and four additional parameters
are needed to fix $a^{\mu}$.

One important property of the group $SL(2,\mathbb{C})$ that is
relevant for the treatment of spin in Lorentz covariant theories is
that $\Lambda(\mathbf{z})$ and $\Lambda (\mathbf{z})^*$ are
inequivalent representations of $SL(2,\mathbb{C})$, which means that
there are no constant matrices $C$ satisfying
\beq
C \Lambda(\mathbf{z}) C^{-1}= \Lambda(\mathbf{z})^*
\label{b.24}
\eeq
for all $\mathbf{z}$. This is distinct from the subgroup $SU(2)$
(rotations) where for $\Lambda = R \in \, SU(2)$ 
\beq
\sigma_2 R \sigma_2 = R^* .
\label{b.25}
\eeq
This observation is related to the appearance of four-component
spinors in Lorentz covariant theories.  The reason for this is that 
\beq
\sigma_2 \X^* \sigma_2 =
\left (
\begin{array}{cc}
x^0 - x^3 & -x^1 + i x^2 \\
-x^1 -i x^2 & x^0 + x^3 \\
\end{array} 
\right )
\label{b.26}
\eeq
represents space reflection in the 2 $\times$ 2 matrix representation
(\ref{b.5}).  Because space reflection (\ref{b.26}) involves both a
similarity transformation and a complex conjugation, the space
reflected vector transforms under an inequivalent complex conjugate
representation of $SL(2,\mathbb{C})$.  In order to realize space
reflection as a linear transformation in Lorentz covariant theories it
is necessary to double the dimension of the representation space by
including the direct sum of a space that transforms with the complex
conjugate representation of $SL(2,\mathbb{C})$.  These considerations
do not apply to Poincar\'e covariant representations because the 
little group, $SU(2)$, is equivalent (\ref{b.25}) to its conjugate 
representation.  This will be discussed in section \ref{section6}.

To show (\ref{b.25}) note
\beq 
\sigma_2 R \sigma_2 = \sigma_2 e^{{\frac {i}  {2}} \pmb{\theta} \cdot
  \pmb{\sigma}} 
\sigma_2 =  e^{{\frac {i}  {2}} \pmb{\theta} \cdot \sigma_2 \pmb{\sigma}\sigma_2  } =
 e^{-{\frac {i}  {2}} \pmb{\theta} \cdot \pmb{\sigma}^*  } = R^* .
\label{b.27}
\eeq
To prove that no matrix $C$ satisfying (\ref{b.24}) exists, assume by
contradiction that there is such a matrix.  Let $\mathbf{z}=
\mathbf{x}$ be real and then let $\mathbf{z}= i\mathbf{y}$ be
imaginary.  Differentiating both sides of equation (\ref{b.24}) with respect to 
$x_i$ and then $y_i$ and setting $z_i=x_i=0$ or $z_i=y_i=0$  gives
\beq 
C \sigma_i C^{-1} =  \sigma_i^* \qquad \mbox{and} \qquad C 
\sigma_i C^{-1} = - \sigma_i^*.
\label{b.28}
\eeq
Adding these equations gives $C \sigma_i C^{-1} = 0$ which is
impossible for product of three invertible matrices, contradicting the
assumed existence of such a $C$.
  
\section{Operators}
\label{section3}

Wigner showed that the Poincar\'e symmetry in a quantum theory is
realized by a unitary ray representation, $U(\Lambda ,a)$, of the
Poincar\'e group.  Bargmann \cite{Bargmann:1954gh} showed that the ray
representation can be replaced by a single-valued unitary
representation of inhomogeneous $SL(2,\mathbb{C})$, satisfying 
\beq
U(\Lambda_2,\A_2) U(\Lambda_1 ,\A_1)=
U(\Lambda_2\Lambda_1, \Lambda_2 \A_1\Lambda_2^{\dagger} + \A_2) .
\label{c.1}
\eeq
The infinitesimal generators of this representation form a source of an
irreducible set of operators on the model Hilbert space.

Because the Poincar\'e group is a ten parameter group there are ten
independent unitary one-parameter groups associated with space
translations (3), time translations (1), rotations (3), and
rotationless Lorentz transformations (3).   The ten parameters can
be chosen as the space-time translation parameters $a^{\mu}$, three
angles of rotation and the rapidities $\rho$ in three independent 
directions.  One can see  from the inhomogeneous 
$SL(2,\mathbb{C})$ representation (\ref{b.12}) that these
define one-parameter groups: 
\beq
(I,A_2)(I,A_1) = (I,A_1+A_2)
\label{c.2}
\eeq
\beq
(e^{i {\frac {\theta_2}  {2}}\hat{\mathbf{e}} \cdot \pmb{\sigma}},0) 
(e^{i {\frac {\theta_1}  {2}}\hat{\mathbf{e}} \cdot \pmb{\sigma}},0) 
 =
(e^{i {\frac {\theta_2+\theta_1}  {2}}\hat{\mathbf{e}} \cdot \pmb{\sigma}},0) 
\label{c.3}
\eeq
\beq
(e^{ {\frac {\rho_2}  {2}}\hat{\mathbf{e}} \cdot \pmb{\sigma}},0) 
(e^{ {\frac {\rho_1}  {2}}\hat{\mathbf{e}} \cdot \pmb{\sigma}},0) 
 =
(e^{ {\frac {\rho_2+\rho_1}  {2}}\hat{\mathbf{e}} \cdot \pmb{\sigma}},0) .
\label{c.4}
\eeq
The unitary representation of these one-parameter groups have
self-adjoint infinitesimal generators $\mathbf{P}$, $H$, $\mathbf{J}$, and 
$\mathbf{K}$, that can be obtained by differentiating with respect to
the appropriate parameter.  Equivalently, the one-parameter groups can
be expressed directly as exponentials of these generators:
\beq 
U[I,(0,\lambda
\hat{\mathbf{a}})] = e^{-i\lambda \hat{\mathbf{a}}\cdot \mathbf{P}}
\label{c.5}
\eeq
\beq
U[I,(a^0,\mathbf{0})] = e^{i \mathbf{a}^0\cdot H}
\label{c.6}
\eeq
\beq
U[\Lambda (i \lambda \hat{\pmb{\theta}}/2) ,0] 
= e^{i \lambda \hat{\pmb{\theta}} \cdot \mathbf{J}}
\label{c.7}
\eeq
\beq
U[\Lambda (\lambda \hat{\pmb{\rho}}/2) ,0] 
= e^{i \lambda \hat{\pmb{\rho}} \cdot \mathbf{K}}.
\label{c.8}
\eeq
These generators are the linear momentum operators, $\mathbf{P}$,
Hamiltonian, $H$, angular momentum operators, $\mathbf{J}$, and 
rotationless Lorentz
boost generators, $\mathbf{K}$.  These designations follow from the commutation 
relations which show that both the linear momentum 
$\mathbf{P}$ and angular momentum $\mathbf{J}$ commute with $H$ and
are thus conserved.

The commutation relations and transformation properties of the generators 
follow from the group representation property.  To construct the 
commutator of the generators of the one-parameter groups 
$g_2(\lambda_2)$ and $g_1(\lambda_1)$ with parameters 
$\lambda_2$ and $\lambda_1$ use the group representation 
property to express the product as:
\beq
U^{\dagger} [g_1(\lambda_1)]
U[g_2(\lambda_2)]
U[g_1(\lambda_1)] = U[g_1(-\lambda_1) g_2(\lambda_2) g_1(\lambda_1)]  ~.
\label{c.9}
\eeq
Taking the second derivative ${\frac {\partial^2}  {\partial
\lambda_1 \partial \lambda_2}}$ of this expression and setting
$\lambda_1=\lambda_2=0$ gives the commutator of the generators of the
unitary one parameter groups $U [g_2(\lambda_2)]$ and $U
[g_1(\lambda_1)]$.  For example, to calculate the commutator of $P^i$
with $K^j$ use the group representation property to express the
product of the three transformations as a single transformation:
\beq
U (\Lambda ,0) U (I,(0,\mathbf{a})) U (\Lambda^{-1}  ,0)=  U (I , \Lambda a) .
\label{c.10}
\eeq
The commutator between $P^i$ and $K^j$ can be determined by 
considering infinitesimal transformations 
\beq
U (\Lambda ,0) \to e^{i \pmb{\rho}\cdot \mathbf{K} } = 
I + i \pmb{\rho}\cdot \mathbf{K} + \cdots 
\label{c.11}
\eeq
\beq
U (I,(0,\mathbf{a})) \to e^{ -i \mathbf{a}\cdot \mathbf{P}} = 
I  -i \mathbf{a}\cdot \mathbf{P} + \cdots
\label{c.12}
\eeq
\beq
U (\Lambda^{-1} ,0) \to e^{-i \pmb{\rho}\cdot \mathbf{K} } = 
I - i \pmb{\rho}\cdot \mathbf{K} + \cdots 
\label{c.13}
\eeq
\beq
\Lambda a = 
( \sinh (\rho) ( \hat{\pmb{\rho}} \cdot \mathbf{a}),
\mathbf{a}  + \hat{\pmb{\rho}} ( \cosh (\rho) -1) 
( \hat{\pmb{\rho}} \cdot \mathbf{a}) ) \to 
(0,\mathbf{a}) +  (\rho (\hat{\pmb{\rho}}\cdot \mathbf{a}),
 \mathbf{0} ) + \cdots .
\label{c.14}
\eeq
To compute the commutator between $P^i$ and $K^j$ expand both 
sides of (\ref{c.10}) using (\ref{c.11}-\ref{c.14}) to leading 
order in $\pmb{\rho}$ and $\mathbf{a}$
\[
U (\Lambda ,0) U (I,(0,\mathbf{a})) U (\Lambda^{-1}  ,0)  =
I -i \mathbf{a} \cdot \mathbf{P} - (i)^2 a^i \rho^j (K^j P^i - P^i K^j) +\cdots =
\]
\beq
U(I,\Lambda a) = I  -i \mathbf{a} \cdot \mathbf{P}
+i \rho^j \delta_{ji} a^i H  + \cdots . 
\label{c.15}
\eeq
Equating the coefficient of $a^i\rho^j$ gives  
\beq
[K^j, P^i] = i \delta_{ij} H .
\label{c.16}
\eeq
The 45 commutation relations involving all ten generators can be computed
using different pairs of unitary one-parameter groups. 

The group representation property (\ref{c.1}) 
also implies transformation properties 
of the infinitesimal generators.  For example if we set $\Lambda'=I$ 
the group representation properties give
\[
U(\Lambda ,a) U(\Lambda',a')U (\Lambda^{-1}  ,-\Lambda^{-1}a)
=
U(\Lambda \Lambda' \Lambda^{-1}, \Lambda a' \Lambda^{\dagger}
-\Lambda \Lambda' \Lambda^{-1} \Lambda'^{\dagger} a 
+a)  
\]
\beq
U(\Lambda ,a) U(I,a')U (\Lambda^{-1}  ,-\Lambda^{-1}a)
=
U(I, \Lambda a' \Lambda^{\dagger} - a 
+a).
\label{c.17}
\eeq
The parameter $a'$ only appears in the translation generators on 
both sides of (\ref{c.17}).
Differentiating with respect to $a_{\mu}'$ and setting $a_{\mu}'=0$
gives
\beq
U(\Lambda,a)  (H,\mathbf{P})^{\mu}  U^{\dagger}(\Lambda ,a) = 
(\Lambda^{-1})^{\mu}{}_{\nu} (H,\mathbf{P})^{\nu} =
(H,\mathbf{P})^{\nu} \Lambda_{\nu}{}^{\mu} .
\label{c.18}
\eeq
It shows that the generators $H$ and $\mathbf{P}$ transform like
components of a four-vector under Lorentz transformations. 
This four vector is the four momentum:
\beq
P^{\mu} = (H,\mathbf{P}).
\label{c.19}
\eeq
Similarly, letting $U(\Lambda' ,a')\to U(\Lambda',0)$ in the first
line of (\ref{c.17}) and differentiating with respect to angle or
rapidity shows that the six Lorentz generators transform as an
antisymmetric tensor operator 
\beq
J^{\mu \nu} =
\left (
\begin{array}{cccc}
0 & -K^x & -K^y & -K^z \\
K^x & 0 & J^z & -J^y \\
K^y & -J^z & 0 &  J^x \\
K^z & J^y & -J^x & 0 \\
\end{array}
\right ).
\label{c.20}
\eeq
The transformation properties of these operators can be compactly summarized 
by the covariant forms of the transformation laws 
\beq
U(\Lambda ,a) P^{\mu} U^{\dagger}(\Lambda ,a) = P^{\nu} \Lambda_{\nu}{}^{\mu}
\label{c.21}
\eeq
\beq
U(\Lambda ,a) J^{\mu \nu} U^{\dagger}(\Lambda ,a) =
(J^{\alpha \beta}- a^\alpha P^\beta + a^\beta P^\alpha)
\Lambda_{\alpha}{}^{\mu} \Lambda_{\beta}{}^{\nu}
\label{c.22}
\eeq
and the commutation relations for the infinitesimal
generators
\beq
[J^{\mu \nu},J^{\alpha \beta}] =
i (\eta^{\mu \alpha} J^{\nu \beta}
-   \eta^{\nu \alpha} J^{\mu \beta}
+ \eta^{\nu \beta} J^{\mu \alpha}
- \eta^{\mu \beta} J^{\nu \alpha})
\label{c.23}
\eeq
\beq
[P^{\mu} ,J^{\alpha \beta}] = i (\eta^{\mu\beta}P^{\alpha} -
\eta^{\mu\alpha}P^{\beta})
\label{c.24}
\eeq
\beq
[P^{\mu},P^{\nu}]=0 .
\label{c.25}
\eeq
The spin is associated with another four vector that is a quadratic polynomial
in the generators, called the Pauli-Lubanski vector 
\cite{lubanski}, defined by
\beq
W^{\mu} = -{\frac {1}  {2}} \epsilon^{\mu \nu \alpha \beta}P_{\nu} J_{\alpha \beta}.
\label{c.26}
\eeq
The commutation relations
\beq
[ W^{\mu} , W^{\nu} ] =  i \epsilon^{\mu \nu \alpha \beta} W_{\alpha} P_\beta
\label{c.27}
\eeq
\beq
[W^{\mu},P^{\nu} ]=0
\label{c.28}
\eeq
follow from (\ref{c.23}-\ref{c.25}).

The Poincar\'e Lie-algebra has two independent polynomial invariants
\cite{Wigner:1939cj} which are the square of the invariant mass (rest
energy) of the system,
\beq
M^2 = - \eta_{\mu \nu} P^{\mu} P^{\nu},
\label{c.29}
\eeq
and the square of the Pauli-Lubanski vector
\beq
W^2 = \eta_{\mu \nu} W^{\mu} W^{\nu} .
\label{c.30}
\eeq
When $M^2 \not= 0$ the spin is related to the invariants $M^2$ and 
$W^2$ by
\beq
\mathbf{j}^2 = W^2/M^2 .
\label{c.31}
\eeq
For massive systems ($M>0$) the invariant
$W^2$ is replaced by the spin $\mathbf{j}^2$
of the system.  The operators are invariant because they commute with all
of the Poincar\'e generators.

\section{Spin and irreducible representations} 
\label{section4}

Wigner \cite{Wigner:1939cj} classified the unitary irreducible
representations of the Poincar\'e group.  His classification was based
on the observation that Lorentz transformations can be used to
transform an arbitrary four vector to one of six standard forms.  This
divides the set of four vectors into six disjoint Lorentz invariant
equivalence classes.  These familiar equivalence classes are time-like
positive time, time-like negative time, light-like positive time,
light-like negative time, space-like, and zero.  Every four vector is a
member of one of these six classes.  Standard vectors, which are
arbitrary but fixed vectors in each class,
are given in Table~\ref{tab1}.  For each standard vector there is a little
group, which is the subgroup of the Lorentz group that leaves that
standard vector invariant.  The little groups for each of the 
standard vectors are given in Table~\ref{tab1}. 
\begin{center}
\begin{table}
\caption{ The little groups for each of the standard vectors.}
\label{tab1}
\vskip 2pt
\begin{tabular}{|c|c|c|}
\hline\noalign{\smallskip}
class & standard vector & little group \\
\noalign{\smallskip}\hline\noalign{\smallskip}
$P^2 = -M^2 <0 ; P^0>0$ & $p_s^{\mu} = (M,0,0,0)$ & SO(3) \\
$P^2 = -M^2 <0 ; P^0<0$ & $p_s^{\mu} = (-M,0,0,0)$ & SO(3) \\
$P^2 =  0 ; P^0>0$ & $p_s^{\mu} = (1,0,0,1)$ & E(2) \\
$P^2 =  0  ; P^0<0$ & $p_s^{\mu} = (-1,0,0,1)$ & E(2) \\
$P^2 =  -M^2 = N^2 >0$  & $p_s^{\mu} = (0,0,0,N)$ & SO(2,1) \\
$P^{\mu}=0$ & $p_s^{\mu} = (0,0,0,0)$ & SO(3,1) \\
\noalign{\smallskip}\hline
\end{tabular}
\end{table}
\end{center}
In Table~\ref{tab1} $SO(3)$ is the group of rotations in three dimensions,
$E(2)$ is the Euclidean group in two dimensions, $SO(2,1)$ is the Lorentz
group in $2+1$ dimensions, and $SO(3,1)$ is the Lorentz group in $3+1$
dimensions.  Irreducible representations of the little groups are used
as labels for the irreducible representation of the Poincar\'e group.
The treatment of each of the six little groups is different and 
is not relevant to our treatment of particle spins.  The interested 
reader is referred to Wigner's original paper \cite{Wigner:1939cj}.

For particles the relevant four vector is the particle's four momentum,
which is a time-like positive-energy four vector for massive particles
or a light-like positive-energy four vector for massless particles.

For a particle of mass $m>0$ the most natural choice for the standard
vector $p_s$ is the rest four momentum $p_s=p_0 =(m,\mathbf{0})$.  The
little group for $p_0$ is the rotation group.  If a particle at rest
is observed in a rotated frame, the particle remains at rest but the
spin of the particle will be rotated relative to the spin observed in the
original frame.  The particle's spin degrees of freedom are associated
with irreducible representations of $SO(3)$, the little group that
leaves $p_s=p_0$  unchanged.

The treatment of spin in relativistic quantum mechanics is slightly
more complicated than it is in non-relativistic quantum mechanics.
The relevant complication is because the commutator of two different
rotationless Lorentz boost generators, $[K^k,K^l] = -i
\epsilon^{klm}J^m$, gives a rotation generator.  This means the
sequences of rotationless Lorentz boosts can generate rotations.  If
we define the spin of a particle to be the spin measured in the
particle's rest frame, then its spin seen by an observer in any other
frame will depend on both the momentum of the particle in the
transformed frame and the specific Lorentz transformation relating the
two frames. To get an unambiguous definition of a spin observable it
is necessary to specify both the frame where the spin is defined (or
measured) and a set of standard Lorentz transformations relating a
frame where the particle has momentum $\mathbf{p}$ to the frame where
the spin is defined (or measured).  The result is that there are an 
infinite number of possible choices of spin observables in
relativistic quantum mechanics.  Some common spin observables are the
canonical spin, the light-front spin, and the helicity.  While all of
the spins that we will consider satisfy $SU(2)$ commutation relations,
the most useful choices are characterized by different simplifying
properties.  The different spin observables are related 
by momentum-dependent
rotations.
 
In this section we discuss the general structure of spin operators in
Poincar\'e invariant quantum mechanics.  We will define spin operators as
operator-valued functions of the infinitesimal generators.  We 
begin by assuming that we are given a fixed standard vector $p_s$ 
with $p_s^2=-m^2$ and $p_s^0>0$.  The
standard vector does not have to be the rest vector, $p_0$. We also assume 
that we are given a
parameterized set of Lorentz transformations, $\Lambda_s(p)$, that transform
the standard vector, $p_s$,  to any other four vector $p$ with $p^2=-m^2$,
\beq
\Lambda_s(p)^{\mu}{}_{\nu}  p_s^{\nu} =p^{\mu} . 
\label{d.1}
\eeq
The choice of $\Lambda_s(p)$ and $p_s$ are arbitrary,  
subject to the constraints $p_s^2=-m^2$, $p_s^0>0$ and (\ref{d.1}). 

For example, one possible choice is $p_s=p_0$ and $\Lambda _s(p)=
\Lambda_c(p)$, the rotationless Lorentz transformation
(\ref{b.19}) with rapidity $\pmb{\rho}=\hat{\mathbf{p}} \sinh^{-1}
(\frac{\vert \mathbf{p}\vert }{m})$, that transforms $p_0$ to $p$.

The next step is to make $\Lambda_s(p)$ into a {\it Lorentz
transformation valued operator} by replacing $p$ in the expression
for $\Lambda_s(p)$ by the four-momentum operator $\underline{P}$ (we use
underlines to indicate operators in this section).  For example, for
$\Lambda_s(p)=\Lambda_c (p)$ given in (\ref{b.19}), the matrix of
operators becomes
\beq
\Lambda_c^{\mu}{}_{\nu} (\underline{P}) :=
\left (
\begin{array}{cccc}
\underline{H}/\underline{M} & \underline{P}_x/\underline{M} & 
\underline{P}_y/\underline{M} & \underline{P}_x/\underline{M} \\
\underline{P}_x/\underline{M} & 1+ {\underline{P}_x\underline{P}_x 
\over \underline{M}(\underline{M}+\underline{H})}  
& {\underline{P}_x\underline{P}_y \over \underline{M}(\underline{M}+
\underline{H})}  &
{\underline{P}_x\underline{P}_z \over \underline{M}
(\underline{M}+\underline{H})} \\
\underline{P}_y/\underline{M} & {\underline{P}_y\underline{P}_x \over 
\underline{M}(\underline{M}+\underline{H})}   &  
1+{\underline{P}_y\underline{P}_y \over \underline{M}
(\underline{M}+\underline{H})}  &
{\underline{P}_y\underline{P}_z \over \underline{M}
(\underline{M}+\underline{H})} \\
\underline{P}_x/\underline{M} & {\underline{P}_z\underline{P}_x 
\over \underline{M}(\underline{M}+\underline{H})}   & 
{\underline{P}_z\underline{P}_y \over 
\underline{M}(\underline{M}+\underline{H})}  &    
1 + {\underline{P}_z\underline{P}_z \over 
\underline{M}(\underline{M}+\underline{H})}
\end{array} 
\right )
\label{d.2}
\eeq

More generally we define 
\beq
\Lambda_{0s}(\underline{P}) =  \Lambda_s(\underline{P}) \Lambda_s^{-1}(p_0).
\label{d.3}
\eeq

Here the first transformation, $\Lambda_s^{-1}(p_0)$, is a constant
matrix that transforms the constant 4-vector $p_0$ to the constant
standard 4-vector $p_s$.  The second transform is a matrix of {\it
  operators} that maps $p_s$ to the operator $\underline{P}$.  The
first matrix is the identity when $p_s=p_0$. The combined
transformation (\ref{d.3}) is still a boost valued operator that
transforms $p_0$ to $P$.  The reason for discussing this more general
case of spins with $p_s \not= p_0$ is that a similar type of spin
arises naturally in composite systems when spins are coupled in the
many-body problem (see (\ref{f.29})).  In the many-body case it is
natural to choose $p_s$ to be the momentum of the particle in the rest
frame of the system rather than in the rest momentum of the particle.
In the many-body case the constant transformation
$\Lambda_s^{-1}(p_0)$ in (\ref{d.3}) is replaced by an operator-valued
transformation that transforms a particle at rest to its momentum in
the rest frame of the system.  This transformation is operator-valued
because the momentum of the particle in the system's rest frame is an
independent variable.  These spin observables have the advantage that
they can be added with ordinary $SU(2)$ Clebsch-Gordan coefficients.
This will be illustrated in section \ref{section6}.

Given the operator $\Lambda^{-1}_{0s}(\underline{P})$ we define 
the $s$-spin operator
\beq
(0,\underline{\mathbf{j}}_s) =
{1 \over \underline{m}} \Lambda_{0s}^{-1}(\underline{P})^{\mu}{}_{\nu} 
\underline{W}^{\nu} =
- {1 \over 2 \underline{m}} \Lambda_{0s}^{-1}(\underline{P})^{\mu}{}_{\nu}
\epsilon^{\nu\alpha \beta \gamma} 
\underline{P}_\alpha \underline{J}_{\beta \gamma}.
\label{d.4}
\eeq
Note that all three components of $\underline{\mathbf{j}}$ are
well-defined Hermetian operators because the mass, all components of
the four momentum, and the Pauli-Lubanski vector are Hermetian and
commute, (\ref{c.25},\ref{c.28}).  The definition (\ref{d.4}) of
$\underline{\mathbf{j}}_s$ depends on both the choice of a standard
vector $p_s$ and a standard boost $\Lambda_s(p)$.

The most familiar choices of $s$ are associated with canonical spin,
helicity, and light-front spin.  For the canonical spin the standard
boost $\Lambda_s(p)= \Lambda_c(p)$ is given by (\ref{b.19}).  For the
helicity the standard boost is $\Lambda_h(p)=
\Lambda_c(p)R(\hat{\mathbf{p}} \leftarrow \hat{\mathbf{z}}) =
R(\hat{\mathbf{p}} \leftarrow \hat{\mathbf{z}})
\Lambda_c(\hat{\mathbf{z}}p) $, where $R(\hat{\mathbf{p}} \leftarrow
\hat{\mathbf{z}})$ is a rotation about the axis
$\hat{\mathbf{z}}\times \hat{\mathbf{p}}$ through an angle $\cos^{-1}
(\hat{\mathbf{z}}\cdot \hat{\mathbf{p}})$.  For the light-front spin 
the $SL(2,\mathbb{C})$ representation of the standard boost is
\beq
\Lambda_f(p) = 
\left (
\begin{array}{cc}
a & 0 \\ 
b+ic & 1/a 
\end{array} 
\right ) =
\left (
\begin{array}{cc}
\sqrt{{p^0+p^3 \over m}} & 0 \\ 
{p_1+ip_2 
\over \sqrt{m(p^0+p^3)}} &
\sqrt{{m \over p^0+p^3 }} \\
\end{array} 
\right ). 
\label{d.5}
\eeq
Each of these choices has simplifying features that are advantageous
for certain problems. These examples will be discussed 
in more detail in the next section.

Note that if the operator $\mathbf{j}_s$ is applied to an eigenstate
of the four momentum with eigenvalue $p_0^{\mu}=(m,0,0,0)$ then on
this state the operator $\Lambda_{0s}(\underline{P})$ becomes the
identity and $\underline{P}_\alpha$ becomes $p_0$.  It follows that
$\mathbf{j}_s$ becomes $j_s^i = {1 \over 2} \epsilon^{ijk}J_{jk}$
which is the total angular momentum.  This is consistent with the
interpretation of the spin as the angular momentum in the particle's
rest frame.  In the relativistic case different spin operators are
distinguished by the transformation used to get to the particles rest
frame.  These spins are normally identified in the particle's rest
frame. 

The spin (\ref{d.4}) looks like it should be a set of four operators
that transform as the components of a four vector because it has the
form of a four-vector operator, $W^{\mu}$, multiplied by the product
of a Lorentz transformation
$\Lambda_{0s}^{-1}(\underline{P})^{\mu}{}_{\nu}$ and a scalar,
$1/\underline{m}$.  Because of this it may be surprising that it has no
zero-component.  The reason that the spin is {\it not} a four-vector
operator is because $\Lambda_s(\underline{P})$ is a Lorentz transform
valued {\it operator} rather than a Lorentz transformation, so it
corresponds to a different Lorentz transformation for each value of
$p$.  To see explicitly how the zero component vanishes assume that
$\mathbf{j}_s$ acts on an eigenstate of the four momentum.  Then the
four-momentum $\underline{P}$ operators are replaced by the components of the 
four-momentum eigenvalue, ${p}$, including the $\underline{P}$ in the
definition of the Pauli-Lubanski vector:

\beq
(0,\underline{\mathbf{j}}_s) \to
{1 \over m} (\Lambda_{0s}^{-1}(p))^{\mu}{}_{\nu} \underline{W}^{\nu} =
-{1 \over 2 m} \Lambda^{-1}_{0s}(p)^{\mu}{}_{\nu}
\epsilon^{\nu\alpha \beta \gamma} p_\alpha \underline{J}_{\beta
  \gamma} ~.
\label{d.6}
\eeq
Using the shorthand notation  $\Lambda =\Lambda_{0s}^{-1}(p)^{\mu}{}_{\nu}$, 
along with the fact the $\epsilon^{\rho\alpha \beta \gamma}$
is a constant tensor, 
\beq
\epsilon^{\rho\alpha \beta \gamma} = 
\epsilon^{\rho'\alpha' \beta' \gamma'} 
\Lambda_{\rho'}{}^{\rho}
\Lambda_{\alpha'}{}^{\alpha}
\Lambda_{\beta'}{}^{\beta'}
\Lambda_{\gamma'}{}^{\gamma}= 
(\Lambda^{-1})^{\rho}{}_{\rho'}
\epsilon^{\rho'\alpha' \beta' \gamma'} 
\Lambda_{\alpha'}{}^{\alpha}
\Lambda_{\beta'}{}^{\beta'}
\Lambda_{\gamma'}{}^{\gamma}
\label{d.7}
\eeq
or equivalently
\beq
\Lambda^{\mu}{}_{\rho} \epsilon^{\rho\alpha \beta \gamma} = 
\epsilon^{\mu\rho \eta \chi}\Lambda_{\rho}{}^{\alpha}
\Lambda_{\eta}{}^{\beta} \Lambda_{\chi}{}^{\gamma} 
\label{d.8}
\eeq
gives 
\beq
\Lambda^{\mu}{}_{\rho} \epsilon^{\rho\alpha \beta \gamma} 
\underline{P}_\alpha \underline{J}_{\beta \gamma}
=
\epsilon^{\mu\rho \eta \chi}\Lambda_{\rho}{}^{\alpha}
\Lambda_{\eta}{}^{\beta} \Lambda_{\chi}{}^{\gamma} 
\underline{P}_\alpha \underline{J}_{\beta \gamma}.
\label{d.9}
\eeq
Since  $\Lambda_{\rho}{}^{\alpha}p_\alpha = - m\delta_{\rho 0}
= (-m,0,0,0)$ it follows 
that 
\beq
\underline{\Lambda}^{\mu}{}_{\rho} \epsilon^{\rho\alpha \beta \gamma}
 \underline{P}_\alpha 
\underline{J}_{\beta \gamma}
=
\underline{m} \epsilon^{\mu 0  \eta \chi}
\underline{\Lambda}_{\eta}{}^{\beta} \underline{\Lambda}_{\chi}{}^{\gamma} 
\underline{J}_{\beta \gamma}
=
\underline{m} (0, \underline{\mathbf{j}}_s),
\label{d.10}
\eeq
which always has a $0$ time component.  The index $\alpha =0$ 
arises because the only non-vanishing component of the 
transformed four momentum is the zero component.
 From this expression we obtain the following equivalent  
formula for the $s$-spin operator:
\beq
\underline{j}^i_s = {1 \over 2} \epsilon_{ijk} \Lambda_{0s}^{-1}
(\underline{P})^{j}{}_{\mu}
\Lambda^{-1}_{0s}
(\underline{P})^{k}{}_{\nu} \underline{J}^{\mu \nu} .
\label{d.11}
\eeq
This spin observable is interpreted as the angular momentum measured
in the particle's rest frame if the particle is transformed to the rest frame
using the boost $\Lambda^{-1}_{0s} (\underline{P})$.  The index $s$ on
the spin operator indicates that it is one of many possible spin
operators that are functions of the Poincar\'e generators.  
Spin operators associated with different choices, $s$, $t$, of boosts
are related by 
\beq
(0,\underline{\mathbf{j}}_t )^{\mu} = 
\Lambda^{-1}_{0t}
(\underline{P})^{\mu}{}_{\nu}
\Lambda_{0s}
(\underline{P})^{\nu}{}_{\rho} (0,\underline{\mathbf{j}}_s )^{\rho}
=
R_{ts} (\underline{P})^{\mu}{}_{\nu}(0,\underline{\mathbf{j}}_s )^{\mu} 
\label{d.12}
\eeq
where $R_{ts}(\underline{P}) := \Lambda^{-1}_{0t}(\underline{P})
\Lambda_{0s} (\underline{P})$ is a rotation valued function of the
momentum operators.  We refer to rotations that relate different spin
observables as generalized Melosh rotations.  The original Melosh
rotation \cite{Melosh:1974cu} is the corresponding rotation that relates the
light-front and canonical spins.

In what follows we no longer use an underscore to indicate operators.
It follows from (\ref{d.4}) that independent of how the individual
components of $ \vec{j}_s $ are constructed they satisfy $SU(2)$
commutation relations because, using (\ref{c.27}), we find
\[
[\mathbf{j}_s^l,\mathbf{j}_s^m] =
{1 \over m^2} \Lambda_{0s}^{-1}(P)^{l}{}_{\mu}
\Lambda^{-1}_{0s}(P)^{m}{}_{\nu}
[W^{\mu},W^{\nu}] =
\]
\beq
{1 \over m^2} \Lambda_{0s}^{-1}(P)^{l}{}_{\mu}
\Lambda_{0s}^{-1}(P)^{m}{}_{\nu}
i \epsilon^{\mu \nu \alpha \beta } W_\alpha P_\beta . 
\label{d.13}
\eeq
Again, because $\epsilon^{\mu \nu \alpha \beta }$
is a constant tensor this commutator is equal to 
\[
[\mathbf{j}_s^l,\mathbf{j}_s^m]
= {i \over m^2}
\epsilon^{l m \alpha \beta }
\Lambda_{0s}^{-1}(P)_{\alpha}{}^{\mu}
\Lambda_{0s}^{-1}(P)_{\beta}{}^{\nu}
W_\mu P_\nu  =
-i {m \over m^2}
\epsilon^{l m \alpha 0 }
\Lambda_{0s}^{-1}(P)_{\alpha}{}^{\nu} W_\nu =
\]
\beq
{i \over m}
\epsilon^{l m n }
\Lambda_{0s}^{-1}(P)_{n}{}^{\nu} W_\nu =
i
\epsilon^{l m n } {\bf j}_s^n .
\label{d.14}
\eeq
We also have
\beq
\mathbf{j}_s^2 =
\eta_{\alpha \beta} {1 \over m^2} \Lambda_{0s}^{-1}(p)^{\alpha}{}_{\mu}
\Lambda_{0s}^{-1}(p)^{\beta}{}_{\nu}
W^{\mu}W^{\nu} = {1 \over m^2} \eta_{\alpha \beta} W^\alpha W^\beta
= \mathbf{j}^2 .
\label{d.15}
\eeq
Thus, no matter which choice of $p_s$ and $\Lambda_{0s}^{-1}(p)$ 
are used
to define the spin, the components are always Hermetian functions of 
the Poincar\'e generators, commute with the
four momentum, satisfy $SU(2)$ commutation relations, and the
square is always the invariant $\mathbf{j}^2= W^2/M^2$.

There are an infinite number of possible spins depending on how one
chooses $\Lambda_{0s}$ and $p_s$.  Which one is measured in an actual
experiment is determined by how the different spins couple to a
classical electromagnetic field.  If this is known for one type of
spin it is easy to determine the corresponding relations for any other
type of spin.  This will be discussed in section \ref{section8}.

We are now in a position to construct the irreducible representation
spaces that we use to describe the states of massive particles.
In addition to the mass and the square of the spin, three
independent components of the four momentum and one component of any
spin vector, for example $\hat{\mathbf{z}}\cdot \mathbf{j}_{s}$,
define a maximal set of commuting Hermetian functions of the
generators.  The simultaneous measurement of these quantities
also determine the state of a particle of that mass and spin.
Once the spin $j^2$ is fixed, the spectrum of both $j^2$
and $\hat{\mathbf{z}}\cdot \mathbf{j}_{s}$ is fixed by the $SU(2)$
commutation relations.  The $SU(2)$ commutation relations imply that
the eigenvalues $\mu$ of $\hat{\mathbf{z}}\cdot \mathbf{j}_{s}$ range
from $-j$ to $j$ in integer steps while the eigenvalues of $j^2$ are
$j(j+1)$ for $j$ integer or half integer.  The spectrum of the three
space components of the linear momentum are fixed to be
$(-\infty,\infty)$ by the covariance relation (\ref{c.18}). The
subscript $s$ indicates that $\mu$ is an eigenvalue of
$\hat{\mathbf{z}}\cdot \mathbf{j}_s$.  Since $\mathbf{j}^2=
\mathbf{j}_s^2$, the total spin does not depend on the choice of
$\mathbf{j}_s$.

For fixed mass $m$ and spin $j$ we define the mass $m$ spin $j$
irreducible representation space to be the space of square integrable
functions
\beq
\psi (\mathbf{p},\mu) = _s\langle (m,j)\mathbf{p}, \mu \vert \psi \rangle  
\label{d.16}
\eeq
with inner product
\beq
\langle \psi \vert \phi \rangle =
\sum_{\mu=-j}^j \int d\mathbf{p} 
\psi^* (\mathbf{p},\mu) \phi (\mathbf{p},\mu).
\label{d.17}
\eeq

The irreducible basis vectors for this space, 
$\vert (m,j)\mathbf{p}, \mu \rangle_s$, are the simultaneous
eigenstates of $M$, $\mathbf{j}^2$, $\mathbf{p}$, and 
$\hat{\mathbf{z}}\cdot \mathbf{j}_{s}$.  We use the
subscript $s$ on the basis vectors to emphasize that the magnetic 
quantum number $\mu$ is an eigenvalue of $\mathbf{j}_s \cdot 
\hat{\mathbf{z}}$ and $\mathbf{j}_s$ defined 
in (\ref{d.16}) depends on the choice of $p_s$ and
$\Lambda_s (p)$.

To show that this Hilbert space is an irreducible representation space 
for the Poincar\'e group we first calculate the unitary representation 
of the Poincar\'e group on this space.  

We begin by considering the action of the little group on the basis
vectors $\vert (m,j) \mathbf{p}_s, \mu \rangle_s$ when $p$ is the
standard vector $p=p_s$.

When $\mathbf{p}_s \not = \mathbf{0}$ the representation of the
rotation group that leaves $p_s$ invariant is related to the 
standard $SO(3)$ 
representations by a constant boost that acts as a similarity transform:
\beq
R_s =  \Lambda^{-1}_s(p_0) R\Lambda_s(p_0) . 
\label{d.18}
\eeq
We consider the action of $U(R_s,0)$ on vectors of the form  
$\vert (m,j) \mathbf{p}_s, \mu \rangle_s$.  Because $R_s$ is an  
element of the little group, it will not change $p_s$.
The result of this operator will be a linear combination of 
states with the same $\mathbf{p}_s$, but different magnetic quantum 
numbers.  Formally
\beq
U(R_s,0) \vert (m,j) \mathbf{p}_s, \mu \rangle_s =
\sum_{\nu=-j}^j \vert (m,j) \mathbf{p}_s, \nu \rangle_s
{_s}\langle (m,j)  \nu \vert U(R_s,0) 
\vert (m,j) \mu \rangle_{p_s} 
\label{d.19}
\eeq
where 
\[
_s\langle (m,j)  \nu \vert U(R_s,0) \vert 
(m,j) \mu \rangle_{p_s} := 
\]
\[
\int~ _s\langle (m,j)\mathbf{p} ,  \nu \vert U(R_s,0)  
\vert (m,j) \mathbf{p}_s ,\mu \rangle_s d\mathbf{p} =
\]
\[
\int~ _s\langle (m,j)\mathbf{p} ,  \nu \vert 
U(\Lambda^{-1}_s(p_0),0) U(R,0)U(\Lambda_s(p_0),0)
\vert (m,j) \mathbf{p}_s ,\mu \rangle_s d\mathbf{p} =
\]
\[
\int {m \over \omega_m (\mathbf{p}_s)} 
\delta(\pmb{\Lambda}_s(p_0) p - \pmb{\Lambda}_s(p_0) p_s)D^j_{\nu \mu}(
R ) d \mathbf{p}  =  
\]
\beq
\int \delta(\mathbf{p} - \mathbf{p}_s)
D^j_{\nu \mu}(
R) d \mathbf{p}  =  D^j_{\nu \mu}(R)
\label{d.20}
\eeq
with $\omega_{m} (\mathbf{p}_s) = \sqrt{\mathbf{p}_s^2+m^2}$. 
 $D^j_{\nu \mu} (R)$ are the ordinary finite dimensional 
unitary irreducible representations \cite{rose}
of the rotation group
\[
D^j_{\mu' \mu} (R) =
\sum_\nu  \frac{[(j+\mu')!(j-\mu')!(j+\mu)!(j-\mu)!]^{1/2}}{
(j+\mu'-\nu)!\nu!(\nu-\mu'+\mu)!(j-\mu-\nu)!} \times
\]
\beq
R_{11}^{j+\mu'-\nu}R_{12}^\nu R_{21}^{\nu-\mu'+\mu}R_{22}^{j-\mu-\nu} ~,
\label{d.21}
\eeq
where $R_{ij}$ are the $SU(2)$ matrix elements 
\beq
R=e^{i \frac{\pmb{\theta}}{2} \cdot \pmb{\sigma}} =
\left (
\begin{array}{cc}
R_{11} & R_{12} \\
R_{21} & R_{22} \\
\end{array}
\right ).
\label{d.22}
\eeq
Equation (\ref{d.19}) can also be understood by considering the
transformation properties of $\mathbf{j}_s$ under rotations.  Using
the definition of the $s$-spin (\ref{d.4}) gives
\[
U^{\dagger}(R_s,0)(0, \mathbf{j_s}) U(R_s,0)
\vert(m,j)\mathbf{p}_s,\mu \rangle_s =
\]
\beq
{1 \over {m}}
(\Lambda_s(p_0) \Lambda^{-1}_s(R_s{p}_s) 
R_s {W}) ^{\nu} \vert(m,j)\mathbf{p}_s,\mu \rangle_s .
\label{d.23}
\eeq
Using the identities (\ref{d.18}) and 
$\Lambda_s(R_s{p_s})= \Lambda_s({p_s}) = I$, (\ref{d.23})  becomes
\[
={1 \over {m}}
(\Lambda_s(p_0)\Lambda^{-1}_s(p_0) R\Lambda_s(p_0) {W}) ^{\nu}
\vert(m,j)\mathbf{p}_s,\mu \rangle_s  
\]
\[
=R {1 \over {m}} (\Lambda_s(p_0) \Lambda^{-1}_s({P}) {W}) ^{\nu}
\vert(m,j)\mathbf{p}_s,\mu \rangle_s 
\]
\beq
=(0,R\mathbf{j}_s) \vert(m,j)\mathbf{p}_s,\mu \rangle_s 
\label{d.24}
\eeq
where we have replaced 
$I = \Lambda^{-1}_s({p}_s) = \Lambda^{-1}_s({P})$
because the operator $P$ acts on an eigenstate with eigenvalue $p_s$.  
Equation (\ref{d.24}) 
shows that $\mathbf{j}_s$ 
transforms like an ordinary three-vector,
$U^{\dagger}(R_s) \mathbf{j}_s U(R_s) = R \mathbf{j}_s$, 
under rotations when applied 
to  $\vert(m,j)\mathbf{p}_s,\mu \rangle_s$,  which is
also consistent with (\ref{d.20}).   

Equation (\ref{d.19}) and (\ref{d.20}) lead to the following 
transformation properties for states with the standard 
momentum with respect to the little group
\beq
U(R_s,0) \vert (m,j) \mathbf{p}_s, \mu \rangle_s =
\sum_{\nu=-j}^j \vert (m,j) \mathbf{p}_s, \nu \rangle_s
D^j_{\nu \mu} (R) .
\label{d.25}
\eeq 
We can also calculate the action of spacetime translations on  
these standard vectors using  (\ref{c.5}) 
\beq
U(I,a) \vert (m,j) \mathbf{p}_s, \mu \rangle_s =
e^{-i p_s \cdot a }  \vert (m,j) \mathbf{p}_s, \mu \rangle_s .
\label{d.26}
\eeq
The last step needed to construct irreducible representations 
is to compute the action of $U(\Lambda_s (p),0)$ on 
the standard states.  First we show that 
$U(\Lambda_s (p) ,0) \vert (m,j) \mathbf{p}_s, \mu \rangle_s $
is an eigenstate of $P^{\mu}$ with eigenvalue $p^{\mu}$.  To show
this use 
(\ref{c.19}) to get
\[
P^{\mu} U(\Lambda_s (p) ,0) \vert (m,j) \mathbf{p}_s, \mu \rangle_s 
= 
\]
\[ 
U(\Lambda_s (p) ,0) U^{\dagger} (\Lambda_s (p) ,0)P^{\mu}  U(\Lambda_s (p) ,0) 
\vert (m,j) \mathbf{p}_s, \mu \rangle_s 
=
\]
\beq
U(\Lambda_s (p) ,0) \Lambda_s (p)^{\mu}{}_{\nu}p_s^{\nu}  \vert (m,j) 
\mathbf{p}_s, \mu \rangle_s 
=
p^{\mu}  U(\Lambda_s (p) ,0) \vert (m,j) \mathbf{p}_s, \mu \rangle_s ,
\label{d.27}
\eeq
which is the desired result.

Next we show that 
$U(\Lambda_s (p) ,0) 
\vert (m,j) \mathbf{p}_s, \mu \rangle_s$ is an 
eigenstate of $\hat{\mathbf{z}} \cdot \mathbf{j}_s$ 
with eigenvalue $\mu$.  Using (\ref{d.1}) we get
\[
\hat{\mathbf{z}} \cdot \mathbf{j}_s U(\Lambda_s (p) ,0) 
\vert (m,j) \mathbf{p}_s, \mu \rangle_s =
\]
\[
U (\Lambda_s (p) ,0) 
U^{\dagger} (\Lambda_s (p) ,0) \hat{\mathbf{z}} \cdot \mathbf{j}_s 
U(\Lambda_s (p) ,0)
\vert (m,j) \mathbf{p}_s, \mu \rangle_s =
\]
\[
U (\Lambda_s (p) ,0) \hat{\mathbf{z}} \cdot
({1 \over {m}}
\Lambda_s(p_0) \Lambda^{-1}_s(\Lambda_s(p){p}_s) 
\Lambda_s (p) {W}) ^{\nu} \vert(m,j)\mathbf{p}_s,\mu \rangle_s 
\]
\[
=U (\Lambda_s (p) ,0) \hat{\mathbf{z}} \cdot
({1 \over {m}}
\Lambda_s(p_0) \Lambda^{-1}_s(p) 
\Lambda_s (p) {W})^{\nu} \vert(m,j)\mathbf{p}_s,\mu \rangle_s =
\]
\beq
U (\Lambda_s (p) ,0) \hat{\mathbf{z}} \cdot
({1 \over {m}}
\Lambda_s(p_0) {W}) ^{\nu} \vert(m,j)\mathbf{p}_s,\mu \rangle_s .
\label{d.28}
\eeq
Inserting $\Lambda^{-1}_s(P)$, which is the identity on the standard
basis state, (\ref{d.28}) becomes 
\[
U (\Lambda_s (p) ,0) \hat{\mathbf{z}} \cdot
( {1 \over {m}}
\Lambda_s(p_0)\Lambda^{-1}_s(P) {W}) ^{\nu} \vert(m,j)\mathbf{p}_s,\mu \rangle_s =
\]
\beq
U (\Lambda_s (p) ,0) \hat{\mathbf{z}} \cdot \mathbf{j}_s
\vert(m,j)\mathbf{p}_s,\mu \rangle_s =
\mu U (\Lambda_s (p) ,0) \vert(m,j)\mathbf{p}_s,\mu \rangle_s  ~,
\label{d.29}
\eeq
which is the desired result.

It follows from (\ref{d.27}) and (\ref{d.29}) that $U (\Lambda_s (p)
,0) \vert(m,j)\mathbf{p}_s,\mu \rangle_s$ is a simultaneous eigenstate
of $\mathbf{j}^2$, $\hat{\mathbf{z}}\cdot \mathbf{j}_s$, $\mathbf{p}$, and
$m$.  Thus it is proportional to $\vert(m,j)\mathbf{p},\mu \rangle_s$.  The
constant factor is fixed up to phase by the requirement that $U
(\Lambda_s (p) ,0)$ is unitary.  If we normalize the states to give
Dirac delta functions in the momentum variables and choose the phase
so that the constant factor is real and positive then the normalization
constant is fixed by
\[
\delta (\mathbf{p}'-\mathbf{p}) = 
_s\langle(m,j)\mathbf{p}',\mu \vert 
U^{\dagger}  (\Lambda ,0) U (\Lambda ,0) 
\vert(m,j)\mathbf{p},\mu \rangle_s 
\]
\beq
= \vert c \vert^2 
\delta (\pmb{\Lambda}{p}'-\pmb{\Lambda}{p})=  \vert c \vert^2
\vert{\partial \mathbf{p} \over \partial \pmb{\Lambda} p} \vert
\delta (\mathbf{p}'-\mathbf{p}) 
\label{d.30}
\eeq
which gives 
\beq
c =  \vert{\partial \pmb{\Lambda} p \over \partial \mathbf{p} } \vert^{1/2}
= \vert {\omega_m (\pmb{\Lambda} p) \over \omega_m (\mathbf{p})}\vert ^{1/2}.
\label{d.31}
\eeq
Thus unitarity and an assumed delta function normalization
imply the transformation property  
\beq
U(\Lambda_s (p) ,0) \vert (m,j) \mathbf{p}_s, \mu \rangle_s =
\vert (m,j) \mathbf{p}, \mu \rangle_s 
\sqrt{{\omega_m (\mathbf{p})\over \omega_m (\mathbf{p}_s) }} .
\label{d.32}
\eeq
We see that on these states the little group rotates the spin 
operator leaving $p_s$ invariant, while $\Lambda_s(p)$ changes 
the momentum from the standard value to any other value 
without changing the $z$-component of the $s$-spin.  

Using the elementary transformations (\ref{d.25},\ref{d.26}) 
and (\ref{d.32}), we can construct the action of
an arbitrary Poincar\'e transformation on any basis state.  To
do this note that for any $p$ any Lorentz transformation can be
decomposed into the following product
\beq
\Lambda = \Lambda_s(\Lambda p)
\Lambda^{-1} _s(\Lambda p)
\Lambda 
\Lambda_s(p)
\Lambda^{-1}_s(p) ~,
\label{d.33}
\eeq
where the $s$-spin Wigner rotation,
\beq
\Lambda^{-1} _s(\Lambda p)
\Lambda 
\Lambda_s(p) := R_{ws} (\Lambda ,p)  
\label{d.34}
\eeq
is an element of the little group associated with $p_s$
since it maps $p_s$ to $p_s$.  Thus
\[
U(\Lambda,a) \vert (m,j) \mathbf{p}, \mu \rangle_s =
U(\Lambda,0)U(I,\Lambda^{-1}a) \vert (m,j) \mathbf{p}, \mu \rangle_s =
\]
\[
e^{-i \Lambda^{-1} a \cdot p} U(\Lambda,0) \vert (m,j) \mathbf{p}, \mu \rangle_s =
\]
\[
e^{-i (\Lambda^{-1} a) \cdot p}
U(\Lambda_s(\Lambda p),0)
U(R_{ws} (\Lambda ,p),0)
U(\Lambda^{-1}_s(p),0)
\vert (m,j) \mathbf{p}, \mu \rangle_s =
\]
\beq
e^{-i a \cdot \Lambda p}
U(\Lambda_s(\Lambda p),0)
U(R_{ws} (\Lambda ,p),0)
\vert (m,j) \mathbf{p}_s, \mu \rangle_s
\sqrt{{\omega_m (\mathbf{p}_s)\over \omega_m (\mathbf{p}) }} .
\label{d.35} 
\eeq
Using 
\beq
\vert (m,j) \mathbf{p}_s, \mu \rangle_s =
U(\Lambda^{-1}_s (p_0),0) \vert (m,j) \mathbf{p}_0, \mu \rangle_s
\sqrt{{\omega_m (\mathbf{p}_0)\over \omega_m (\mathbf{p}_s) }}
\label{d.35a}
\eeq
and the fact that 
$\Lambda_s(p_0) R_{ws} (\Lambda ,p)\Lambda^{-1}_s(p_0)$ is a rotation,  
(\ref{d.35}) becomes:
\[
\sum_{\nu=-j}^j e^{-i a \cdot \Lambda p}
U(\Lambda_s(\Lambda p),0)
\vert (m,j) \mathbf{p}_s, \nu \rangle_s
D^j_{\nu_s \mu_s}[\Lambda_s(p_0) R_{ws} (\Lambda ,p)\Lambda^{-1}_s(p_0)]
\sqrt{{\omega_m (\mathbf{p}_s)\over \omega_m (\mathbf{p}) }} 
\]
\[
=\sum_{\nu=-j}^j e^{-i a \cdot \Lambda p}
\vert (m,j) \pmb{\Lambda}p, \nu \rangle_s
D^j_{\nu \mu}[\Lambda_s(p_0) R_{ws} (\Lambda ,p)\Lambda^{-1}_s(p_0)]
\sqrt{{\omega_m (\mathbf{p}_s)\over \omega_m (\mathbf{p}) }}
\sqrt{{\omega_m (\pmb{\Lambda}p)\over \omega_m (\mathbf{p}_s) }} 
\]
\beq
=\sum_{\nu=-j}^j e^{-i a \cdot \Lambda p}
\vert (m,j) \pmb{\Lambda}p, \nu \rangle_s
D^j_{\nu \mu}[\Lambda_s(p_0) R_{ws} (\Lambda ,p)\Lambda^{-1}_s(p_0)]
\sqrt{{\omega_m (\pmb{\Lambda}p) \over \omega_m (\mathbf{p}) }} .
\label{d.37}
\eeq
Thus the general form of any finite Poincar\'e transformation in this
representation of the Hilbert space is
\[
U(\Lambda,a) \vert (m,j) \mathbf{p}, \mu \rangle_s =
\]
\beq
\sum_{\nu=-j}^j e^{-i a \cdot \Lambda p}
\vert (m,j) \pmb{\Lambda}p, \nu \rangle_s
D^j_{\nu \mu}[\Lambda_s(p_0) R_{ws} (\Lambda ,p)\Lambda^{-1}_s(p_0)]
\sqrt{{\omega_m (\pmb{\Lambda}p) \over \omega_m (\mathbf{p}) }} .
\label{d.38}
\eeq
By construction it is apparent
that it is possible to start from the highest weight, $\mu=j$, spin
state with standard momentum $p=p_s$ and generate  all of the basis
vectors in the Hilbert space using only Poincar\'e transformations.
This establishes the irreducibility of this representation.

It is useful to introduce for the Wigner functions of the Poincar\'e group 
a notation that is similar to the 
notation used for Wigner functions of the rotation group:
\[
{\cal D}^{m,j}_{s:\mathbf{p}' \mu'; \mathbf{p} \mu}[\Lambda ,a]:=
\]
\[
_s\langle  (m,j) \mathbf{p}', \mu' 
\vert  U(\Lambda,a) \vert (m,j) \mathbf{p}, \mu \rangle_s =
\]
\beq
e^{-i a \cdot \Lambda p} \delta (\mathbf{p}' - \pmb{\Lambda}p) 
{\cal D}^j_{\mu' \mu}[\Lambda_s(p_0) R_{ws} (\Lambda ,p)\Lambda^{-1}_s(p_0)]
\sqrt{{\omega_m (\pmb{\Lambda}p) \over \omega_m (\mathbf{p}) }}. 
\label{d.39}
\eeq
Note that the Poincar\'e group Wigner functions are basis dependent.

Using this notation (\ref{d.38}) can be  written as
\beq
U(\Lambda,a) \vert (m,j) \mathbf{p}, \mu \rangle_s =
\sum_{\nu=-j}^j \int d\mathbf{p}'
\vert (m,j) \mathbf{p'}, \nu \rangle_s
{\cal D}^{m,j}_{s:\mathbf{p}' \nu; \mathbf{p} \mu}[\Lambda ,a] .
\label{d.40}
\eeq
A consequence of definition (\ref{d.39}) is that these Poincar\'e group 
Wigner functions are explicit unitary representations of the 
Poincar\'e group.  They satisfy the group representation property 
\[
\int d\mathbf{p}'' \sum_{\mu''=-j}^j 
{\cal D}^{m,j}_{s:\mathbf{p}' \mu'; \mathbf{p}'' \mu''}[\Lambda_2 ,a_2]
{\cal D}^{m,j}_{s:\mathbf{p}'' \mu''; \mathbf{p} \mu}[\Lambda_1 ,a_1]= 
\]
\beq
{\cal D}^{m,j}_{s:\mathbf{p}' \mu'; \mathbf{p} \mu}
[\Lambda_2 \Lambda_1 ,\Lambda_2 a_1+ a_2]
\label{d.41}
\eeq
and unitarity
\beq
\int d\mathbf{p}'' \sum_{\mu''=-j}^j 
{\cal D}^{m,j*}_{s:\mathbf{p}'' \mu''; \mathbf{p}' \mu'}[\Lambda ,a]
{\cal D}^{m,j}_{s:\mathbf{p}'' \mu''; \mathbf{p} \mu}[\Lambda ,a]
= \delta (\mathbf{p}'-\mathbf{p})\delta_{\mu' \mu} .
\label{d.42}
\eeq

In dealing with electromagnetic interactions where the coupling of the
spin to a magnetic field is known for one type of spin, say the
$s$-spin, and the dynamics is given in a basis with a different type
of spin, say the $t$-spin, the transformation from a basis associated
with the standard vector $p_s$ and standard boost $\Lambda_s(p)$ to
the basis associated with the pair, $p_t$ and $\Lambda_t(p)$, is
needed.  The corresponding spin operators are
\beq
(0,\underline{\mathbf{j}}_s) =
{1 \over \underline{m}} \Lambda_{0s}^{-1}(\underline{P})^{\mu}{}_{\nu} 
\underline{W}^{\nu} 
\label{d.43}
\eeq
and 
\beq
(0,\underline{\mathbf{j}}_t) =
{1 \over \underline{m}} \Lambda_{0t}^{-1}(\underline{P})^{\mu}{}_{\nu} 
\underline{W}^{\nu} ~,
\label{d.44}
\eeq
where 
\beq
\Lambda^{-1}_{0s}(\underline{P}) = \Lambda_s(p_0) \Lambda^{-1}_s(\underline{P}) 
\label{d.45}
\eeq
and
\beq
\Lambda^{-1}_{0t}(\underline{P}) = 
\Lambda_t(p_0) \Lambda^{-1}_t(\underline{P}) .
\label{d.46}
\eeq
When $p=p_0$ (i.e. $\mathbf{p}= \mathbf{0}$) we have the identity
$\Lambda^{-1}_{0s}(p_0) =\Lambda^{-1}_{0t}(p_0) =I$.  This 
means that
\beq
\mathbf{j}_s \vert (m,j) \mathbf{p}_0 , \mu \rangle_s  =
\mathbf{j}_t \vert (m,j) \mathbf{p}_0 , \mu \rangle_t  =
(\mathbf{W}/m) \vert (m,j) \mathbf{p}_0 , \mu \rangle . 
\label{d.47}
\eeq 
This is because the spin operators (\ref{d.43}) and (\ref{d.44}) 
are defined so 
they are identical when they are applied to 
zero-momentum,  $\mathbf{p}=\mathbf{p}_0 =\mathbf{0}$:
\beq
\vert (m,j) \mathbf{0} , \mu \rangle_s =
\vert (m,j) \mathbf{0} , \mu \rangle_t .
\label{d.48}
\eeq
It follows that 
\[
\vert (m,j) \mathbf{p} , \mu \rangle_s =
U(\Lambda_s(p),0) \vert (m,j) \mathbf{p}_s , \mu \rangle_s
\sqrt{{\omega_m(\mathbf{p}_s ) \over \omega_m(\mathbf{p} ) }}
\]
\[
=U(\Lambda_s(p),0) U(\Lambda^{-1}_s(p_0),0)
\vert (m,j) \mathbf{0} , \mu \rangle_s
\sqrt{{\omega_m(\mathbf{0} ) \over \omega_m(\mathbf{p} ) }}=
\]
\[
=U(\Lambda_s(p),0) U(\Lambda^{-1}_s(p_0),0)
\vert (m,j) \mathbf{0} , \mu \rangle_t
\sqrt{{\omega_m(\mathbf{0} ) \over \omega_m(\mathbf{p} ) }}
\]
\[
=U(\Lambda_s(p),0) U(\Lambda^{-1}_s(p_0),0)
U(\Lambda_t(p_0),0)
\vert (m,j) \mathbf{p}_t , \mu \rangle_t
\sqrt{{\omega_m(\mathbf{p}_t ) \over \omega_m(\mathbf{p} ) }}
\]
\[
=U(\Lambda_s(p),0) U(\Lambda^{-1}_s(p_0),0)
U(\Lambda_t(p_0),0)U(\Lambda^{-1}_t(p),0)
\vert (m,j) \mathbf{p} , \mu \rangle_t 
\]
\[
=U(\Lambda_s(p)\Lambda^{-1}_s(p_0)
\Lambda_t(p_0)\Lambda^{-1}_t(p),0)
\vert (m,j) \mathbf{p} , \mu \rangle_t 
\]
\[
= \sum_{\mu'=-j}^j
\int 
\vert (m,j) \mathbf{p}' , \mu' \rangle_t
d\mathbf{p}'
{\cal D}^{mj}_{t:\mathbf{p}' \mu';\mathbf{p};\mu} 
[\Lambda_s(p)\Lambda^{-1}_s(p_0)
\Lambda_t(p_0)\Lambda^{-1}_t(p),0]  
\]
\beq
=
\sum_{\mu'=-j}^j\vert (m,j) \mathbf{p} , \mu' \rangle_t
D^j_{\mu'\mu} [\Lambda^{-1}_{0t}(p)
\Lambda_{0s}(p)] ~,
\label{d.49}
\eeq
where for the last step we used (\ref{d.39}) and 
$\Lambda_s(p)\Lambda^{-1}_s(p_0) \Lambda_t(p_0)\Lambda^{-1}_t(p)p=p$.
This shows these bases differ by a momentum-dependent generalized
Melosh rotation (\ref{d.12}).  

The relation (\ref{d.49}) is 
\beq
\vert (m,j) \mathbf{p} , \mu \rangle_s 
=
\sum_{\mu'=-j}^j
\vert (m,j) \mathbf{p} , \mu' \rangle_t
D^j_{\mu'\mu} [\Lambda^{-1}_{0t}(p)
\Lambda_{0s}(p)]~ .
\label{d.50}
\eeq

This section illustrated the general structure of positive-mass
positive-energy irreducible representations of the Poincar\'e group.
We started with a complete set of commuting Hermetian operators
constructed as functions of the Poincar\'e generators.  The spectrum
of all of the commuting observables was fixed once the spectrum of $m$
and $\mathbf{j}^2$ was fixed.  A representation of the Hilbert space
was defined by square integrable functions of the eigenvalues of these
commuting observables over their spectra.  On each space associated
with a fixed mass and spin we constructed an irreducible unitary
representation of the Poincar\'e group.

In this construction we found that there are many different
observables that behave like spins.  They are all non-linear functions
of the Poincar\'e generators satisfying $SU(2)$ commutation relations.
It is also possible to change the choice of independent continuous
variables.  Different choices of commuting continuous variables 
(linear momentum, four velocity, light-front components of the momentum) 
along with the appropriate choice of spin variable are relevant in dynamical
models based on Dirac's forms of dynamics \cite{Dirac:1949cp}.

The construction of both the irreducible representation and the
representation space can be done for many-body systems in the same way
that it was done for single particles.  The idea is to use the
elementary transformations (\ref{d.25},\ref{d.26}) and (\ref{d.32}), with
eigenstates of commuting observables constructed from the many-body
generators.  To do this it is necessary to decompose states with total
$p=p_s$ into irreducible representations of the little group.  The
main difference is that the mass will generally have a continuous
spectrum and there may be multiple copies of representations of 
given mass and spin.

\section{Examples}
\label{section5}

In this section we discuss the three most common spin observables
and discuss the properties that distinguish each of them.

In the previous section we introduced a large number of different
types of observables which we identified as spins.  Each of these were
functions of the Poincar\'e generators satisfying $SU(2)$ commutation
relations and commuting with the linear momentum.  All have the same
square, which is $W^2/M^2$, the ratio of the two Casimir operators for the
Poincar\'e group.  Each of the spins used in applications has some
particular property that makes them useful.  In general different types
of spin are characterized by the choice of boost used to relate the
spin of a particle (system) with momentum $\mathbf{p}$ to the spin in
a standard frame.  Specifically the magnetic quantum number remains
unchanged when {\it this} boost is applied to a standard frame eigenstate of
the $z$-component of spin and momentum; (\ref{d.32}).  
For the examples in this section
we assume that the standard vector is the rest vector, $p_0$.

\subsection{Canonical spin}

The boost used to define the canonical spin is the rotationless boost
(\ref{b.19}).  Under rotations
\beq
U(R,0) \vert \mathbf{p},\mu \rangle = \sum_{\nu=-j}^j
\vert R\mathbf{p},\nu \rangle D^j_{\nu \mu}[\Lambda_c^{-1}(Rp)R \Lambda_c(p)] . 
\label{e.1}
\eeq
The special property of the canonical boost is that the Wigner rotation 
(\ref{d.34}) of any rotation $R$ is $R$:
\beq
\Lambda_c^{-1}(Rp)R \Lambda_c(p)=R .
\label{e.2}
\eeq
Using this in (\ref{e.1}) gives
\beq
U(R,0) \vert \mathbf{p},\mu \rangle = \sum_{\nu=-j}^j
\vert R\mathbf{p},\nu \rangle D^j_{\nu \mu}[R] . 
\label{e.3}
\eeq
where the argument of the $D$ function is independent of $\mathbf{p}$.
This is useful when applied to a system of particles with different
momenta.  Under rotations all of the particles transform with the same
rotation, independent of their individual momenta.  This allows the
spins to be coupled with ordinary $SU(2)$ Clebsch-Gordan coefficients.
For the other types of spins the arguments of the $D$-functions involve
Wigner rotations with different values of $p$.  In order to couple the
spins it is normally necessary first to convert them to canonical
spins so all spins rotate the same way.

The identity (\ref{e.2}) is most easily proved in the $SU(2)$ representation,
(\ref{b.17}).  In this representation $\Lambda_c(p)=
e^{{1 \over 2}\mathbf{z} \cdot \pmb{\sigma}}$ with
$\mathbf{z}=\hat{\mathbf{p}} \sinh^{-1} (\vert \mathbf{p} \vert/m)$.
For this proof we let boldface $\mathbf{R}$ denote a three-dimensional
rotation and $R$ denote the corresponding $SU(2)$ rotation. It follows
that 
\beq 
R \Lambda_c (p) R^{\dagger} = R e^{\mathbf{z} \cdot
\pmb{\sigma}} R^{\dagger}= e^{\mathbf{z} \cdot ({R} \pmb{\sigma}
{R}^{\dagger}) }= e^{\mathbf{z} \cdot ( \mathbf{R}^{-1} \pmb{\sigma})
} = e^{\mathbf{R} \mathbf{z} \cdot \pmb{\sigma} }= \Lambda_c
(\mathbf{R} p) ~,
\label{e.4} 
\eeq 
where we have used (\ref{b.7}) for rotations
($\Lambda \to R$): 
\beq 
x^{\mu} (\mathbf{R}^{-1} \sigma)_{\mu} =
(\mathbf{R}x)^{\mu} \sigma_{\mu} = R (x^{\mu} \sigma_{\mu})
R^{\dagger}= x^{\mu} R ( \sigma_{\mu}) R^{\dagger}.
\label{e.5} 
\eeq 
Equation (\ref{e.4}) and $R^{\dagger}=R^{-1}$ imply
the desired result (\ref{e.2}).  

\subsection{Helicity}

The helicity \cite{JacobWick:1959} is the operator $\hat{\mathbf{p}} \cdot
\mathbf{j}_c$ where $\mathbf{j}_c$ is the canonical spin.  To relate
this to the formalism derived in section \ref{section4} 
we let $R(\hat{\mathbf{z}}
\to \hat{\mathbf{p}})$ denote the rotation about an axis perpendicular
to the plane containing $\hat{\mathbf{z}}$ and $\hat{\mathbf{p}}$ that
rotates $\hat{\mathbf{z}}$ into the direction $\hat{\mathbf{p}}$. 
The helicity boost is defined by
\beq
\Lambda_h(p) = \Lambda_c(p) R(\hat{\mathbf{z}} \to \hat{\mathbf{p}}) =
R(\hat{\mathbf{z}} \to \hat{\mathbf{p}}) \Lambda_c(p_z)  ~,
\label{e.6} 
\eeq
where $p_z$ is the 4-vector with 3-magnitude $\vert \mathbf{p}\vert$
in the $z$ direction.

Helicity eigenstates are related to canonical spin eigenstates by
\beq
\vert \mathbf{p} , \mu\rangle_h :=
U(R(\hat{\mathbf{z}} \to \hat{\mathbf{p}}))
\vert \vert 
 \mathbf{p} \vert \hat{\mathbf{z}}, \mu
\rangle_c = \sum_{\nu=-j}^j
\vert \mathbf{p} , 
\nu\rangle_c D^j_{\nu \mu}[R(\hat{\mathbf{z}} \to \hat{\mathbf{p}})] .
\label{e.7} 
\eeq 
This equation shows that the generalized Melosh rotation 
(\ref{d.12},\ref{d.50}) 
relating the canonical and helicity spins is $R(\hat{\mathbf{z}} \to
\hat{\mathbf{p}})$.

The helicity spin, $\mathbf{j}_h$, defined using the helicity boost in 
(\ref{d.4}) satisfies
\beq
\hat{\mathbf{z}} \cdot  \mathbf{j}_h = \hat{\mathbf{p}} \cdot
\mathbf{j}_c ~,
\label{e.8} 
\eeq
which means that the z-component of the helicity spin is the helicity.

The helicity-spin Wigner rotation (\ref{d.34}) is 
\beq
R_{wh}(\Lambda ,p) = 
R^{-1}(\hat{\mathbf{z}} \to \hat{\mathbf{p}}) \Lambda_c^{-1}(\Lambda p)
\Lambda  \Lambda_c(p) R(\hat{\mathbf{z}} \to \hat{\mathbf{p}}) ~,
\label{e.9} 
\eeq
which is a rotation about the $z$ axis.  Thus 
\beq
U(\Lambda,0) \vert \mathbf{p} , \mu \rangle_h := 
\vert \pmb{\Lambda}{p} , \mu \rangle_h e^{i \mu \phi} 
\sqrt{{\omega_m (\pmb{\Lambda} p) \over \omega_m (\mathbf{p}) }} ~,
\label{e.10} 
\eeq
where $\phi$ is the angle of rotation of the Wigner rotation.
Thus the helicity eigenvalue is Lorentz invariant.  

The identification of $
\hat{\mathbf{z}} \cdot  \mathbf{j}_h$ with 
$\hat{\mathbf{p}} \cdot  \mathbf{j}_c$ follows from the calculation
\[
\hat{\mathbf{p}} \cdot \mathbf{j}_c \vert \mathbf{p}' , \mu \rangle_h=
\hat{\mathbf{p}}' \cdot \mathbf{j}_c U(R(\hat{\mathbf{z}} \to
\hat{\mathbf{p}}')) \vert \vert  \mathbf{p}'\vert 
\hat{\mathbf{z}} , \mu \rangle_c =
\]
\[
\hat{\mathbf{p}}' \cdot U(R(\hat{\mathbf{z}}\to \hat{\mathbf{p}}'))
U(R^{-1}(\hat{\mathbf{z}}\to \hat{\mathbf{p}}')) \mathbf{j}_c
U(R(\hat{\mathbf{z}} \to \hat{\mathbf{p}}'))\vert \vert 
 \mathbf{p}'\vert \hat{\mathbf{z}}, \mu
\rangle_c =
\]
\[
U(R(\hat{\mathbf{z}}\to \hat{\mathbf{p}}')) 
(\hat{\mathbf{p}}' \cdot R(\hat{\mathbf{z}} \to \hat{\mathbf{p}}') 
 \mathbf{j}_c ) \vert
\vert  \mathbf{p}'\vert \hat{\mathbf{z}} , \mu \rangle_c =
\]
\[
U(R(\hat{\mathbf{z}}\to \hat{\mathbf{p}}'))
(R^{-1}(\hat{\mathbf{z}} \to \hat{\mathbf{p}}')
\hat{\mathbf{p}}') \cdot \mathbf{j}_c \vert 
\vert  \mathbf{p}'\vert \hat{\mathbf{z}} , \mu \rangle_c =
\]
\[
U(R(\hat{\mathbf{z}}\to \hat{\mathbf{p}}'))
\vert \mathbf{p}\vert \hat{\mathbf{z}}\cdot 
\mathbf{j}_c \vert \vert \mathbf{p}'\vert \hat{\mathbf{z}}  , \mu \rangle_c =
\]
\beq
\mu  
U(R(\hat{\mathbf{z}}\to \hat{\mathbf{p}}')) 
\vert  \vert \mathbf{p}'\vert \hat{\mathbf{z}}  , \mu \rangle_c =
\mu \vert \mathbf{p}' , \mu \rangle_h =  
\hat{\mathbf{z}} \cdot  \mathbf{j}_h
\vert  \mathbf{p}', \mu \rangle_h   ~.
\label{e.11} 
\eeq

\subsection{Light-front spin}

In the $SL(2,\mathbb{C})$ representation the light-front boosts are
represented by the three-parameter {\it subgroup} of lower triangular 
matrices with real entries on the diagonal.  Considering the 
transformation properties (\ref{b.7}) for the four momentum  
\beq
\left (
\begin{array}{cc}
a & 0 \\
b+ic & 1/a
\end{array}
\right )
\left (
\begin{array}{cc}
m & 0 \\
0  & m
\end{array}
\right )
\left (
\begin{array}{cc}
a & b - ic  \\
0  & 1/a
\end{array}
\right ) = 
\left (
\begin{array}{cc}
ma^2 & ma(b -ic)  \\
ma(b+ic)  & m(b^2+c^2)/a^2
\end{array}
\right ) 
\label{e.12} 
\eeq
we can identify the parameters of the light-front boost as follows
\beq
ma^2 = p^0+ p^3 \qquad a = \sqrt{{p^0+ p^3 \over m}}
\label{e.13} 
\eeq
\beq
ma(b -ic)= p^1-i p^2 \qquad b-ic = {p^1-i p^2 \over ma}  ~.
\label{e.14} 
\eeq
Because the light-front boosts form a subgroup,  any sequence of
light-front boosts is the unique light-front boost parameterized by the final
momentum of the sequence.  
This means that the Wigner rotation (\ref{d.34}) of a light-front boost
is the {\it identity} so the light-front spins remain unchanged
under the three parameter group of light-front boosts.

Unlike helicities,  the light-front spin is not invariant with respect
to rotations.

\section{Adding spins}
\label{section6}

Multiparticle systems can be described by tensor products of 
single-particle systems.   The Hilbert space is the tensor product,
\beq
{\cal H} = \otimes_i {\cal H}_{m_i j_i},
\label{f.1}
\eeq
where ${\cal H}_{m_i j_i}$ are the mass $m_i$ spin $j_i$ single-particle
irreducible representation spaces constructed in section \ref{section4}.

There is a natural representation $U_0(\Lambda,a)$ of the Poincar\'e
group on this space which is the tensor product of the irreducible
representations constructed in section \ref{section4}
\beq
U_0(\Lambda,a) = \otimes_i U_{m_i j_i}(\Lambda ,a).
\label{f.2}
\eeq
Dynamically this representation describes a system of free particles.
In this representation the infinitesimal generators are sums
of the single-particle generators.  In the $s$-spin basis this 
representation has the explicit form
\[
U_0(\Lambda,a) \vert (m_{1},j_{1}) \mathbf{p}_{1} ,\mu_{1} \rangle_s \otimes
\cdots \otimes  
\vert (m_N,j_N) \mathbf{p}_N ,\mu_{N} \rangle_s =
\]
\[
\sum_{\mu_1'\cdots \mu_N'} \int d\mathbf{p}'_1 \cdots d\mathbf{p}'_N  
\vert (m_1,j_1) \mathbf{p}_1' ,\mu_{1}' \rangle_s \otimes
\cdots \otimes  
\vert (m_N,j_N) \mathbf{p}_N' ,\mu_{N}' \rangle_s \times
\]
\[
{\cal D}^{m_1,j_1}_{s:\mathbf{p}_1' \mu_{1}'; \mathbf{p}_1 \mu_{1}}[\Lambda ,a]
\cdots
{\cal D}^{m_N,j_N}_{s:\mathbf{p}_N' \mu_{N}'; \mathbf{p}_N \mu_{N}}[\Lambda ,a]=
\]
\[
e^{-i a \cdot \sum_i \Lambda p_i}
\sum_{\nu_1\cdots \nu_N}
\vert (m_1,j_1) \pmb{\Lambda}p_1, \nu_{1} \rangle_s \otimes 
\cdots
\vert (m_N,j_N) \pmb{\Lambda}p_N, \nu_{N} \rangle_s
\]
\beq
\times \prod_i
D^j_{\nu_{i} \mu_{i}}[\Lambda_s(p_0) R_{ws} (\Lambda ,p_i)\Lambda^{-1}_s(p_0)]
\sqrt{{\omega_{m_i} (\pmb{\Lambda}p_i) \over \omega_{m_i} (\mathbf{p}_i) }} .
\label{f.3}
\eeq
Just like in the case of ordinary rotations, the tensor product of
irreducible representations of the Poincar\'e group is reducible.
Poincar\'e group Clebsch-Gordan coefficients are coefficients of a
unitary transformation that transforms the tensor product into
irreducible blocks labeled by many-body mass and spin eigenvalues.
For non-interacting systems the many-body mass is just the invariant
mass of the many-body system.  The structure of the Clebsch-Gordan
coefficients depends on the choice of basis used to define vectors in
the irreducible blocks.  They are derived below.

We start by evaluating the coefficients of the unitary transformation
that transforms a tensor product of two irreducible representations
into a superposition of irreducible representations.

Basis states for the tensor product state are simultaneous eigenstates
of the mass, spin, linear momentum and magnetic quantum number for
each particle
\beq
\vert (m_1,j_1) \mathbf{p}_1 ,\mu_{1} \rangle_s \otimes 
\vert (m_2,j_2) \mathbf{p}_2 ,\mu_{2} \rangle_s .
\label{f.4}
\eeq
As in the previous section the subscript $s$ indicates choice of 
spin operator.

Infinitesimal generators for the combined system are sums of
generators for the individual constituent particles.  The four
momentum of the combined system is $P=p_1+p_2$. This is a sum of
timelike positive-time vectors so it is a timelike positive-time
vector.  Following what was done for single-particles we
look at representations of the little group for a standard momentum
vector.  For many-body systems we choose the standard vector 
to be the zero of the total three-momentum vector.

Tensor product eigenstates with the standard vector for the
two-particle system $P=p_s= p_0= (M,\mathbf{0})$ have the form
\beq
\vert (m_1,j_1) \mathbf{k} ,\mu_{1} \rangle_s \otimes 
\vert (m_2,j_2) -\mathbf{k} ,\mu_{2} \rangle_s
\label{f.5}
\eeq
where $\mathbf{p}_1= -\mathbf{p}_2 := \mathbf{k}$.
We also define
\beq
k_1 = (\omega_{m_1} (\mathbf{k}) , \mathbf{k})   
\qquad
k_2 = (\omega_{m_2} (\mathbf{k}) ,-\mathbf{k}).   
\label{f.6}
\eeq
where $\omega_{m_i} (\mathbf{k}) = \sqrt{\mathbf{k}^2+m_i^2}$.

It is useful to decompose the vector $\mathbf{k}$ 
into orbital angular momentum components using 
spherical harmonics:
\[
\vert (m_1,j_1,m_2,j_2) k , l , \mu_l , \mu_{1}, \mu_{2} \rangle_s  :=
\]
\beq
\int \vert (m_1,j_1) \mathbf{k} ,\mu_{1} \rangle_s \otimes 
\vert (m_1,j_2) -\mathbf{k} ,\mu_{2} \rangle_s \, d\hat{\mathbf{k}} \,
Y^{l}_{\mu_l} (\hat{\mathbf{k}}) .
\label{f.7}
\eeq
To construct irreducible representations consider the 
transformation properties of (\ref{f.7}) under 
rotations (the little group associated with $p_0$).

Let $U(R)=U_1(R,0)\otimes U_2(R,0)$.  Applying this 
operator to (\ref{f.7}) gives 
\[
U(R,0) \vert (m_1,j_1,m_2,j_2) k , l , \mu_{l} , \mu_{1}, \mu_{2} \rangle_s  :=
\]
\[
U(R,0) \int \vert (m_1,j_1) \mathbf{k} ,\mu_{1} \rangle_s \otimes 
\vert (m_2,j_2) -\mathbf{k} ,\mu_{2} \rangle_s 
d\hat{\mathbf{k}}
Y^{l}_{\mu_l} (\hat{\mathbf{k}}) 
=
\]
\[
\sum_{\mu_1' \mu_2'} 
\int \vert (m_1,j_1) R\mathbf{k} ,\mu_{1}' \rangle_s \otimes 
\vert (m_2,j_2) -R\mathbf{k} ,\mu_{2}' \rangle_s
\,d\hat{\mathbf{k}} \,
Y^{l}_{\mu_l} (\hat{\mathbf{k}}) \times
\]
\beq
D^{j_1}_{\mu_{1}'\mu_{1}} [\Lambda_s^{-1} (R k_1)R \Lambda_s (k_1)]
D^{j_2}_{\mu_{2}'\mu_{2}} [\Lambda_s^{-1} (R k_2)R \Lambda_s (k_2)]  . 
\label{f.8}
\eeq
Changing variables $\mathbf{k}\to R^{-1} \mathbf{k}$ (\ref{f.8}) becomes 
\[
= \sum_{\mu_1' \mu_2'} 
\int \vert (m_1,j_1) \mathbf{k} ,\mu_{1}' \rangle_s \otimes 
\vert (m_2,j_2) -\mathbf{k} ,\mu_{2}' \rangle_s
\,d\hat{\mathbf{k}}\,
Y^{l}_{\mu_l} (R^{-1}\hat{\mathbf{k}}) \times
\]
\beq
D^{j_1}_{\mu_{1}'\mu_{1}} [\Lambda_s^{-1} (k_1)R \Lambda_s (R^{-1}k_1)]
D^{j_2}_{\mu_{2}'\mu_{2}} [\Lambda_s^{-1} (k_2)R \Lambda_s (R^{-1}k_2)].   
\label{f.9}
\eeq
Noting that
\[
Y^{l}_{\mu_l}(R^{-1}\hat{\mathbf{k}})= 
\langle R^{-1}\hat{\mathbf{k}} \vert l,\mu_l \rangle = 
\langle \hat{\mathbf{k}} \vert U(R) \vert l,\mu_l \rangle =
\sum_{m_l'}  \langle \hat{\mathbf{k}} \vert l, \mu_l' \rangle 
\langle l,\mu_l'\vert  U(R) \vert l,\mu_l \rangle =
\]
\beq
\sum_{m_l'} Y^{l}_{\mu_l'}(\hat{\mathbf{k}}) D^l_{\mu_l'\mu_l} [R]
\label{f.10}
\eeq
Eq. (\ref{f.9})  becomes 
\[
=
\sum_{\mu_1' \mu_2' \mu_l'} 
\int \vert (m_1,j_1) \mathbf{k} ,\mu_{1}' \rangle_s \otimes 
\vert (m_2,j_2) -\mathbf{k} ,\mu_{2}' \rangle_s
\,d\hat{\mathbf{k}}\,
Y^{l}_{\mu_l'} (\hat{\mathbf{k}}) \times
\]
\beq
D^{j_1}_{\mu_{1}'\mu_{1}} [\Lambda_s^{-1} (k_1)R \Lambda_s (R^{-1}k_1)]
D^{j_2}_{\mu_{2}'\mu_{2}} [\Lambda_s^{-1} (k_2)R \Lambda_s (R^{-1}k_2)]   
D^l_{\mu_l' \mu_l} [R]. 
\label{f.11}
\eeq
The important observation is that the spins and orbital angular momenta
all transform with {\it different} rotations so they cannot be consistently
added with ordinary Clebsch-Gordan coefficients.  The rotations of
the spins are Wigner rotations of the rotations that appear in the
orbital angular momentum.  The canonical spin, that uses the
rotationless boost (\ref{b.17},\ref{b.18},\ref{b.19}), $\Lambda_s(p)=
\Lambda_c(p)$, with $p_s=p_o$ has the unique feature, (\ref{e.2}), that
the Wigner rotations of any rotation is the rotation.  This means that
for canonical spins ($s=c$)
\[ 
\Lambda_c^{-1} (R k_1)R \Lambda_c (k_1) = R
\] 
\beq 
\Lambda_c^{-1} (R k_2)R \Lambda_c (k_2) = R
\label{f.12} 
\eeq 
or equivalently
\[ 
\Lambda_c^{-1} (k_1)R \Lambda_c (R^{-1}k_1) = R
\] 
\beq 
\Lambda_c^{-1} (k_2)R \Lambda_c (R^{-1}k_2) = R
\label{f.13} 
\eeq 
so all three $SU(2)$-Wigner functions in (\ref{f.11}) have 
the same arguments, independent of $\mathbf{k}_i$.  Thus for 
{\it canonical} spin, $s=c$, we have 
\[
U(R,0) \vert (m_1,j_1,m_2,j_2) k , l , \mu_l , \mu_{1}, \mu_{2} \rangle_c  :=
\]
\[
=
\sum_{\mu_1' \mu_2' \mu_l'}
\int \vert (m_1,j_1) \mathbf{k} ,\mu_{1}' \rangle_c \otimes 
\vert (m_2,j_2) -\mathbf{k} ,\mu_{2}' \rangle_c
\,d\hat{\mathbf{k}}\,
Y^{l}_{\mu_l'} (\hat{\mathbf{k}}) \times
\]
\beq
D^{j_1}_{\mu_{1}'\mu_{1}} [R]
D^{j_2}_{\mu_{2}'\mu_{2}} [R]   
D^l_{\mu_l' \mu_l} [R]. 
\label{f.14}
\eeq
which has the property that the spins and orbital angular momenta {\it all}
rotate with the same rotation.

Recall that we initially defined all types of spins so that they agree
in the particle's rest frame (\ref{d.47}).  The state (\ref{f.7}) is a
rest state of the two-body system.  Following what was done in the 
one-body case we assume all of the two-body $s$-spins agree with the canonical 
spin state (\ref{f.14}) in the two-body rest frame.  Since we 
want to treat the case of coupling any type of spins we use 
generalized Melosh rotations (\ref{d.12},\ref{d.50}) to express the 
single-particle canonical spin state (\ref{f.14}) in terms of single-particle 
$s$-spin states:
\beq
\vert (m_1,j_1) \mathbf{k} ,\mu_{1} \rangle_c=
\sum_{\mu_{1}'=-{j_1}}^{j_1} \vert (m_1,j_1) \mathbf{k} ,\mu_{1}' \rangle_s
D^{j_1}_{\mu_{1}'\mu_{1}} [\Lambda_s^{-1} (k_1) \Lambda_c (k_1)]
\label{f.15}
\eeq
\beq
\vert (m_2,j_2) -\mathbf{k} ,\mu_{2} \rangle_c=
\sum_{\mu_{2}'=-{j_2}}^{j_2} \vert (m_2,j_2) -\mathbf{k} ,\mu_{2}' \rangle_s
D^{j_2}_{\mu_{2}'\mu_{2}} [\Lambda_s^{-1} (k_2) \Lambda_c (k_2)].
\label{f.16}
\eeq
By using (\ref{f.15}) and (\ref{f.16}) in (\ref{f.14}) 
the two-body rest canonical spin
state can be expressed in terms of the single-particle $s$-spin states
as
\[
\vert (m_1,j_1,m_2,j_2) k , l , \mu_l , \mu_{1}, \mu_{2} \rangle_c  :=
\]
\[
\sum_{\mu_{1}'\mu_{2}'} 
\int \vert (m_1,j_1) \mathbf{k} ,\mu_{1}' \rangle_s \otimes 
\vert (m_2,j_2) -\mathbf{k} ,\mu_{2}' \rangle_s 
d\hat{\mathbf{k}}
Y^{l}_m (\hat{\mathbf{k}}) \times 
\]
\beq 
D^{j_1}_{\mu_{1}'\mu_{1}} [\Lambda_s^{-1} (k_1) \Lambda_c (k_1)]
D^{j_2}_{\mu_{2}'\mu_{2}} [\Lambda_s^{-1} (k_2)\Lambda_c (k_2)]  .
\label{f.17}
\eeq
This  state is identical to the state in (\ref{f.14}) and
it necessarily has the same transformation property under rotations. 

Using the property 
\[
\sum_{\mu_{1}'\mu_{2}'}  D^{j_1}_{\mu_1\mu_1'} [ R]
D^{j_2}_{\mu_2\mu_2'} [ R] \langle j_1, \mu_1', j_2, \mu_2' 
\vert j_{12}, \mu_{12} \rangle
= 
\]
\beq
\sum_{j_{12}\mu_{12}'}  \langle j_1, \mu_1, j_2, \mu_2 \vert j_{12},
\mu_{12}' 
\rangle
D^{j_{12}}_{\mu_{12}'\mu_{12}} [ R]
\label{f.18}
\eeq
of the $SU(2)$ Clebsch-Gordan coefficients, the spins and orbital
angular momenta can be coupled to a
total spin that transform irreducibly under rotations.  Thus we are 
led to define the rest states of the system by
\[
\vert k, j (m_1,j_1,m_2,j_2, l , s_{12} ) \mathbf{0} ,  \mu  \rangle_c :=
\]
\[
\sum' \vert (m_1,j_1,m_2,j_2) k , l , \mu_{l}' , \mu_{1}', \mu_{2}' \rangle_c
\langle j_1, \mu_{1}', j_2, \mu_{2}' \vert s_{12}, \mu_s \rangle
\langle l, m, s_{12}, \mu_s \vert j, \mu \rangle  =
\]
\[
\sum'\sum''  \int \vert (m_1,j_1) \mathbf{k} ,\mu_{1}' \rangle_s \otimes 
\vert (m_2,j_2) -\mathbf{k} ,\mu_{2}' \rangle_s
\,d\hat{\mathbf{k}}\,
Y^{l}_{m} (\hat{\mathbf{k}}) \times
\]
\[
D^{j_1}_{\mu_1'\mu_1''} [\Lambda_s^{-1} (k_1) \Lambda_c (k_1)]
D^{j_2}_{\mu_2'\mu_2''} [\Lambda_s^{-1} (k_2) \Lambda_c (k_2)] \times
\]
\beq
\langle j_1, \mu_1'', j_2, \mu_2'' \vert s_{12}, \mu_s \rangle
\langle l, m, s_{12}, \mu_s \vert j, \mu \rangle .  
\label{f.19}
\eeq
It follows from (\ref{f.14}) and (\ref{f.18}) that these 
vectors transform as spin-$j$ irreducible representations 
with respect to rotations
\[
U(R,0) \vert k, j (m_1,j_1,m_2,j_2, l , s_{12} ) \mathbf{0} ,  \mu
\rangle_c =
\]
\beq
\sum_{\mu=-j}^j\vert k, j (m_1,j_1,m_2,j_2)  \mu' , l , s_{12} \rangle_c
D^j_{\mu'\mu}[R] ~.
\label{f.20}
\eeq
Note that in this expression $k$ is a function of the invariant mass
\[
M_0= \sqrt{m_1^2+\mathbf{k}^2}+ \sqrt{m_2^2+\mathbf{k}^2} =
\]
\beq
\omega_{m_1} (\mathbf{k}) + \omega_{m_2} (\mathbf{k}).
\label{f.21}
\eeq
This is an irreducible basis for the rest states. Following 
again what was done in the one-particle case, having decomposed the 
rest states into irreducible representations of the rotation
group (little group of $p_0$), we define $s$-spin states with arbitrary 
momentum by applying 
$U(\Lambda_s(P))=U_1(\Lambda_s(P))\otimes U_2(\Lambda_s(P))$ to the rest states.
Thus we define the  $s$-states of total
momentum $P$ following the construction used in (\ref{d.32})
\[
\vert k, j (m_1,j_1,m_2,j_2 , l , s_{12}) \mathbf{P} ,  \mu  \rangle_s
:=
\]
\beq
U(\Lambda_s (P),0)\vert k, j (m_1,j_1,m_2,j_2 , l , s_{12}) \mathbf{0} ,  
\mu  \rangle_c 
\sqrt{\frac{M_0}{\sqrt{M_0^2 +\mathbf{P}^2}}}  ~,
\label{f.22}
\eeq
where the square root factors (\ref{d.30}-\ref{d.31}) imply a 
$\delta(\mathbf{P}-\mathbf{P}')$
normalization for unitarity.  To calculate the Clebsch-Gordan coefficients we 
express the irreducible basis state (\ref{f.19}) in terms of tensor 
products of the single-particle basis states.  

Using (\ref{f.19}) in (\ref{f.22}) gives
\[
\vert k, j (m_1,j_1,m_2,j_2 , l , s_{12}) \mathbf{P} ,  \mu  \rangle_s =
\]
\[
\sum_{\mu_1', \mu_2',\mu_1'', \mu_2'', m, \mu_s}  \int U_1(\Lambda_s (P),0) 
\vert (m_1,j_1) \mathbf{k} ,\mu_1' \rangle_s \otimes 
U_2(\Lambda_s (P),0) \vert (m_2,j_2) -\mathbf{k} ,\mu_2' \rangle_s
\,d\hat{\mathbf{k}}\,   \times 
\]
\[
Y^{l}_{m} (\hat{\mathbf{k}}) 
D^{j_1}_{\mu_1'\mu_1''} [\Lambda_s^{-1} (k_1) \Lambda_c (k_1)]
D^{j_2}_{\mu_2'\mu_2''} [\Lambda_s^{-1} (k_2) \Lambda_c (k_2)] \times 
\]
\[
\langle j_1, \mu_1'', j_2, \mu_2'' \vert s_{12}, \mu_s \rangle
\langle l, m, s_{12}, \mu_s \vert j, \mu \rangle 
\sqrt{\frac{M_0}{\sqrt{M_0^2 +\mathbf{P}^2}}}  =
\]
\[
\sum_{\mu_1 ,\mu_2 ,\mu_1', \mu_2',\mu_1'', \mu_2'', m, \mu_s}  \int  
\vert (m_1,j_1) \mathbf{p}_1 ,\mu_1 \rangle_s \otimes 
\vert (m_2,j_2) \mathbf{p}_2 ,\mu_2 \rangle_s
\,d\hat{\mathbf{k}}\,
Y^{l}_{m} (\hat{\mathbf{k}}) \times
\]
\[
D^{j_1}_{\mu_1 \mu_1'}[R_{ws} (\Lambda_s(P) ,k_1)]
\sqrt{{\omega_{m_1} (\pmb{p_1}) \over \omega_{m_1} (\mathbf{k}) }}
D^{j_2}_{\mu_2 \mu_2'}[R_{ws} (\Lambda_s(P) ,k_2)]
\sqrt{{\omega_{m_2} (\pmb{p}_2) \over \omega_{m_2} (\mathbf{k}) }}
\times
\]
\[
D^{j_1}_{\mu_1'\mu_1''} [\Lambda_s^{-1} (k_1) \Lambda_c (k_1)]
D^{j_2}_{\mu_2'\mu_2''} [\Lambda_s^{-1} (k_2) \Lambda_c (k_2)] \times 
\]
\beq
\langle j_1, \mu_1'', j_2, \mu_2'' \vert s_{12}, \mu_s \rangle
\langle l, m, s_{12}, \mu_s \vert j, \mu \rangle   
\sqrt{\frac{M_0}{\sqrt{M_0^2 +\mathbf{P}^2}}} ~,
\label{f.23}
\eeq
where
\beq
p_i = p_i (P,k_i) = \Lambda_s(P) k_i .
\label{f.24}
\eeq

This equation expresses a two-particle $s$-spin state as a 
linear combination of tensor products of single s-spin states.  
The overlap with the single-particle $s$-spin states gives
\[
_s\langle  (m_1,j_1) \mathbf{p}_1 ,\mu_1  (m_2,j_2) \mathbf{p}_2 ,\mu_2 
\vert k, j (m_1,j_1,m_2,j_2) \mathbf{P} ,  \mu , l , s_{12} \rangle_s =
\]
\[
\sum_{\mu_1',\mu_2',\mu_1'',\mu_2'',\mu_s,m} \int  
\delta (\mathbf{p}_1 - \mathbf{p}_1(P,k))
\delta (\mathbf{p}_2 - \mathbf{p}_2(P,k))
\,d\hat{\mathbf{k}}\,
Y^{l}_{m} (\hat{\mathbf{k}}) \times
\]
\[
D^{j_1}_{\mu_1 \mu_1'}[R_{ws} (\Lambda_s(P) ,k_1)]
D^{j_1}_{\mu_1'\mu_1''} [\Lambda_s^{-1} (k_1) \Lambda_c (k_1)] \times 
\]
\[
D^{j_2}_{\mu_2 \mu_2'}[R_{ws} (\Lambda_s(P) ,k_2)]
D^{j_2}_{\mu_2'\mu_2''} [\Lambda_s^{-1} (k_2) \Lambda_c (k_2)]
\times 
\]
\[
\langle j_1, \mu_1'', j_2, \mu_2'' \vert s_{12}, \mu_s \rangle
\langle l, m, s_{12}, \mu_s \vert j, \mu \rangle \times
\]
\beq
\sqrt{{\omega_{m_1} (\pmb{p_1}) \over \omega_{m_1} (\mathbf{k}) }}
\sqrt{{\omega_{m_2} (\pmb{p}_2) \over \omega_{m_2} (\mathbf{k}) }}
\sqrt{\frac{M_0}{\sqrt{M_0^2 +\mathbf{P}^2}}}.   
\label{f.25}
\eeq
This expression is one form of the Poincar\'e group 
Clebsch-Gordan coefficient {\it in the
$s$-basis}.  Changing variables from $\mathbf{p}_1$ and $\mathbf{p}_2$
to $\mathbf{P}$ and $\mathbf{k}$ inverts all of the Jacobians (square
root factors) and eliminates the angular integral
\[
_s\langle  (m_1,j_1) \mathbf{p}_1 ,\mu_1  (m_2,j_2) \mathbf{p}_2 ,\mu_2 
\vert k, j (m_1,j_1,m_2,j_2) \mathbf{P} ,  \mu , l , s_{12} \rangle_s 
\]
\[
\sum_{\mu_1',\mu_2',\mu_1'',\mu_2'',\mu_s,m}
\delta (\mathbf{P} - \mathbf{p}_1-\mathbf{p}_2)
{\delta (k -k(\mathbf{p}_2,\mathbf{p}_2)) \over k^2}
Y^{l}_{m} (\hat{\mathbf{k}}(\mathbf{p}_1,\mathbf{p}_2) \times
\]
\[
D^{j_1}_{\mu_1 \mu_1'}[R_{ws} (\Lambda_s(P) ,k_1)]
D^{j_1}_{\mu_1'\mu_1''} [\Lambda_s^{-1} (k_1) \Lambda_c (k_1)] \times 
\]
\[
D^{j_2}_{\mu_2 \mu_2'}[R_{ws} (\Lambda_s(P) ,k_2)]
D^{j_2}_{\mu_2'\mu_2''} [\Lambda_s^{-1} (k_2) \Lambda_c (k_2)]
\times 
\]
\[
\langle j_1, \mu_1'', j_2, \mu_2'' \vert s_{12}, \mu_s \rangle
\langle l, m, s_{12}, \mu_s \vert j, \mu \rangle \times
\]
\beq
\sqrt{{\omega_{m_1} (\mathbf{k}) \over \omega_{m_1} (\mathbf{p}_1) }}
\sqrt{{\omega_{m_2} (\mathbf{k}) \over \omega_{m_2} (\mathbf{p}_2) }}
\sqrt{\frac{\sqrt{M_0^2  +\mathbf{P}^2}}{{M_0 }}} .   
\label{f.26}
\eeq
These are the formal expressions for the Poincar\'e group
Clebsch-Gordan coefficients in the $s$-basis.  This construction
is based on our convention that all different types of 
one-body spins are identified in the one-body rest frame and the 
different types of many-body spins are identified in the 
many-body rest frame.  The quantum numbers $l$
and $s_{12}$ are degeneracy quantum numbers that separate different 
irreducible representations with the same mass and spin.

The Poincar\'e group Clebsch-Gordan coefficients have the 
same relations with the Poincar\'e group Wigner functions as
the rotation group Clebsch-Gordan coefficients have 
with the rotation group Wigner functions:
\[
\int \sum_{\mu_1',\mu_2'}   d\mathbf{p}_1' \mathbf{p}_2'
{\cal D}^{m_1,j_1}_{s:\mathbf{p}_1 \mu_1; \mathbf{p}_1' \mu_1'}[\Lambda ,a]
{\cal D}^{m_2,j_2}_{s:\mathbf{p}_2 \mu_2; \mathbf{p}_2' \mu_2'}[\Lambda ,a]
\times
\]
\[
_s\langle  (m_1,j_1) \mathbf{p}'_1 ,\mu'_1  (m_2,j_2) \mathbf{p}'_2 ,\mu'_2 
\vert k, j (m_1,j_1,m_2,j_2) \mathbf{P} ,\mu , l , s_{12} \rangle_s = 
\]
\[
\int \sum_{\mu'} d\mathbf{P}' k^2 dk {} 
_s\langle  (m_1,j_1) \mathbf{p}_1 ,\mu_1  (m_2,j_2) \mathbf{p}_2 ,\mu_2 
\vert k, j (m_1,j_1,m_2,j_2) \mathbf{P}' , \mu' , l , s_{12} \rangle_s
\times 
\]
\beq
{\cal D}^{m(k),j}_{s:\mathbf{P}' \mu'; \mathbf{P} \mu}[\Lambda ,a] .
\label{f.27}
\eeq

If we compare (\ref{f.19}) to (\ref{f.17}) we see that they 
differ by a pair of $SU(2)$ Clebsch-Gordan coefficients. 
It has the same structure as a non-relativistic state where 
two single-particle spins are added to an orbital angular 
momentum to get a total spin.  In the relativistic case
the spins that are added in this way differ from the single
particle $s$-spins by the rotations
\beq
R_{ws} (\Lambda_s(P) ,k_i) \Lambda_s^{-1}(k_i)  \Lambda_c (k_i) =
\Lambda_s^{-1}(p_i) \Lambda_s(P) \Lambda_c(k_i) ~,
\label{f.28}
\eeq
which are the composition of a Melosh rotation (\ref{d.12},\ref{d.50}) 
from the canonical spin to the $s$-spin followed by a Wigner rotation 
(\ref{d.34}) for the
$s$-boost.

It is useful to identify the corresponding relativistic spin operators 
that can be
added, using the ordinary rules of angular momentum addition, to the
orbital angular momentum to get the total two-body spin.  We define
the single-particle $s$-constituent spin operator for particle $i$ by
\beq
\mathbf{j}_{iss} :=  
\Lambda_c^{-1} (k_i)\Lambda_s^{-1} (P) \Lambda_s(p_i) 
\mathbf{j}_{is} = {1 \over m_i} \Lambda_c^{-1} (k_i)\Lambda_s^{-1} (P)
W_i .
\label{f.29}
\eeq
These single-particle constituent spin operators are actually
{\it many-body} operators because they depend on the total momentum of the
system.  In Eq. (\ref{f.29}) the quantities $k_i$, $p_i$,
$P$, $\mathbf{j}_s$, $W$, and $m_i$ are interpreted as operators.  The
transformation $\Lambda_c^{-1} (k_i)\Lambda_s^{-1} (P) \Lambda_s(p_i)$
relating $\mathbf{j}_{ss}$ to $\mathbf{j}_{s}$ is a momentum-dependent
rotation.

The transformation $\Lambda_s(P)\Lambda_c(k_i)$ is a
boost from the rest frame of particle $i$ to its final momentum by
first boosting to the standard frame of the many-body system followed
by a boost to the final momentum of the particle.  It has the same
form as the boost in (\ref{d.3}), $\Lambda_s({P})\Lambda_s^{-1}(p_0)$, 
with the constant boost $\Lambda_s^{-1}(p_0)$ replaced by
$\Lambda_c(k_i)$.  This has the consequence that the constituent spins
are defined so that the zero-momentum vector of the {\it many-body
system} is the standard vector.

The constituent spin operators defined in (\ref{f.29}) have the
property that they can be added to the orbital angular momentum to get
the total $s$-spin of the combined system:
\beq
\mathbf{j}_s = \mathbf{l} + \mathbf{j}_{1ss} + \mathbf{j}_{2ss}
= 
\mathbf{l} + 
\Lambda_c^{-1} (k_1)\Lambda_s^{-1} (P) \Lambda_s(p_1)
\mathbf{j}_{1s} + 
\Lambda_c^{-1} (k_2)\Lambda_s^{-1} (P) \Lambda_s(p_2) \mathbf{j}_{2s} ~,
\label{f.30}
\eeq
where again the Lorentz transformations above are interpreted as matrices of
operators.  These constituent spins add like ordinary non-relativistic
spins, however they differ from the corresponding single-particle
spins by momentum-dependent rotations.  The rotations in Eq.~(\ref{f.30})
can be factored
into the product of a generalized Melosh rotation and an $s$-spin Wigner
rotation:
\[
\mathbf{j}_s = \mathbf{l} +  
[\Lambda_c^{-1}(k_1) \Lambda_s(k_1)][
\Lambda_s^{-1}(k_1)
\Lambda_s^{-1} (P) \Lambda_s(p_1)] \mathbf{j}_{1s}
+
\]
\beq
[\Lambda_c^{-1}(k_2) \Lambda_s(k_2)]
[\Lambda_s^{-1}(k_2)
\Lambda_s^{-1} (P) \Lambda_s(p_2)] \mathbf{j}_{2s} .
\label{f.31}
\eeq
This illustrates how to add single particle spins in a composite
relativistic system.  It is more complicated than the way that they
are added in non-relativistic systems.

For the canonical spins basis $s=c$ the Melosh rotations are the
identity, so the combined rotations in (\ref{f.31}) reduce to a
canonical-spin Wigner rotation. For light-front
spins, since the light-front boosts form a subgroup, the Wigner rotations
of the light-front boosts become the identity and  the
combined rotations in (\ref{f.31}) reduce to a Melosh rotation.  This
is the origin of introducing the Melosh rotation.  For the helicity basis
the Melosh rotation is $R(\hat{\mathbf{z}}\to \hat{\mathbf{p}})$ and
the Wigner rotations are diagonal and only contribute a phase.  For a
general $s$-spins the Clebsch-Gordan coefficients have both
generalized Melosh rotations and $s$-spin Wigner rotations.

It is instructive to examine the transformation properties of the 
constituent $s$-spins under Lorentz transformations.
To do this first note the 
transformation property of  
$k_i := \Lambda^{-1}_s(P)p_i$ is 
\[
U^{\dagger} (\Lambda ,0)k_i U (\Lambda ,0) =
\Lambda^{-1}_s(\Lambda P)\Lambda p_i
= \Lambda^{-1}_s(\Lambda P)\Lambda \Lambda_s(P) \Lambda^{-1}_s(P) p_i
=
\]
\beq
\Lambda^{-1}_s(\Lambda P)\Lambda \Lambda_s(P) k_i  ~.
\label{f.32}
\eeq
This shows that the operators $k_i$ are {\it not} four-vectors; instead they 
Wigner rotate under Lorentz transformations. 
We compare this to the transformation properties 
of $\mathbf{j}_{ss}$ given by (\ref{f.29}):
\beq
U^{\dagger} (\Lambda )\mathbf{j}_{ss} U(\Lambda) =
{1 \over m_i} \Lambda_c^{-1} (
\Lambda_s^{-1}(\Lambda P) \Lambda \Lambda_s (P) k_i) \Lambda_s^{-1} (\Lambda P)
\Lambda W ~.
\label{f.33}
\eeq
Using the property (\ref{e.2}) of canonical boosts, (\ref{f.33}) becomes
\[
{1 \over m_i} 
\Lambda_s^{-1}(\Lambda P) \Lambda \Lambda_s (P)
\Lambda_c^{-1} ( k_i)
\Lambda_s^{-1}(P) \Lambda^{-1} \Lambda_s  (\Lambda P) 
\Lambda_s^{-1} (\Lambda P)
\Lambda W =
\]
\[
{1 \over m_i} 
\Lambda_s^{-1}(\Lambda P) \Lambda \Lambda_s (P)
\Lambda_c^{-1} ( k_i) 
\Lambda_s^{-1} (P)  W =
\]
\beq
\Lambda_s^{-1}(\Lambda P) \Lambda \Lambda_s (P)
\mathbf{j}_{ss} .
\label{f.34}
\eeq
This is identical to the transformation property
(\ref{f.32}) of $k_i$. 
It is precisely because the constituent spins and relative
momentum have the same transformation properties with respect
to rotations that allows them to be combined.  The difference between
the constituent spins and the single-particle spins is that because
the standard vector is the zero-momentum vector of the system, the
Wigner rotations all involve the same boost $\Lambda_s(P)$ rather than 
the different single-particle boosts, $\Lambda_s(p_i)$. 

It is interesting to compare the constituent spin defined in
Eq. (\ref{f.29}) to the spin defined in (\ref{d.3}) and (\ref{d.5})
for the case that $p_s$ is the rest frame of the system:
\beq
\mathbf{j}_{ss} :=  
{1 \over m_i} \Lambda_c^{-1} (k_i)\Lambda_s^{-1} (P) W_i ~,
\label{f.35}
\eeq
\beq 
\mathbf{j}_s = {1 \over {m}_i} 
\Lambda_s(p_0) \Lambda^{-1}_s({P}) {W}_i. 
\label{e.36}
\eeq
We see that both correspond to single-particle spins with 
a {\it standard vector that is different than the single-particle rest vector}.
The only difference is that the boost from the standard frame
to the rest frame, $\Lambda_c^{-1} (k_i)$, involves another variable,
while the corresponding boost $\Lambda_s(p_0)$ involves a constant.

As a final remark, when coupling spins with Poincar\'e group
Clebsch-Gordan coefficients, the observation that the spins must be first
converted to canonical spins in the system rest frame before being
added means that the Clebsch-Gordan coefficients in any $s$-spin basis
are related to the Clebsch-Gordan coefficients in the canonical spin
basis by applying generalized Melosh rotations to the single-particle
spins and the total spin.  Thus the Clebsch-Gordan coefficients in 
different spin bases are related by 
\[
_t\langle  (m_1,j_1) \mathbf{p}_1 ,\mu_1  (m_2,j_2) \mathbf{p}_2 ,\mu_2 
\vert k, j (m_1,j_1,m_2,j_2) \mathbf{P} ,  \mu , l , s_{12} \rangle_t =  
\]
\[
\sum_{\mu_1',\mu_2',\mu'}
D^{j_1}_{\mu_1\mu_1'}[\Lambda_t^{-1}(p_1)
\Lambda_{s}(p_1)] 
D^{j_2}_{\mu_2\mu_2'}[\Lambda_t^{-1}(p_2) \Lambda_{s}(p_2)] \times
\]
\[ 
_s\langle  (m_1,j_1) \mathbf{p}_1 ,\mu_1'  (m_2,j_2) \mathbf{p}_2 ,\mu_2' 
\vert k, j (m_1,j_1,m_2,j_2) \mathbf{P} ,  \mu' , l , s_{12} \rangle_s
\times 
\]
\beq 
D^j_{\mu'\mu}[\Lambda_s^{-1}(P)
\Lambda_{t}(P)]. 
\label{f.37}
\eeq

When the spins are successively added using pairwise coupling in any
basis, the intermediate state Melosh rotations {\it all} identically
cancel.  It follows, for example, if we use $s$-basis Poincar\'e group
Clebsch-Gordan coefficients to successively couple products of
$s$-basis unitary representations of the Poincar\'e group to a direct
integral of $s$-base unitary representations, the results are
identical to what one would get using $c$-basis Poincar\'e group
Clebsch-Gordan coefficients and Melosh rotating the initial and final
spin states to the $s$-basis.  Thus {\it the effect of combining
  spins in composite systems is independent of the choice of spin
  basis up to the initial and final Melosh rotations.}  For example,
this means that the spin structures of composite systems using
canonical spin or light-front spin are identical up to a trivial
overall change of basis.

This means that for systems of free particles there
is no loss in generality in coupling using only canonical spins
(or any other type of spin).

\section{Two and four component spinors}
\label{section7}

In Poincar\'e invariant quantum mechanics spin-$\frac{1}{2}$ particles
are described using two-component spinors while in the Dirac equation
they are described by four-component spinors.  In any experiment there
are only two spin states that can be measured.  In this section we
discuss the relation between these two equivalent treatments of spin.

The difference between these two treatments of spin is that 
the two-component spinor description  uses
irreducible representations of the Poincar\'e group to describe
particles while the four-component spinor description 
 uses finite-dimensional representations of
the Lorentz group.

The connection between these two representations is most easily illustrated
by taking apart a Wigner rotation and absorbing the momentum-dependent
boosts into the state vectors.  For a spin $j$ particle the action of
the Lorentz group on an irreducible $s$-basis state is (\ref{d.38})
\beq
U(\Lambda, 0) \vert (m,j) \mathbf{p}, \mu \rangle_s
 = 
\sum_{\mu'} \vert (m,j) \pmb{\Lambda}p , \mu' \rangle_s 
\sqrt{\frac{\omega_m (\pmb{\Lambda} p)}{\omega_m (\mathbf{p})}} 
D^j_{\mu' \mu}[\Lambda_s^{-1} (\Lambda p)\Lambda \Lambda_s ({p})].
\label{g.1}
\eeq
In what follows we use the fact that the $SU(2)$ Wigner functions,
$D^j_{\mu' \mu}[R]$, which are $2j+1$ dimensional representations of
$SU(2)$ are also $2j+1$ dimensional representations of
$SL(2,\mathbb{C})$ when the $SU(2)$ matrix elements, $R$, are replaced
by the corresponding $SL(2,\mathbb{C})$ matrix elements, $\Lambda$.

To show this first note that the group representation property for the
$D^j_{\mu' \mu}[R]$ can be written as
\beq
0 =\sum_{\mu''=-j}^j D^j_{\mu \mu''}[e^{{i\over 2}\pmb{\theta}_1 \cdot
    \pmb{\sigma}}] 
D^j_{\mu'' \mu'}
[e^{{i\over 2}\pmb{\theta}_2 \cdot \pmb{\sigma}}] - 
D^j_{\mu \mu'}[e^{{i\over 2}\pmb{\theta}_1 \cdot \pmb{\sigma}} 
e^{{i\over 2}\pmb{\theta}_2 \cdot \pmb{\sigma}}].
\label{g.2}
\eeq 
The right hand side of Eq. (\ref{g.2}) is an entire function of
the three components of the two real angles, $\pmb{\theta}_1$ and
$\pmb{\theta}_2$.  This is because the $D^j_{\mu' \mu}[R]$ are
homogeneous polynomials in the matrix elements of $R$ with real
coefficients (\ref{d.21}), so they are entire functions of $R$, and
the $SU(2)$ rotations, $R= e^{{i\over 2}\pmb{\theta} \cdot
\pmb{\sigma}}$, are entire (exponential) functions of the angles.  It
follows that $D^j_{\mu' \mu}[e^{{i\over 2}\pmb{\theta} \cdot \pmb{\sigma}}]$ 
is an entire function of the angles.
Since Eq. (\ref{g.2}) is identically zero for all real
$\pmb{\theta}_1$ and $\pmb{\theta}_2$, by analytic continuation it is
identically zero for all complex angles, $\pmb{\theta}_i \to
\mathbf{z}_i$.  Since the most general $SL(2,\mathbb{C})$ matrix,
$\Lambda = e^{{\mathbf{z} \over 2} \cdot \pmb{\sigma}}$ (\ref{b.15}),
is an analytic continuation, $\pmb{\theta} \to -i \mathbf{z}$, 
to a complex angle of an
$SU(2)$ matrix, $R =e^{i{\pmb{\theta} \over 2} \cdot \pmb{\sigma}}$, it
follows that
\beq
\sum_{\mu''=-j}^j D^j_{\mu \mu''}[e^{{\mathbf{z}_1 \over 2} \cdot \pmb{\sigma}}] D^j_{\mu'' \mu'}
[e^{{\mathbf{z}_2 \over 2} \cdot \pmb{\sigma}}] = 
D^j_{\mu \mu'}[e^{{\mathbf{z}_1 \over 2} \cdot \pmb{\sigma}} 
e^{{\mathbf{z}_2 \over 2} \cdot \pmb{\sigma}}] .
\label{g.3}
\eeq
This shows that $D^j_{\mu' \mu}[\Lambda]$, given by (\ref{d.21}),   
is a $2j+1$ dimensional representation of $SL(2,\mathbb{C})$.
While these representations are irreducible, they are 
no longer unitary.

Using the group representation property (\ref{g.3}) with respect to 
$SL(2,\mathbb{C})$ we can split up the Wigner
function in (\ref{g.1}) into a product of three distinct parts:
\beq
D^j_{\mu' \mu}[\Lambda_s^{-1} (\Lambda p)\Lambda \Lambda_s ({p})] =
\sum_{\mu_1 \mu_2} D^j_{\mu' \mu_1}[\Lambda_s^{-1} (\Lambda p)]
D^j_{\mu_1 \mu_2}[\Lambda ]
D^j_{\mu_2 \mu}[\Lambda_s ({p})] .
\label{g.4}
\eeq
If we use (\ref{g.4}) in (\ref{g.1}) and 
right multiply by the inverse of the last matrix we obtain
\[
\sum_{\mu'} U(\Lambda, 0) \vert (m,j) \mathbf{p}, \mu' \rangle_s 
\sqrt{\omega_m (\mathbf{p})}
D^j_{\mu' \mu}[\Lambda_s^{-1} (p)] 
= 
\]
\beq
\sum_{\mu'\mu''} \vert (m,j) \pmb{\Lambda}p \mu'' \rangle_s 
\sqrt{\omega_m (\pmb{\Lambda} p)} 
D^j_{\mu'' \mu'}[\Lambda_s^{-1} (\Lambda p)]
D^j_{\mu' \mu}[\Lambda] .
\label{g.5}
\eeq
This is completely equivalent to (\ref{g.1}).  This leads us to 
define a {\it Lorentz covariant} basis 
state by 
\beq
\vert (m,j) \mathbf{p}, b \rangle := \sum_{\mu'}
\vert (m,j) \mathbf{p}, \mu' \rangle_s \sqrt{\omega_m (\mathbf{p})}
D^j_{\mu' b}[\Lambda_s^{-1} (p)] .
\label{g.6}
\eeq
Here we use the index notation $b$ to emphasize that it is not a magnetic
quantum number, even though it has $2j+1$ values.  The Hilbert space
resolution of the identity in this representation is 
\[
I = \int \sum_{\mu=-j}^j \vert (m,j) \mathbf{p}, \mu \rangle_s~ d \mathbf{p} ~
_s\langle (m,j) \mathbf{p}, \mu \vert =
\]
\beq
\int \sum_{b,b'=-j}^j \vert (m,j) 
\mathbf{p}, b \rangle {d\mathbf{p} \over \omega 
(\mathbf{p})} D^j_{bb'}[\Lambda_s (p)\Lambda_s^{\dagger} (p) ]   
\langle (m,j) \mathbf{p}, b' \vert  ~.
\label{g.7}
\eeq

The matrix $D^j_{bb'}[\Lambda_s (p)\Lambda_s^{\dagger} (p) ]$ looks like it
depends on $s$, but because $\Lambda_s (p)= \Lambda_c (p) R_{cs}(p)$
(\ref{d.12},\ref{d.50}),
the Melosh rotations, $R_{cs}(p)$, cancel giving
\beq
\Lambda_s (p)\Lambda_s^{\dagger} (p) = 
\Lambda_c (p)\Lambda_c(p)^{\dagger}  =
\Lambda_c^2 (p)
= {1 \over m} p^{\mu} \sigma_{\mu} ~,
\label{g.8}
\eeq
which is a positive (has positive eigenvalues) Hermetian kernel 
(for timelike $p$) that is
independent of $s$.  Here we used the fact that a general boost can be
expressed as a rotation followed by a canonical boost (\ref{b.22}),
the fact the canonical boosts are positive Hermetian $SL(2,\mathbb{C})$
matrices and the identity $\Lambda_c^2 = \cosh (\rho) I +
\hat{\mathbf{p}}\cdot \pmb{\sigma} \sinh (\rho) =
p^{\mu}\sigma_{\mu}$.  Using (\ref{g.8}) the resolution of the
identity (\ref{g.7}) can be expressed in the following manifestly 
covariant form
\beq
I = \int \sum_{bb'}\vert (m,j) \mathbf{p}, b \rangle 
d^4p \theta(p^0) \delta (p^2 + m^2) 
D^j_{bb'}[p^{\mu}\sigma_{\mu}/m ]   
\langle (m,j) \mathbf{p}, b' \vert .
\label{g.9}
\eeq
Wightman \cite{Wightman:1980} uses symmetric tensor products of the
spin $1/2$ representations of this form as representations of the
irreducible representation of the Poincar\'e group.

This means that if we define covariant wave functions
\beq
\psi (p,b) := \langle (m,j) \mathbf{p}, b \vert \psi \rangle 
\label{g.7a}
\eeq
the Hilbert space scalar product is 
\beq
\langle \psi \vert \phi \rangle =
\int \sum \psi^* (p,b) 
d^4p \theta(p^0) \delta (p^2 + m^2) 
D^j_{bb'}[p^{\mu}\sigma_{\mu}/m ]   
\phi (p,b')   
\label{g.7b}
\eeq
and
\beq
U(\Lambda, 0)\vert (m,j) \mathbf{p}, b \rangle = 
\vert (m,j) \pmb{\Lambda}{p}, b' \rangle
D^j_{b' b}[\Lambda] 
\label{g.10}
\eeq
is unitary with respect to the inner product (\ref{g.7b}).  Here the
wave functions are really equivalence classes of functions that agree
on the mass shell.  The presence of non-trivial scalar products is a
generic feature of covariant unitary representations of the Poincar\'e group.
Note the mass is selected by the kernel of inner product, which
carries all of the dynamical information in this representation.  In
quantum field theory the kernel of the non-trivial scalar products are
the Wightman functions (i.e. vacuum expectation values of products of
fields) which also carry all of the dynamical information.

While Eqs. (\ref{g.7}-\ref{g.10}) contain exactly the same
information as (\ref{g.1}), the Wigner function of the little group is
replaced by a {\it momentum-independent} $2j+1$ dimensional
representation of $SL(2,\mathbb{C})$.  The covariant representation has the
advantage that it is independent of the choice of the standard vector
or standard boost that are used in the construction of irreducible
representations of the Poincar\'e group.  The disadvantage is that the
finite dimensional irreducible representations of the Lorentz group
are not unitary and do not admit a linear representation of space
reflection; however the norm associated with the inner product
(\ref{g.7b}) with the non-trivial kernel is non-negative.

To understand origin of the spin doubling in Lorentz covariant
theories recall that (\ref{b.25}) implies 
$R=\sigma_2 R^* \sigma_2$ for any $SU(2)$ rotation $R$.
 Using this identity the Wigner rotation $R$ in (\ref{g.1}) can be
replaced by $\sigma_2 R^* \sigma_2$.  Making this replacement and
repeating steps (\ref{g.4}-\ref{g.7}) we define a new covariant state
\beq
\vert (m,j) \mathbf{p}, \dot{b} \rangle := \sum_{\mu}
\vert (m,j) \mathbf{p}, \mu \rangle_s \sqrt{\omega_m (\mathbf{p})}
D^j_{\mu \dot{b}}[\sigma_2 \Lambda_s^{-1*} (p) \sigma_2]  ~,
\label{g.11}
\eeq
which has the transformation property 
\beq
U(\Lambda, 0) \vert (m,j) \mathbf{p}, \dot{b} \rangle =
\sum_{\dot{b}'} \vert (m,j) \pmb{\Lambda}{p}, \dot{b}' \rangle
D^j_{\dot{b}'\dot{b}}[\sigma_2 \Lambda^* \sigma_2]. 
\label{g.12}
\eeq
In $SU(2)$ $i\sigma_2$ corresponds to a rotation about the $y$ axis 
by $\pi$ so (\ref{g.12}) can equivalently be written as
\beq
U(\Lambda, 0) \vert (m,j) \mathbf{p}, \dot{b} \rangle =
\sum_{\dot{b}'} \vert (m,j) \pmb{\Lambda}{p}, \dot{b}' \rangle
D^j_{\dot{b}'\dot{b}}[R_y(\pi)  \Lambda^* R_y^{-1}(\pi)]. 
\label{g.13}
\eeq 
We use the dot on the $b$ index to distinguish the 
complex-conjugate representation from the original representation.
The relevant observation (see the discussion following (\ref{b.27})) 
is that while the complex-conjugate representation of $SU(2)$ is
equivalent (related by a constant similarity transformation) to
$SU(2)$, this is no longer true for $SL(2,\mathbb{C})$.  The states
(\ref{g.6}) and (\ref{g.11}) transform under
different inequivalent representations of $SL(2,\mathbb{C})$.

From a strictly mathematical point of view it is possible to use
either of the two inequivalent representations, but there is also
physical relation between these two representations.  They are related
by space reflection as discussed in (\ref{b.26}).  The difficulty
arises because $D^j_{bb'}[p^{\mu}\sigma_{\mu}/m ]$ appears in the
kernel of the scalar product, $ d^4p \theta(p^0) \delta (p^2 + m^2)
D^j_{bb'}[p^{\mu}\sigma_{\mu}/m ]$, rather than in the wave function.
This means that space reflection not only transforms the wave
function, it also changes the scalar product by replacing
$D^j_{bb'}[p^{\mu}\sigma_{\mu}/m ]$ by $D^{j}_{\dot{b}
  \dot{b}'}[p^{\mu}\sigma_2 \sigma^*_{\mu}\sigma_2/m ]$ (which is also
positive for timelike $p$).

The simplest way to allow space reflection to be represented by a
linear operator in the Lorentz covariant representation is to use a
Hilbert space representation where the kernel of the covariant 
scalar product contains direct sum of both representations:
\[
d^4p \theta(p^0) \delta (p^2 +
m^2) D^j_{bb'}[p^{\mu}\sigma_{\mu}/m ]\to 
\]
\beq
d^4p \theta(p^0) \delta (p^2 +
m^2) 
\left (
\begin{array}{cc}
D^j_{bb'}[p^{\mu}\sigma_{\mu}/m ] & 0 \\
0 & D^{j}_{\dot{b} \dot{b}'}[p^{\mu}\sigma_2\sigma^*_{\mu}\sigma_2/m ] 
\end{array}
\right ).
\label{g.14}  
\eeq
Then space reflection can be represented by a
linear operator.  This is the origin of the 4-component treatment of
spin $1/2$.

The key observation is the identity 
\beq
\Lambda \Lambda_s (p) = \Lambda_s (\Lambda p)
\Lambda_s^{-1} (\Lambda p) \Lambda \Lambda_s (p) =
\Lambda_s (\Lambda p) R_{ws} (\Lambda ,p) ~,
\label{g.25}
\eeq
which shows that $p$-dependent boosts, $\Lambda_s (p)$, convert Lorentz
transformations, $\Lambda$, to Wigner rotations, $R_{ws} (\Lambda ,p)$.

To relate this to the transformation properties of two-component 
spinors under $SL(2,\mathbb{C})$ we define four different 
types of 2 component spinors 
\beq
\xi_a , \, \xi^a,  \,
\xi_{\dot{a}}, \, \xi^{\dot{a}}.
\label{g.26}
\eeq
These two component spinors are characterized by their $SL(2,\mathbb{C})$ 
transformation properties:
\beq
\Lambda_a{}^b = e^{i \mathbf{z} \cdot \pmb{\sigma}}
\label{g.27}
\eeq
\beq
\Lambda_{\dot{a}}{}^{\dot{b}} = (e^{i \mathbf{z} \cdot \pmb{\sigma}})^*
\label{g.28}
\eeq
\beq
\Lambda^b{}_a = \sigma_2 e^{i \mathbf{z} \cdot \pmb{\sigma}} \sigma_2 
= ((e^{i \mathbf{z} \cdot \pmb{\sigma}})^t)^{-1}
\label{g.29}
\eeq
\beq
\Lambda^{\dot{a}}{}_{\dot{b}} = \sigma_2 (e^{i \mathbf{z} 
\cdot \pmb{\sigma}})^* \sigma_2 =
((e^{i \mathbf{z} \cdot \pmb{\sigma}})^{\dagger})^{-1}
\label{g.30}
\eeq
\beq
\xi_a \to \xi'_a = \Lambda_a{}^b \xi_b
\label{g.31}
\eeq
\beq
\xi_{\dot{a}} \to \xi'_{\dot{a}} = \Lambda_{\dot{a}}{}^{\dot{b}} \xi_{\dot{b}}
\label{g.32}
\eeq
\beq
\xi^a \to \xi^{a\prime}  = \Lambda^a{}_b \xi^b
\label{g.33}
\eeq
\beq
\xi^{\dot{a}} \to \xi^{\dot{a}\prime} = \Lambda^{\dot{a}}{}_{\dot{b}} \xi^{\dot{b}}
\label{g.34}
\eeq
\beq
\xi_a = (\sigma_{2})_{ab} \xi^b \qquad \xi_{\dot{a}} = (\sigma_{2})_{\dot{a}\dot{b}} 
\xi^{\dot{b}} ~.
\label{g.34.a}
\eeq
The reason for introducing the upper- and lower-index 2-component 
Lorentz spinors is that the products
\beq
\sum_a \xi^a \chi_a   
\qquad
\mbox{and} 
\qquad
\sum_{\dot{a}} \xi^{\dot{a}} \chi_{\dot{a}}   
\label{g.35}
\eeq
are Lorentz invariant.  This follows from the identity
$\Lambda^{-1} = \sigma_2 \Lambda^t \sigma_2$ which holds
for any $SL(2,\mathbb{C})$ matrix.  The proof is 
elementary:
\beq
\Lambda^{-1} = e^{-\mathbf{z}\cdot \pmb{\sigma}} =
e^{\mathbf{z}\cdot \sigma_2\pmb{\sigma}^t \sigma_2} =
\sigma_{2} (e^{\mathbf{z}\cdot \pmb{\sigma}})^t\sigma_2
= \sigma_2 \Lambda^t \sigma_2 .
\label{g.36}
\eeq
Using (\ref{g.36}) 
\beq
\sum_a \xi^a \chi_a  \to 
\sum_a \xi^{\prime a}  \chi'_a = 
(\Lambda^{t})^{-1} \xi \cdot \Lambda \chi =
\xi \cdot \Lambda^{-1} \Lambda \chi =
\sum_a \xi^a \chi_a 
\label{g.37}
\eeq
\beq
\sum_{\dot{a}} \xi^{\dot{a}} \chi_a  \to 
\sum_{\dot{a}} \xi^{\prime \dot{a}}  \chi'_{\dot{a}} = 
(\Lambda^{\dagger})^{-1} \xi \cdot \Lambda^* \chi =
\xi \cdot (\Lambda^*)^{-1} \Lambda^* \chi =
\sum_{\dot{a}} \xi^{\dot{a}} \chi_{\dot{a}} . 
\label{g.38}
\eeq
The matrix $\sigma_2$ acts like a metric tensor - it can be
used to raise and lower indices.  The sum over an upper and 
lower undotted or dotted index is Lorentz invariant.

The difference with an ordinary metric is that $\sigma_2$ is
antisymmetric so the invariants $\xi^a \xi_a = \xi^{\dot{a}}
\xi_{\dot{a}}=0$ always vanish.  In the literature $\sigma_2$ is 
sometimes replaced by the real antisymmetric matrix 
$\epsilon = i \sigma_2$ and its inverse $\epsilon^{-1}= 
- \epsilon$, which is the $SL(2,\mathbb{C})$ representation of 
a rotation about the $y$-axis by $\pi$. 

To motivate the choice of the spinor representation of space reflection 
note the four-vector $\X$ transforms like a mixed spin tensor 
\beq
\X \to \X'=\Lambda \X \Lambda^{\dagger} =
(\Lambda \otimes \Lambda^*) \X ~,
\label{g.39}
\eeq
which suggests the notation
\beq
\X_{a\dot{b}} \to \X_{a\dot{b}}' =    
\Lambda_a{}^c \Lambda_{\dot{c}}{}^{\dot{d}} \X_{c\dot{d}} ~,
\label{g.40}
\eeq
where we have assumed that repeated spinor indices are summed from 
1 to 2. Space reflection, given by (\ref{b.26}), is represented by 
\beq
\X \to \X' = \sigma_2 \X^* \sigma_2 =
-(\sigma_2 \otimes \sigma_2) \X^*  ~.
\label{g.41}
\eeq
The Lorentz transformation properties of the reflected 
vector 
$\X'$ are 
\beq
\X' \to \X'' 
=  -(\sigma_2 \otimes \sigma_2) (\Lambda^* \otimes \Lambda) \X^* 
=(\sigma_2 \Lambda^* \sigma_2) \otimes 
(\sigma_2 \Lambda \sigma_2) (-\sigma_2 \otimes \sigma_2) \X^* ~.
\label{g.42}
\eeq
Eq. (\ref{g.42}) shows that the reflected
four-vector $\X'$ transforms like a mixed-spin tensor with upper indices
\beq
X_{a\dot{b}} \to X^{\dot{a}b} .
\label{g.43}
\eeq

To determine the spinor representation of space reflection we note
that a positive energy light-like four vector can be represented as the
tensor product of a two-spinor and its complex conjugate
\beq
X_{a\dot{b}} = \xi_a(\mathbf{x}) \xi^*_{\dot{b}}(\mathbf{x}) ~,
\label{g.44}
\eeq
where 
\beq
\mathbf{x} = {1 \over 2}\mbox{Tr} (\pmb{\sigma} 
\xi_a(\mathbf{x}) \xi^*_{\dot{b}}(\mathbf{x})) 
\eeq
are the space components of the light-like four 
vector.  The space reflection operator on $X$ in (\ref{g.41}) on 
this four vector
is 
\beq
X_{a\dot{b}} \to  \xi_a(-\mathbf{x}) \xi^*_{\dot{b}}(-\mathbf{x}) = 
-(\sigma_2 \xi^*(\mathbf{x})^{\dot a})\otimes (\sigma_2
\xi^{**}(\mathbf{x})^b) ~.
\label{g.45}
\eeq
This is consistent with the following spinor representation of space 
reflection
\beq
\xi_a(\mathbf{x})  \to 
\xi_a(-\mathbf{x})= (i\sigma_2 \xi^*(\mathbf{x}))^{\dot{a}}
\label{g.44a}
\eeq
\beq
\xi_{\dot a}(\mathbf{x})  \to 
\xi_{\dot a}(-\mathbf{x})=
(i\sigma_2 \xi^*(\mathbf{x}))^a.
\label{g.45a}
\eeq
Because space reflection changes a spinor that transforms under
one representation of $SL(2,\mathbf{C})$ to one that transforms
under the conjugate representation it cannot be represented by 
a linear transformation in terms of Lorentz covariant spinors.
The two different kinds of Lorentz covariant spinors are called right and 
left handed spinors because they are related by space reflection. 

In order to represent space reflection by a linear transformation 
it is enough to replace a single spinor by the direct sum of
a right and left handed spinor.  This $4$ spinor has the Lorentz 
transformation properties: 
\beq
\xi \to 
\left (
\begin{array}{c}
\xi_a \\
\chi^{\dot{b}}
\end{array} 
\right ) 
\qquad 
\left (
\begin{array}{c}
\xi \\
\chi
\end{array} 
\right ) 
\to 
\left (
\begin{array}{c}
\xi' \\
\chi'
\end{array} 
\right ) =
\left (
\begin{array}{cc}
\Lambda & 0\\
0 & (\Lambda^{\dagger})^{-1}\\
\end{array} 
\right ) 
\left (
\begin{array}{c}
\xi \\
\chi
\end{array} 
\right ) .
\label{g.46}
\eeq
With this choice both spinors have the same transformation under
$SU(2)$ rotations because $(R^{\dagger})^{-1}=R$.  Space reflection
becomes a linear transformation that interchanges the right and left
handed spinors and multiplies by $\pm i\sigma_2$ (i.e. raises or
lowers the spin indices).   The
new components allow for a linear realization of 
space reflection.

We define the doubled representation of the Lorentz group and 
space reflection operator by
\beq
S(\Lambda) = \left (
\begin{array}{cc}
\Lambda & 0\\
0 & (\Lambda^{\dagger})^{-1}\\
\end{array} 
\right ) 
\qquad 
\cal{P} 
= \left (
\begin{array}{cc}
0 & I \\
I & 0 \\
\end{array} 
\right ). 
\label{g.47}
\eeq
The doubling also occurs for higher spins.  For 
$2(2j+1)$ component spinors the $SL(2,\mathbb{C})$ matrices
$\Lambda$ are replaced by $D^j_{\mu \mu'}[\Lambda]$: 
\beq
S(\Lambda) = \left (
\begin{array}{cc}
D^j[\Lambda] & 0\\
0 & D^j[\Lambda^{\dagger}]^{-1}\\
\end{array} 
\right ) .
\label{g.48}
\eeq

If we restrict $\Lambda$ to $SU(2)$ then  
\beq
S(\Lambda) \to S(R) =  \left (
\begin{array}{cc}
D^j[R] & 0\\
0 & D^j[(R^{\dagger})^{-1}]\\
\end{array} 
\right ) =
\left (
\begin{array}{cc}
D^j[R] & 0\\
0 & D^j[R]\\
\end{array} 
\right ). 
\label{g.49}
\eeq
This means that a four-component  spin-up state is a direct sum of
two identical $SU(2)$ spin up states, similarly for spin down states.

The structure of field operators is based on the dual roles
played by the Lorentz group and little group of the 
Poincar\'e group.  

Fields are operator densities that transform linearly with respect
to a finite dimensional representation $S(\Lambda)$ of the 
Lorentz group. 
\beq
U (\Lambda ,a) \Psi_a (x) U^{\dagger}(\Lambda ,a) =
S(\Lambda^{-1})_{aa'} \Psi_{a'}(\Lambda x+a) .
\label{g.50}
\eeq  

Free fields are linear in operators that create and/or
annihilate particles.  The operator $a^{\dagger}_s(\mathbf{p},\mu)$
creates the one-particle state with $s$-spin, $\vert (m,j)
\mathbf{p},\mu\rangle_s$, out of the vacuum
\beq
a_s^{\dagger}(\mathbf{p},\mu) \vert 0 \rangle =
\vert (m,j) \mathbf{p},\mu\rangle_s .
\label{g.51}
\eeq
The creation operator has the same Poincar\'e transformation 
properties as the single-particle basis states; the spins
transform with a representation of the little group of the Poincar\'e group:
\[
U(\Lambda ,a)a^{\dagger}_s(\mathbf{p},\mu) U^{\dagger}(\Lambda ,a) =
e^{-i \Lambda p \cdot a} 
a^{\dagger}_s(\pmb{\Lambda }{p},\nu)
\sqrt{\omega(\pmb{\Lambda} p)\over \omega(\mathbf{p})} 
D^j_{\nu \mu}(\Lambda_{0s}^{-1} (\Lambda p) \Lambda  \Lambda_{0s} (p)) =
\]
\beq
e^{-i \Lambda p \cdot a} 
\sqrt{\omega(\pmb{\Lambda} p)\over \omega(\mathbf{p})} 
D^j_{\mu \nu}(\Lambda_{0s}^t (p) \Lambda^t \Lambda_{0s}^{-1t} (\Lambda p) ) 
a^{\dagger}_{s}(\pmb{\Lambda }{p},\nu).
\label{g.52}
\eeq
Taking adjoints gives the transformation properties of the
annihilation operator
\[
U(\Lambda ,a)a_s(\mathbf{p},\mu) U^{\dagger}(\Lambda ,a) =
e^{i \Lambda p \cdot a} 
a_s(\pmb{\Lambda }{p},\nu)
\sqrt{\omega(\pmb{\Lambda} p)\over \omega(\mathbf{p})} 
D^{j*}_{\nu \mu}(\Lambda_{0s}^{-1} (\Lambda p) \Lambda  \Lambda_{0s} (p)) =
\]
\beq
e^{i \Lambda p \cdot a} 
D^{j}_{\mu \nu}(\Lambda_{0s}^{-1} (p) \Lambda^{-1}  \Lambda_{0s} (\Lambda p))
a_s(\pmb{\Lambda }{p},\nu)
\sqrt{\omega(\pmb{\Lambda} p)\over \omega(\mathbf{p})} .
\label{g.53}
\eeq
Note that for the equations with the Wigner functions on the left,
the argument of the Wigner function in the creation operator 
is  
$\Lambda_{0s}^t (p) \Lambda^t \Lambda_{0s}^{-1t} (\Lambda p)$
while the argument of the Wigner function in the annihilation 
operator is  $(\Lambda_{0s}^{-1} (p) \Lambda^{-1}  \Lambda_{0s} (\Lambda p))$.
If we use (\ref{g.36}) we have 
\beq
\Lambda_{0s}^t (p) \Lambda^t \Lambda_{0s}^{-1t} (\Lambda p) =
\sigma_2 (\Lambda_{0s}^{-1} (p) \Lambda^{-1} \Lambda_{0s} (\Lambda p)
\sigma_2 .
\label{g.53b}
\eeq

In general the field has a representation of the form
\beq
\Psi_a (x) = \int d \mathbf{p} 
(U_s(\mathbf{p},\mu) a_s(\mathbf{p},\mu) e^{-i \omega(\mathbf{p})t+i 
\mathbf{p}\cdot \mathbf{x}} + 
V_s(\mathbf{p},\mu) b^{\dagger}_s(\mathbf{p},\mu) e^{i \omega(\mathbf{p})t
-i \mathbf{p}\cdot \mathbf{x}}) .
\label{g.54}
\eeq
The structure of the complex coefficients $U_s(\mathbf{p},\mu)$ and 
$V_s(\mathbf{p},\mu)$ are determined by comparing the coefficients of the 
creation and annihilation operator in 
\[
U(\Lambda ,a)\Psi_a(x)  U^{\dagger}(\Lambda ,a) =
S(\Lambda^{-1})_{aa'} \Psi_{a'}(\Lambda x +a) =
\]
\[
\int d \mathbf{p} 
 [ U_s(\mathbf{p},\mu) e^{i \Lambda p \cdot a} 
a_s(\pmb{\Lambda }{p},\nu)
\sqrt{\omega(\pmb{\Lambda} p)\over \omega(\mathbf{p})} 
D^{j}_{\nu \nu'}(R_y(\pi)) \times 
\]
\[
D^{j}_{\nu' \nu''}(\Lambda_{0s}^{-1} (\Lambda p) \Lambda  \Lambda_{0s} (p))
D^{j}_{\nu'' \mu}(R^{-1}_y(\pi)) 
e^{-i \omega(\mathbf{p})t+i \mathbf{p}\cdot 
\mathbf{x}} +  
\]
\beq
V_s(\mathbf{p},\mu) 
e^{-i \Lambda p \cdot a} 
b^{\dagger}_s(\pmb{\Lambda }{p},\nu)
\sqrt{\omega(\pmb{\Lambda} p)\over \omega(\mathbf{p})} 
D^j_{\nu \mu}(\Lambda_{0s}^{-1} (\Lambda p) \Lambda  \Lambda_{0s} (p))
e^{i \omega(\mathbf{p})t-i \mathbf{p}\cdot \mathbf{x}} ]   ~,
\label{g.55}
\eeq
where we have used (\ref{g.53b}) in (\ref{g.55}). 
Comparison of these equivalent expressions,
after the variable change $p' = \Lambda p$ in equation (\ref{g.55}), 
including the Jacobian from the change of variables
in the momentum integral, gives
\beq
S(\Lambda)_{ab}U_b(\mathbf{p}, \mu )\sqrt{\omega (\mathbf{p})}=
U_a(\pmb{\Lambda} p, \nu )\sqrt{\omega (\pmb{\Lambda} p)}
D^j_{\nu\mu}(\Lambda_{0s}^{-1}(\Lambda p)\Lambda \Lambda_{0s}(p))  
\label{g.56}
\eeq
\beq
S(\Lambda)_{ab}V_b(\mathbf{p}, \mu )\sqrt{\omega (\mathbf{p})}=
V_a(\pmb{\Lambda} p, \nu )\sqrt{\omega (\pmb{\Lambda} p)}
D^{j*}_{\nu\mu}(\Lambda_{0s}^{-1}(\Lambda p)\Lambda \Lambda_{0s}(p)) ~.  
\label{g.57}
\eeq
It is useful to define new quantities
\beq
u_a (\mathbf{p}, \mu ) := U_a(\mathbf{p}, \mu )\sqrt{\omega (\mathbf{p})}
\label{g.58}
\eeq
\beq
v_a (\mathbf{p}, \mu ) := V_a(\mathbf{p}, \nu ) D^j_{\nu\mu}(R_y(\pi))
\sqrt{\omega (\mathbf{p})} ~.
\label{g.59}
\eeq
In terms of these new quantities the covariance 
relations take on the form
\beq
S(\Lambda)_{ab}u_b(\mathbf{p}, \mu )=
u_a(\pmb{\Lambda} p, \nu )
D^j_{\nu\mu}(\Lambda_{0s}^{-1}(\Lambda p)\Lambda \Lambda_{0s}(p))  
\label{g.60}
\eeq
\beq
S(\Lambda)_{ab}v_b(\mathbf{p}, \mu )=
v_a(\pmb{\Lambda} p, \nu )
D^j_{\nu\mu}(\Lambda_{0s}^{-1}(\Lambda p)\Lambda \Lambda_{0s}(p)).  
\label{g.61}
\eeq
To determine the $\mathbf{p}$ dependence of 
$u_b(\mathbf{p}, \mu )$ or $v_a(\mathbf{p}, \mu )$ we set 
$p=p_0$, $\Lambda=\Lambda_{0s} (p)$.  In this case
the Wigner rotation is the identity   
\beq
\Lambda_{0s}^{-1}(\Lambda p)\Lambda \Lambda_{0s}(p)\to 
\Lambda_{0s}^{-1}(\Lambda_{0s}(p) p_0)\Lambda_{0s}(p)
\Lambda_{0s}(p_0) =I ~,
\label{g.62}
\eeq
which gives
\beq
u_a(\mathbf{p}, \nu ) = 
S(\Lambda_{0s}(p))_{ab}u_b(\mathbf{0}, \mu )
\label{g.63}
\eeq
\beq
v_a(\mathbf{p}, \nu ) = 
S(\Lambda_{0s}(p))_{ab}v_b(\mathbf{0}, \mu ).
\label{g.64}
\eeq
From these expressions we see that $u_a(\mathbf{p}, \nu )$
and $v_a(\mathbf{p}, \nu )$ are representations of an $s$-boost
from the rest frame, multiplied by a constant matrix that maps the 
2(2j+1) component spinors to (2j+1) component spinors. 
It is instructive to note the similarity with equation (\ref{g.25}), which
also uses a Lorentz boost to intertwine finite-dimensional representations 
of the Lorentz group with unitary representations of the rotation
group.  In the field theory case, when $\Lambda$  is a rotation 
\beq
S(\Lambda) = \left (
\begin{array}{cc}
D^j(\Lambda) & 0\\
0 & D^j(\Lambda^{\dagger})^{-1})\\
\end{array} 
\right )\to 
\left (
\begin{array}{cc}
I & 0\\
0 & I\\
\end{array} 
\right )D^j(R) ,
\label{g.65}
\eeq 
both the left and right handed spinor have identical transformation 
properties and can be factored out.  Similarly the spin operator becomes
\beq
\mathbf{J} = 
-i {d \over d\lambda }\left (
\begin{array}{cc}
D^j(e^{i\lambda \pmb{\sigma}/2}) & 0\\
0 & D^j(e^{i\lambda \pmb{\sigma}/2)}\\
\end{array} 
\right )_{\lambda =0} \to 
\left (
\begin{array}{cc}
I & 0\\
0 & I\\
\end{array} 
\right )(-i {d \over d\lambda }D^j(e^{i\lambda \pmb{\sigma}/2}))_{\lambda =0}.
\label{g.66}
\eeq 

Using these relations we get a standard looking representation
for a free field
\[
\Psi_a (x) = \int {d \mathbf{p} \over \sqrt{\omega (\mathbf{p})}} 
(u_a(\mathbf{p},\mu) a_s(\mathbf{p},\mu) e^{-i \omega(\mathbf{p})t+i 
\mathbf{p}\cdot \mathbf{x}} +
\]
\[ 
v_a(\mathbf{p},\mu) D^{1/2}_{\mu \nu} (R^{-1}_y(\pi))
b^{\dagger}_s(\mathbf{p},\nu) e^{i \omega(\mathbf{p})t
-i \mathbf{p}\cdot \mathbf{x}}) =
\]
\[
\int {d \mathbf{p} \over \sqrt{\omega (\mathbf{p})}} 
(S(\Lambda_{0s}(p))_{ab} u_b(\mathbf{0},\mu) a_s(\mathbf{p},\mu) e^{-i 
\omega(\mathbf{p})t+i 
\mathbf{p}\cdot \mathbf{x}} +
\]
\beq 
S(\Lambda_{0s}(p))_{ab} v_b(\mathbf{0},\mu) D^{1/2}_{\mu \nu} (R^{-1}_y(\pi))
b^{\dagger}_s(\mathbf{p},\nu) e^{i \omega(\mathbf{p})t-i \mathbf{p}\cdot \mathbf{x}}) .
\label{g.68}
\eeq
This has the standard form of a free Dirac field operator, up to
normalization, except normally the $y$-rotation is absorbed in the
definition of the $v_b (\mathbf{0},\mu)$. 

In this section we started from a Poincar\'e covariant description of
a particle as developed in section \ref{section4}, absorbed the
momentum-dependent boosts from the Wigner rotation into the wave
function, doubled the representation of $SL(2,\mathbb{C})$ to
represent space reflection linearly, constructed fields that
transform covariantly under the same doubled representation of the
Lorentz group and arrived at the standard form of a free Dirac field.
The Dirac equation was never used in this derivation, even though the
resulting free field is a solution of the Dirac equation.  The Hilbert
space in this case has scalar product with a momentum-dependent
kernel.  This same construction trivially generalizes to higher spin
fields and states.

The important observation is that Poincar\'e covariant two-component
spinors contain exactly the same information as Lorentz covariant 4-component
spinors.   In the field theory the boosts in (\ref{g.6}) and
(\ref{g.11}) appear in the spinors $u_a(\mathbf{p},\mu)$ and
$v_a(\mathbf{p},\mu)$ which intertwine finite dimensional
representations of the Lorentz group with irreducible representations
of the little group of the Poincar\'e group.  By using the Lorentz
covariant representations all of the dependence on the spin (s)
representation of the particle states disappears.  This is because
$s$-dependence in the creation and annihilation operators cancels with
the $s$-dependence in the coefficient functions $u_a(\mathbf{p},\mu)$
and $v_a(\mathbf{p},\mu)$.

\section{Spin and dynamics}
\label{section8}

The $N$-particle representation of the Poincar\'e group given in
(\ref{f.2}-\ref{f.3}) describes the dynamics of a system of $N$ free particles.
The mass operator for this representation is the invariant mass of
$N$ free particles and the spin is the $s$-spin of $N$ free particles.
These are both functions of the Poincar\'e generators, which are sums
of the one-body generators.

In a dynamical model one expects that both the mass and spin will
be interaction dependent.  This is because the mass and spin operators are
functions of the generators, some of which are interaction-dependent
\cite{Dirac:1949cp} 
in dynamical models.  Because
\beq
M^2 = H^2 -\mathbf{P}^2 
\label{h.1}
\eeq
it is clear that the mass operator acquires an interaction dependence 
through the Hamiltonian.

The $s$-spin (\ref{d.6}) is a function of the mass, $\Lambda_s(P)$, and
the Pauli-Lubanski vector, $W^{\mu}$.  Each of these terms also
involves interactions, and while it is possible to satisfy the
commutation relations with interactions that lead to a non-interacting
spin (they are called generalized Bakamjian-Thomas models
\cite{Bakamjian:1953kh} \cite{Keister:1991sb}), for systems of more
than two particles this condition is not compatible with the
additional requirements imposed by cluster properties of the
generators.  The origin of this problem is the treatment of the
relative orbital angular momentum of two interacting subsystems.  The
dynamical and kinematic masses of these subsystems are different (in
fact they are represented by non-commuting operators) which implies a
dynamical dependence on the relative orbital angular momentum of these
subsystems.  The interaction dependence in the orbital angular
momentum leads to an interaction dependence in the spin.  This leads
to the question of how to understand the relation between the spin of
an interacting system and the spin of the constituent subsystems.

In this section we argue that it is enough to understand how the total
and single-particle spins are related in a non-interacting system.
To establish this result we show that it is always possible to find a
unitary transformation, $A$, that (1) preserves the $S$ matrix and (2)
leads to an equivalent model with a non-interacting spin.  While the
transformed Poincar\'e generators will no longer satisfy cluster
properties, the transformed $S$-matrix must satisfy cluster properties
and in this transformed model the relation between the single-particle
spins and the system spin is the same as it is for a system of $N$
free particles.  Since the $S$-matrix is the only observable, there is
no loss of generality in working with models where the spins are
coupled as they are in a non-interacting relativistic system.

The same unitary transformation must be applied to operators for the
equivalence to hold for matrix elements of operators, like current
matrix elements.  Furthermore, as we have argued in section
\ref{section6}, in the process of adding the spins and angular momenta
with Poincar\'e group Clebsch-Gordan coefficients, the intermediate
generalized Melosh rotations (\ref{d.12},\ref{d.50}) all cancel up to
the overall single-particle and system Melosh rotations, so there is
no loss of generality in using the canonical (or any other type of)
spin to add the single-particle spins and orbital angular momenta to get the
system spins. The generalized Melosh rotations can be used to
transform the system spin and single-particle spins to any 
other type of $s$-spin.

This means that in order to understand the relation between
single-particle spins and the system spin in $S$-matrix or bound
states observables in dynamical models, it is sufficient to study
dynamical models where the system spin is the canonical spin of the
corresponding non-interacting system.  When this system is embedded in
a larger system there will generally be violations of cluster
properties with observable consequences.

To construct the desired unitary transformation we introduce another
function of the Poincar\'e generators that is conjugate to the linear
momentum and commutes with the canonical spin.  We first consider the
case of a single-particle in a canonical spin basis.  In this
representation the desired operator is represented by $\mathbf{X}_c =
i\pmb{\nabla}_p$ where the partial derivative with respect to the
linear momentum is computed by holding the canonical spin constant
(because different spins are related by momentum-dependent Melosh
rotations (\ref{d.12},\ref{d.50}), holding different spins constant
leads to different ``position'' operators).

In this single-particle (irreducible) representation we define the 
operator $\mathbf{X}_{c}$ by the equation
\beq
_c\langle (m,j) \mathbf{p},\mu \vert \mathbf{X}_{c} \vert \psi \rangle 
:= i \frac{\partial}{\partial \mathbf{p}}{}
_c \langle (m,j) \mathbf{p},\mu  \vert \psi
\rangle  .
\label{h.2}
\eeq
This looks like a non-relativistic position operator except in the
relativistic case the partial derivative is computed holding the 
$z$-component $\mu$ of the canonical spin constant.  In addition
it has no simple transformation properties with respect to the
Poincar\'e group.
Since the single-particle representation is irreducible,  
the operator $\mathbf{X}_c$ is expressible 
as a function of the infinitesimal generators.

To determine the relation of $\mathbf{X}_c$ to the infinitesimal
generators we consider the action of Lorentz transformations
on states in this single-particle basis.  Since both boosts and rotations
change the momenta, the operator $\mathbf{X}_c$ will appear in
expressions for both the boost and rotation generators:
\[ 
_c\langle (m,j) \mathbf{p} ,\mu \vert {K}^i \vert \psi \rangle =
\]
\[
-i {\partial \over \partial \rho} 
_c\langle (m,j) \mathbf{p} ,\mu \vert e^{iK_i\rho} \vert \psi \rangle_{\rho=0} =
\]
\beq
-i {\partial \over \partial \rho} (\int \sum_{\mu'} d \mathbf{p}'
{\cal D}^{m,j}_{c:\mathbf{p},\mu;\mathbf{p}',\mu'}
[\Lambda_c (\rho \hat{\mathbf{x}}_i),0)]
 d \mathbf{p}' \psi (\mathbf{p}',\mu'))_{\rho=0}
\label{h.3}
\eeq
\[ 
_c\langle (m,j) \mathbf{p} ,\mu \vert {J}^i \vert \psi \rangle =
\]
\[
-i {\partial \over \partial \theta}
 _c\langle (m,j) \mathbf{p}
,\mu \vert e^{iJ_i \theta } \vert \psi \rangle_{\theta=0} = 
\]
\beq
-i {\partial \over \partial \theta} (\int \sum_{\mu'} d \mathbf{p}'
{\cal D}^{m,j}_{c:\mathbf{p},\mu;\mathbf{p}',\mu'}
[(R (\theta \hat{\mathbf{x}}_i ,0),0)]
d \mathbf{p}' \psi (\mathbf{p}',\mu'))_{\theta=0} .
\label{h.4}
\eeq
Specifically, using the chain rule, the derivatives with respect to the
rapidity or angle can be replaced by derivatives with respect to the
momentum, which in this representation is identified with the operator,
$\mathbf{X}_c$.  Straightforward calculations lead to the following
relations between $\mathbf{X}_c$, $\mathbf{K}$ and
$\mathbf{J}$:
\beq
\mathbf{J} = \mathbf{j}_c + \mathbf{X}_c \times \mathbf{P} 
\label{h.5}
\eeq
and 
\beq
\mathbf{K} = \frac{1}{2} \{ H,\mathbf{X}_c\}+ \frac{1}{M+H} 
(\mathbf{p} \times \mathbf{j}_c)~.
\label{h.6}
\eeq
The second equation can be inverted to  express the
operator $\mathbf{X}_c$ in terms of the Poincar\'e generators:
\beq
\mathbf{X}_c = \frac{1}{2} \{ \frac{1} {H} ,\mathbf{K} \} -
\frac{\mathbf{P} \times (H\mathbf{J} + \mathbf{P} \times \mathbf{K})}{MH(M+H)}.
\label{h.7}
\eeq
The operator $\mathbf{X}_c$ is called the Newton-Wigner
\cite{Newton:1949cq} position operator.  There are similar operators
\cite{Keister:1991sb} that are partial derivatives with respect to
momentum holding various $s$-spins constant.  All of these ``position
operators'' are well-defined functions of the Poincar\'e generators,
but none of them have the physical interpretation of a position
observable.

Equation (\ref{h.7}) leads to the following expression for the canonical
spin $\mathbf{j}_c$
in term of $\mathbf{P}$, $\mathbf{J}$ and $\mathbf{X}_c$.
\beq
\mathbf{j}_c = \mathbf{J} - \mathbf{X}_c \times \mathbf{P}.
\label{h.8}
\eeq
This looks just like the standard non-relativistic expression showing
that the total angular momentum is the sum of an orbital part angular
momentum and a spin.

While we derived these formulas by considering properties of a single
particle, because both $\mathbf{X}_c$ and $\mathbf{j}_c$ are 
functions of the infinitesimal generators, relations (\ref{h.6}) and
(\ref{h.7}) between $\mathbf{X}_c$ and the Poincar\'e generators hold
for any representation of the Poincar\'e group.

For a system of $N$ non-interacting particles equation (\ref{h.8}) 
is replaced by 
\beq
\mathbf{j}_{c0} = \mathbf{J}_0 - \mathbf{X}_{c0} \times \mathbf{P}_0 ~,
\label{h.9}
\eeq
where the $0$ means that the operators are functions of the {\it
non-interacting} generators, which are sums of the single-particle
generators.

Next we consider an interacting system, and to be specific we
assume an instant-form dynamics where both the linear, $\mathbf{P}$,
and angular momentum, $\mathbf{J}$, do not have interactions.
In an instant form dynamics the interactions appear in the 
Hamiltonian and rotationless boost generators.  For a system with 
interactions $\mathbf{X}$ defined by
(\ref{h.7}) becomes interaction dependent due to the interactions in
$H$, $M$, and $\mathbf{K}$ unless we carefully engineer
the interactions to cancel in (\ref{h.7}). This will be done in what
follows.  More generally (\ref{h.8}) implies that the canonical spin
of this system becomes interaction-dependent when $\mathbf{X}_c$ 
is interaction dependent.  When the system is at
rest the orbital angular momentum containing the interaction
dependence disappears - however for a system satisfying cluster
properties, subsystems move relative to each other and the system.  In
these cases the relative  orbital angular momenta of the subsystems
acquire an interaction dependence.  This is the origin of the
interaction dependence of the spins.

The desired unitary transformation is constructed from multichannel
wave operators.  We briefly summarize the construction of these
operators; a more complete discussion can be found in \cite{fcwp}.
Asymptotically a scattering state in a channel $\alpha$ looks like a
number of mutually non-interacting bound clusters.  Each bound cluster
will have a mass, spin, total momentum, and spin projection.
Relativistically these clusters transform like free particles with the
mass and spin of the bound subsystem.  We write these states in the
form 
\beq 
\vert \phi_{\alpha_i, m_i,s_i} \mathbf{p}_i,\mu_i \rangle_c ~,
\label{h.10}
\eeq 
where $\alpha_i$ is a label for the $i^{th}$ bound cluster in
scattering channel $\alpha$.

The vector (\ref{h.10}) can be considered as a mapping from the square
integrable functions of $\mathbf{p}_i$ and $\mu_i$, called ${\cal
  H}_{\alpha_i}$ to the Hilbert space for the particles in the bound
cluster.  Here ${\cal H}_{\alpha_i}$ is a mass $m_i$ spin $s_i$
irreducible representation space for the Poincar\'e group.  The
asymptotic states in a  reaction with $n_{\alpha}$
asymptotic clusters having wave packets $f_i (\mathbf{p}_i,\mu_i)$ have
the form
\beq
\vert \Psi_{\alpha}\rangle = 
\prod_i  
\int d\mathbf{p}_i \sum_{\mu_i} 
\vert \phi_{\alpha_i, m_i,s_i} \mathbf{p}_i,\mu_i \rangle_c  
f_i (\mathbf{p}_i,\mu_i).
\label{h.11}
\eeq
We write (\ref{h.11}) formally as
\beq
\vert \Psi_{\alpha}\rangle := \Phi_{\alpha} \vert f_{\alpha} \rangle ,
\label{h.12}
\eeq
where $\Phi_{\alpha}$ is a mapping, called the channel injection operator,
from the channel Hilbert space
\beq
{\cal H}_{\alpha}:= \otimes {\cal H}_{\alpha_i}
\label{h.13}
\eeq
to the $N$-particle Hilbert space.

The non-interacting dynamics of the bound clusters is given by
the tensor product of the irreducible unitary representations
of the Poincar\'e group associated with mass and spin of each cluster: 
\beq
\langle \mathbf{p}_1, \mu_1, \cdots ,\mathbf{p}_{n_\alpha},\mu_{n_\alpha}
\vert  
U_{\alpha}(\Lambda ,a) \vert f_\alpha \rangle =
\prod  \int \sum {\cal D}^{m_i,j_i}_{\mu_i \mathbf{p}_i ;\nu_i \mathbf{p}_i'}[\Lambda,a] 
d \mathbf{p}_i' \langle \mathbf{p}_i', \nu_i \vert f_{i}\rangle ~,
\label{h.14}
\eeq
which in the notation (\ref{h.12}) becomes 
\beq
\Phi_{\alpha} U_\alpha (\Lambda, a) \vert f_{\alpha} \rangle ~.
\label{h.15}
\eeq
To treat multichannel scattering and bound states on the same footing 
we define the asymptotic Hilbert space as the direct sum of the  
channel spaces, including the $N$-body bound state channels, 
\beq
{\cal H}_{as} := \oplus {\cal H}_{\alpha}. 
\label{h.16}
\eeq

The asymptotic injection operator that maps the asymptotic 
Hilbert space ${\cal H}_{as}$ to the N-particle Hilbert space ${\cal H}$ 
is defined by
\beq
\Phi = \sum_{\alpha} \Phi_{\alpha } 
\label{h.17}
\eeq
and the free asymptotic dynamics by
\beq
\Phi U_{as} (\Lambda ,a) = 
\sum_{\alpha} \Phi_{\alpha} U_\alpha (\Lambda, a) ~.
\label{h.18}
\eeq
In this notation scattering states, $\vert \Psi_{\pm}\rangle$,   
are defined by the strong limits
\beq
\lim_{t \to \pm \infty} 
\Vert U(I,(t,\pmb{0}))\vert \Psi_{\pm}\rangle -
\Phi U_{as}(I,(t,\pmb{0})) \vert f \rangle\Vert=0 ,  
\label{h.19}
\eeq
where $\vert f \rangle$ represents a wave packet in the asymptotic 
Hilbert space, ${\cal H}_{as}$.

Wave operators are mappings from the asymptotic Hilbert space to the 
$N$-particle  Hilbert space defined by 
\beq
\Omega_{\pm} := \lim_{t \to \pm \infty} 
U(I,(-t,\pmb{0}))\Phi U_{as}(I,(t,\pmb{0})).
\label{h.20}
\eeq
The  wave  operators are asymptotically complete when they are unitary 
mappings from ${\cal H}_{as}$ to the $N$-particle Hilbert space
(recall that
the asymptotic space includes system bound states.)  The wave operators are
relativistically invariant when they satisfy
\beq
U(\Lambda, a) \Omega_{\pm}= \Omega_{\pm} U_{as}(\Lambda ,a) .
\label{h.21}
\eeq
Wave operators that do not satisfy these properties are considered
pathological, and in what follows we assume that the wave operators
are both asymptotically complete and relativistically invariant.

The scattering operator is defined as the unitary mapping 
\beq
S = \Omega^{\dagger}_{+}\Omega_{-}
\label{h.22}
\eeq
on ${\cal H}_{as}$.
In an instant-form dynamics $\mathbf{P}=\mathbf{P}_0$.  It follows from 
(\ref{h.21}) that 
\beq 
\mathbf{P}_0 \Omega_{\pm}= \Omega_{\pm} \mathbf{P}_{as} 
\label{h.23}
\eeq
\beq
\mathbf{X} \Omega_{\pm}= \Omega_{\pm} \mathbf{X}_{as}. 
\label{h.24}
\eeq
The first equation means that the mixed-basis matrix elements of the wave 
operators in eigenstates of $\mathbf{P}_0$ and $\mathbf{P}_{as}$
have the form 
\beq
\langle \mathbf{P}, \cdots \vert \Omega_{\pm} \vert \mathbf{P}_{as}, \cdots
\rangle = \delta(\mathbf{P}-\mathbf{P}_{as}) \langle \cdots \vert 
\hat{\Omega}_{\pm}(\mathbf{P}) \vert \cdots \rangle . 
\label{h.25}
\eeq
Equation (\ref{h.24}) means that if $\langle \mathbf{P}, \cdots \vert
= _I\langle \mathbf{P}, \cdots \vert$
are irreducible eigenstates associated with the dynamical
representation $U(\Lambda ,a)$, then the reduced matrix elements
$_I\langle \cdots \vert \hat{\Omega}_{\pm}(\mathbf{P}) \vert \cdots
\rangle $ are independent of $\mathbf{P}$.  

Since in an instant form dynamics $\mathbf{P}_0$ is also the
translation generator for the non-interacting system, if $\langle
\mathbf{P}, \cdots \vert= _0\langle \mathbf{P}, \cdots \vert$ are
irreducible eigenstates associated with the non-interacting
representation $U_0(\Lambda ,a)$, then equation (\ref{h.23}) still
holds, but in this case the reduced matrix elements of the wave
operators will have an explicit momentum dependence.  On the other
hand, since the $S$ matrix only depends on the asymptotic momentum we
have
\[
\langle \mathbf{P}_{as}, \cdots
\vert  S \vert \mathbf{P}'_{as}, \cdots
\rangle =
\]
\[
\delta ( \mathbf{P}_{as}- \mathbf{P}_{as}') \langle \cdots
\vert  \hat{S}(\mathbf{P})  \vert \cdots
\rangle =
\]
\[
\langle \mathbf{P}_{as}, \cdots 
\vert \Omega^{\dagger}_{+} \Omega_{-}  \vert \mathbf{P}'_{as}, \cdots
\rangle =
\]
\[
\int \delta (\mathbf{P}_{as}-\mathbf{P}'_{as})
\langle \cdots \vert \hat{\Omega}^{\dagger}_{+} 
\vert \cdots \rangle_I {} _I \langle \cdots 
\hat{\Omega}_{-}
\vert \cdots \rangle  =
\]
\beq
\int \delta (\mathbf{P}_{as}-\mathbf{P}'_{as})
\langle \cdots \vert \hat{\Omega}^{\dagger}_{+}(\mathbf{P}) 
\vert \cdots \rangle_0 {} _0 \langle \cdots 
\hat{\Omega}_{-}(\mathbf{P}) 
\vert \cdots \rangle   ~,
\label{h.26}
\eeq
where we used both the interacting and free-particle irreducible bases
as intermediate states.  The $S$ matrix elements are independent of the 
choice of basis used in the $N$-particle Hilbert space.  The third line 
of equation (\ref{h.26}) implies that 
$\langle \cdots
\vert  \hat{S}(\mathbf{P})  \vert \cdots
\rangle$ is independent of $\mathbf{P}$:
\[
\langle \cdots \vert \hat{\Omega}^{\dagger}_{+}(\mathbf{P}) 
\vert \cdots \rangle_0 {} _0 \langle \cdots 
\hat{\Omega}_{-}(\mathbf{P}) 
\vert \cdots \rangle = 
\]
\beq
\langle \cdots \vert \hat{\Omega}^{\dagger}_{+}(\mathbf{0}) 
\vert \cdots \rangle_0 {} _0 \langle \cdots 
\hat{\Omega}_{-}(\mathbf{0}) 
\vert \cdots \rangle .
\label{h.27}
\eeq
Given this information we define new wave operators $\bar{\Omega}_{\pm}$ 
in a free-particle irreducible basis by
\beq
_0\langle \mathbf{P}, \cdots \vert \bar{\Omega}_{\pm} 
\vert \mathbf{P}_{as}, \cdots
\rangle = \delta({P}-\mathbf{P}_{as}) _0\langle \cdots \vert 
\hat{\Omega}_{\pm}(\mathbf{0}) \vert \cdots \rangle  ~,
\label{h.28}
\eeq
where we have set $\mathbf{P}$ to zero in the reduced matrix element 
in the mixed representation involving a non-interacting irreducible 
basis and the asymptotic basis.

These new wave operators have the following important properties
\beq
S = \bar{\Omega}^{\dagger}_+ \bar{\Omega}^{\dagger}_- =
{\Omega}^{\dagger}_+ {\Omega}^{\dagger}_- 
\label{h.29}
\eeq
and 
\beq
\mathbf{X}_0 \bar{\Omega}^{\dagger}_{\pm} = \bar{\Omega}^{\dagger}_{\pm}
\mathbf{X}_{as}.
\label{h.30}
\eeq
The unitarity of the wave operators means that 
\beq
A = \bar{\Omega}^{\dagger}_- {\Omega}_- = \bar{\Omega}^{\dagger}_+ {\Omega}_+
\label{h.31}
\eeq
is an $S$-matrix preserving unitary operator.  Using the unitary 
operator (\ref{h.31}) 
we define the 
equivalent dynamical representation of the Poincar\'e group
by
\beq
\bar{U}(\Lambda,a) := A U(\Lambda,a) A^{\dagger}.
\label{h.32}
\eeq
Because the dynamics is instant form we have  
\beq
\mathbf{P}_0 A = A \mathbf{P}_0 \qquad 
\mathbf{J}_0 A = A \mathbf{J}_0
\label{h.33}
\eeq
and by construction 
\beq
\mathbf{X}_0 A = A \mathbf{X} .
\label{h.34}
\eeq
It follows that the transformed canonical spin 
\beq
\bar{\mathbf{j}}_c = \mathbf{J}_0 - \mathbf{X}_0 \times \mathbf{P}_0 =
\mathbf{j}_{c0} 
\label{h.35}
\eeq
has no interactions.  This shows that $\bar{U}(\Lambda,a)$ is a
dynamical unitary representation of the Poincar\'e group that gives
the same $S$-matrix and bound state observables as the original
representation $U(\Lambda,a)$, and in addition has a non-interacting
spin.

This is the desired result.  To see that this also applies to other
forms of the dynamics we note that once we have a mass operator that
commutes with $\mathbf{P}_0$, $\mathbf{X}_{c0}$ and $\mathbf{j}_{c0}$,
the kernel of that operator in an irreducible free particle basis is
the product of three momentum conserving delta functions, a delta
function in the total canonical spin, a delta function in the
$z$-component of the total canonical spin, an a reduced kernel in the
non-interacting mass and kinematically invariant variables.  These
kinematically invariant variables are just the degeneracy variables
that appear in the various Clebsch-Gordan coefficients.  Replacing in the
delta functions linear momentum and canonical spin by the
four-velocity and canonical spin or light-front components of the four
momentum and light-front spin, give $S$-matrix equivalent models in
each of Dirac's forms of dynamics.  A similar construction can be used
to prove the existence of scattering equivalent dynamical models
in each of Dirac's forms of dynamics.

The conclusion of this section is that if one wants to understand the
relation between the spins of single particles and spin of the system
there is no loss of generality with treating the spins as
non-interacting spins.  This provides a justification for a number of
applications of the Bakamjian-Thomas type of dynamics \cite
{Glockle:1986zz} \cite{Witala:2008va} \cite{Lin:2008va}
\cite{Witala:2009zzb} \cite{Elster:2008ca} \cite{Witala:2011va}.

\section{Few-body problems}
\label{section9}

Generalized Bakamjian-Thomas models are a class of relativistic
quantum mechanical models of interacting particles where the spin is
identical to the spin of a system of non-interacting particles.  In
the previous section we demonstrated that any relativistic dynamical
model was related to an equivalent Bakamjian-Thomas model by an
$S$-matrix preserving unitary transformation.  While the equivalent
Bakamjian-Thomas unitary representation of the Poincar\'e group will
not asymptotically break up into tensor-product representations, the
$S$ matrix, which is unchanged from the original model, must satisfy
cluster properties if the original model satisfies cluster properties.
Thus, for the purpose of understanding bound-state or $S$-matrix
observables, there is no loss of generality in using Bakamjian-Thomas
models of the system.  The important consequence is that in these
models the relation of the total spin of a composite system to 
the spins of its constituent particles is identical to that relation for 
$N$ non-interacting  relativistic particles. 

For this reason it is instructive to consider the structure of
Bakamjian-Thomas few-body models.  The important property of this
class of models is that the two and three-body interactions must
commute with the non-interacting three-body spin.  This ensures that
the dynamical spin has no interactions.  The simplest way to realize
this property is to couple the spins and orbital angular momenta using
the Poincar\'e group Clebsch-Gordan coefficients
(\ref{f.25}-\ref{f.26}) to construct a Poincar\'e irreducible
free-particle basis.  In this basis the interactions must be diagonal
in the square of the spin and commute with and be independent of the
magnetic quantum number.

By inspecting the structure of the Poincar\'e group Clebsch-Gordan
coefficients (\ref{f.25}-\ref{f.26}) one can see that the spin is
constructed using ordinary $SU(2)$ Clebsch-Gordan coefficients; but
the angular momenta being added are the constituent spins (\ref{f.29})
  and orbital angular momenta in relative momentum operators
  (\ref{f.32}) that Wigner rotate with the constituent spins
    (\ref{f.29}).  If the potential is expressed in a basis of
    eigenstates of the constituent spins, the projection of these
    spins on an axis, and the orbital angular momentum three vectors
    that Wigner rotate with the constituent spins, then all that is
    required is that the potential be a rotationally invariant in this
    basis.

In this basis the dynamical problem can be solved using standard
methods that take advantage of the rotational invariance; either using
standard partial-waves methods or direct 3-dimensional integration 
in
the same manner that they are used in non-relativistic calculations
\cite{Golak:2010wz} \cite{Glockle:2009ss}.

The relevant momenta and constituent spins variables
are related to the single-particle spins and momenta by boosting 
all of them to the rest frame of the non-interacting 
system, and then converting the resulting $s$ spins to canonical
spins.  The relevant momentum variables (\ref{f.32}) are 
\beq 
\mathbf{q}_i = \pmb{\Lambda}_c^{-1}(P) p_i 
\label{i.1}
\eeq
and the relevant spins (\ref{f.29}) are 
\beq 
\mathbf{j}_{iss} = \Lambda_c^{-1}(q_i)\Lambda_s^{-1}(P)\Lambda_s(p_i) 
\mathbf{j}_{is}    ~.
\label{i.2}
\eeq
These identifications are important for the relativistic
transformation properties in the Bakamjian-Thomas representation.  If
one works in the three-body rest frame then the $\mathbf{q}_i$ are
just the single-particle momenta and the constituent spins become the
single particle constituent spins.  In this frame the relativistic
invariance requirement on the spins reduces to the requirement that
the potentials are rotationally invariant functions of the three
momenta and single-particle canonical spins.  To transform out of the
rest frame it is necessary to make the identifications (\ref{i.1}) and
(\ref{i.2}).

It is interesting to note that the desired rotational covariance can be
realized by treating the spins and orbital angular momenta in a purely
non-relativistic manner in the rest frame, constructing two-body
relative momenta using Galilean boost to two body-rest frames.  It is
equally possible to realize the rotational invariance by adding the
spins and orbital angular momenta using successive coupling with the
Poincar\'e group Clebsch-Gordan coefficients of section \ref{section6}.  In
both cases, if one starts with the momenta and spins in (\ref{i.1})
and (\ref{i.2}), these choices amount to a variable change. 
These choices have nothing to do with relativity - they are 
simply alternative variable choices that make the rotational 
invariance of the interactions easy to recognize.
 
This is consistent with  the observation that the only symmetry
that needs to be respected in the rest frame is the symmetry 
associated with the little group, which is the rotation group for 
positive mass systems.  

However, there are other considerations that go beyond the Poincar\'e
symmetry.  Most notably are cluster properties.  Cluster properties
provide the justification for tests of special relativity on isolated
subsystems.  In the three-body problem it is natural  first to treat
the two-body problem using the Bakamjian-Thomas method.  A spectator
particle can be included by taking the tensor product of the two-body
Bakamjian-Thomas representation of the Poincar\'e group with the
one-body irreducible representation associated with the spectator.
The resulting tensor-product unitary representation of the Poincar\'e
group does not have a kinematic spin, but it does satisfy cluster
properties.  A scattering equivalent three-body Bakamjian-Thomas model
is obtained by considering this model in the non-interacting
three-body rest frame, replacing all of the single-particle momenta
and spins by the momenta (\ref{i.1}) and constituent spins
(\ref{i.2}).  This implies a specific and simple relation between the
two-body Bakamjian-Thomas interactions in the two-body problem and the
corresponding Bakamjian-Thomas interactions in the three-body problem.
This connection is realized by embedding the two-body interactions in
the three-body problem using Poincar\'e group Clebsch-Gordan
coefficients.  While similar considerations apply to larger systems,
for these systems the equivalent Bakamjian-Thomas model that satisfies
$S$-matrix cluster properties necessarily includes many-body
interactions that are generated from the subsystem interactions \cite{fcwp}.

\section{Coupling to electromagnetic fields} 
\label{section10}

Given all of the different kinds of spin operators introduced in this
paper, one has to confront the question of relating theory to
experiment.  Normally the spin is measured by considering how it
couples to a classical electromagnetic field.  Formally, in the
one-photon-exchange approximations this involves a coupling of the
form
\beq
e \int d\mathbf{x} J^{\nu} (x) A_{\mu}(x) d \mathbf{x}.
\label{j.1}
\eeq
The connection with the theory is through matrix elements of the 
current of the form
\beq
_s\langle (m', j') \mathbf{p}',\mu' ,\cdots \vert J^{\mu}(x) \vert 
(m, j) \mathbf{p},\mu ,\cdots \rangle_s  .
\label{j.2}
\eeq
The Poincar\'e transformation properties of this matrix element means that 
it can be expressed in terms of invariants and geometric quantities that
arise strictly from the transformation properties of the current and 
initial and final states.  Once this operator is known in one basis 
the relations between the different bases discussed in this paper
can be used to calculate the current in any other basis.  
It is only necessary to know the generalized Melosh rotations relating 
two different spin bases.  Thus using (\ref{d.12},\ref{d.50}) we get
\[
_s\langle (m', j') \mathbf{p}',\mu' ,\cdots \vert J^{\mu}(x) \vert 
(m, j) \mathbf{p},\mu ,\cdots \rangle_s = 
\]
\[
\sum D^{j}_{\mu'\mu''} [\Lambda_s^{-1} (p') \Lambda_t (p')]
_t\langle (m', j') \mathbf{p}',\mu'' ,\cdots \vert J^{\mu}(x) \vert 
(m, j) \mathbf{p},\mu''' ,\cdots \rangle_t \times
\]
\beq
D^{j}_{\mu'''\mu} [\Lambda_t^{-1} (p) \Lambda_s (p)] .
\label{j.3}
\eeq

\section{Summary}
\label{section11}

In this paper we presented a general discussion of the treatment of
spin in relativistic few-body systems.  The goal of this work was to
understand the relation between the spin of a dynamical system and
the spin of its elementary constituents. This is relevant for
understanding scattering experiments where, for example, a polarized
target breaks up into constituents and one is interested in the
relation of the polarization of the target to the polarization of the
constituents. Other examples involve using electromagnetic probes that
interact with the currents of the individual constituent particles. Our
intention is to include sufficient generality so models with different
 treatments of spin can be compared.

There are many good references on single-particle spins for
relativistic systems, and also many references on Clebsch-Gordan
coefficients for the Poincar\'e
group \cite{coester}\cite{moussa}\cite{Keister:1991sb}, which can
be used to add spins and orbital angular momenta in relativistic
systems, but most of them focus on the canonical spins, and are
relevant for a system of two free particles. This work discusses the
addition of a more general class of spins along with the impact of the
dynamics on the spin coupling.  We also discussed the connection
between two and four component spinors in this context.

The new feature of spin in relativistic quantum mechanics is that
spins undergo momentum-dependent rotations under the action of Lorentz
transformations.  This means that there is no unambiguous way to
compare the spins of particles with different momenta and the addition
of spins becomes more complicated than it is in non-relativistic
quantum mechanics.  In section \ref{section4} we pointed out that one way to
define a spin operator is to use an arbitrary but fixed set of
Lorentz transformations to refer the particles to a common frame where
the spins can be compared.  We constructed a number of functions of
the single-particle Poincar\'e generators corresponding to different 
arbitrary but fixed Lorentz transformations and showed that the resulting 
spin operators all satisfied $SU(2)$ commutation relations.  We also 
showed that the different choices of spin operators were related by 
momentum-dependent rotations, which we called generalized Melosh rotations.
The exercise is not academic - at least three different kinds of spins are
commonly used in applications.  These include the canonical spin, the 
light-front spin, and the helicity spin and they are all related by 
different generalized Melosh rotations. 
     
We then showed that the canonical spin played a special role in adding
spins.  This is because only for the canonical spins are Wigner
rotations of rotations  equal to the original rotation. This
means that particles with different momenta have identical rotational
properties when rotated  which allowed them to be added using
ordinary $SU(2)$ Clebsch-Gordan coefficients.  The coupling
coefficient for other types of spins are constructed by first using
generalized Melosh rotations to convert to canonical spins.  Next the
canonical spins are added using $SU(2)$ Clebsch Gordan coefficients,
and finally the resulting canonical spin is converted back to the
initial type of spin using another generalized Melosh rotation.  The
different generalized Melosh rotations used in  this construction
involve different momenta (i.e the momentum of each particle and the
total momentum of the subsystem).  We also remarked that in the
process of successive pairwise coupling all of the intermediate
generalized Melosh rotations cancel.  All that remains are the
generalized Melosh transformations on the single-particle spins and
the final total spin.  This led us to point out that there is no loss
of generality in performing all of the spin additions using canonical
spins.  The resulting coupling coefficients can then be converted to
coupling coefficients for any other type of spin using the appropriate
generalized Melosh rotation.  An important observation 
resulting from this construction is that there are a number of intermediate
spins that couple to the final total spin using ordinary SU(2)
Clebsch-Gordan coefficients.  We called these spins constituent spins.
It is important to note  that the constituent spins are actually
many-body operators that are related to the true single-particle spins
by dynamical rotations (both Winger rotations and generalized Melosh
rotations).  The angles of these rotations depend on the momentum 
distribution of the composite system as well as on  the total momentum 
of the system.  
 
All of this discussion assumed that all the spins are  associated with a
non-in\-te\-ra\-cting systems of particles.  For interacting systems the
internal orbital angular mo\-me\-nta associated with subsystems depend on
the mass eigenvalues of the subsystems, rather than the invariant mass
of the constituents in each subsystem.  This would suggest that
modifications are required to couple the particle spins and internal
orbital angular momenta in interacting systems.  In section \ref{section8} we
argued that this was not the case.  We showed that is was always
possible to find an S-matrix preserving unitary transformation that
removes the interactions from the spin at the expense of modifying the
internal momentum distribution of the wave function.  In general we do
not know the momentum distribution of the wave function without doing
a full dynamical calculation, however it follows that there is no loss
of generality in coupling the spins, treating them all as
kinematic quantities. Quantitative predictions will be sensitive to
the momentum distribution in the wave functions due to the presence of
dynamical rotations in the spin and orbital angular momentum
coupling coefficients.

We also considered the choices of vectors that should be used to
describe the internal relative orbital angular momenta for systems of
particles.  The most important requirement is that they must be defined by
boosting the single particle momenta to a common frame - normally the
rest frame of the non-interacting system.  The resulting vectors are
no longer 4-vectors, but they have the desirable property that they
all can be Melosh rotated (if necessary) so they undergo the same
Wigner rotations as the constituent spins.  This allows them to be
coupled with the constituent spins using ordinary Clebsch-Gordan
coefficients to get the total spin.  When everything is expressed in terms
of these momentum vectors and the corresponding constituent spins 
the coupling proceeds as in the non-relativistic case.  

In relativistic quantum theory both two and four component spinors
arise in applications.  In section \ref{section7} we pointed out that two
component spinors arise by considering positive mass-positive energy
irreducible representations of the Poincar\'e group while
four-component spinors are associated with finite-dimensional
representations of the Lorentz group.  We demonstrated the relation
between these two groups by taking apart a Wigner rotation, thus
removing the momentum-dependent boosts.  The resulting spin no longer
depends on the choice of boost, but because the $SL(2,\mathbb{C})$
representation of the boosts and their complex conjugates are
inequivalent, and both representations are related by space reflection,
it is natural to use a doubled representation when space reflection is
an important symmetry.  In making contact with the particle spins the
boosts must be reintroduced - this choice appears in both the Dirac
spinors and the creation and annihilation operators.

This work was supported in part by the U.S. Department of Energy, under 
contract DE-FG02-86ER40286 and by the Polish National Science Center 
under Grant No. DEC-2011/01/B/ST2/00578.

\end{document}